\documentclass[a4paper,11pt,oneside,titlepage,openany]{book}
\usepackage[numbers,square,comma,sort&compress]{natbib}
\usepackage[intlimits]{amsmath}
\usepackage{amssymb}
\usepackage{graphicx}
\usepackage{slashed}
\usepackage[colorlinks=true]{hyperref}
\usepackage{fancyhdr}
\usepackage{subfigure}
\usepackage{bibunits}
\usepackage{bibentry}

\nobibliography* 
\newcommand{\beq}{\begin{equation}}
\newcommand{\eeq}{\end{equation}}
\newcommand{\ii}{\mathrm{i}}
\DeclareMathOperator{\Tr}{Tr}
\DeclareMathOperator{\Log}{Log}
\newcommand{\dotp}{\cdot\!}

\newcommand{\fig}[1]{\ref{fig:#1}}
\newcommand{\eq}[1]{(\ref{eq:#1})}
\newcommand{\mg}{\ensuremath{\mathcal G}}

\newcommand{\xx}[4]{\left|\chi_{#1}({#3},{#2})\right\rangle \left\langle\chi_{#1}({#4},{#2})\right|}
\newcommand{\gga}[4]{\left|\Gamma_{#1}(#3,#2)\right\rangle\left\langle\Gamma_{#1}(#4,#2)\right|}
\newcommand{\gar}[3]{\left|\Gamma_{#1}(#3,#2)\right\rangle}
\newcommand{\gal}[3]{\left\langle\Gamma_{#1}(#3,#2)\right|}
\newcommand{\xr}[3]{\left|\chi_{#1}(#3,#2)\right\rangle}
\newcommand{\xl}[3]{\left\langle\chi_{#1}(#3,#2)\right|}

\newcommand{\dip}{\ensuremath{D(p,P)}}
\newcommand{\diq}{\ensuremath{D(q,P)}}
\newcommand{\glp}{\ensuremath{\left\langle\Gamma(p,P)\right|}}
\newcommand{\grp}{\ensuremath{\left|\Gamma(p,P)\right\rangle}}

\newcommand{\grq}{\ensuremath{\left|\Gamma(q,P)\right\rangle}}
\newcommand{\xlp}{\ensuremath{\left\langle\chi(p,P)\right|}}
\newcommand{\xrp}{\ensuremath{\left|\chi(p,P)\right\rangle}}

\newcommand{\gilp}{\ensuremath{\left\langle\Gamma_i(p,P)\right|}}
\newcommand{\girp}{\ensuremath{\left|\Gamma_i(p,P)\right\rangle}}
\newcommand{\gilq}{\ensuremath{\left\langle\Gamma_i(q,P)\right|}}
\newcommand{\girq}{\ensuremath{\left|\Gamma_i(q,P)\right\rangle}}
\newcommand{\xilp}{\ensuremath{\left\langle\chi_i(p,P)\right|}}
\newcommand{\xirp}{\ensuremath{\left|\chi_i(p,P)\right\rangle}}
\newcommand{\xilq}{\ensuremath{\left\langle\chi_i(q,P)\right|}}

\newcommand{\xg}{\ensuremath{\left\langle\chi_j(q,P)|\Gamma_i(q,P)\right\rangle}}
\newcommand{\xilqi}{\ensuremath{\left\langle\chi_i(q,P_i)\right|}}
\newcommand{\xirpi}{\ensuremath{\left|\chi_i(p,P_i)\right\rangle}}

\fancypagestyle{plain}{%
\fancyhf{} % clear all header and footer fields
\fancyfoot[RO]{\thepage}

}

\title{Properties of quarks and mesons in the Dyson-Schwinger/Bethe-Salpeter approach}
\author{Martina Blank}

\begin{document}
% further preliminaries
\definecolor{darkred}{rgb}{0.5,0,0}
\definecolor{darkblue}{rgb}{0,0,0.5}
\hypersetup{linktocpage=true,linkcolor=darkred, urlcolor=darkblue,citecolor=darkred}

\setlength{\headheight}{25.16252pt}

%end of preliminaries
\frontmatter
\thispagestyle{empty}
\begin{titlepage}
\begin{center}
\LARGE{\bfseries Martina Blank\\
\vspace*{0.5cm}
\LARGE{Properties of quarks and mesons in the Dyson-Schwinger/Bethe-Salpeter approach}\\
\vspace*{1.5cm}
\LARGE Dissertation}\\
\vspace*{1cm}
\Large zur Erlangung des akademischen Grades \\
\Large Doktorin der Naturwissenschaften (Dr.rer.nat.)\\
\vspace*{1.3cm}
\includegraphics[width=4cm]{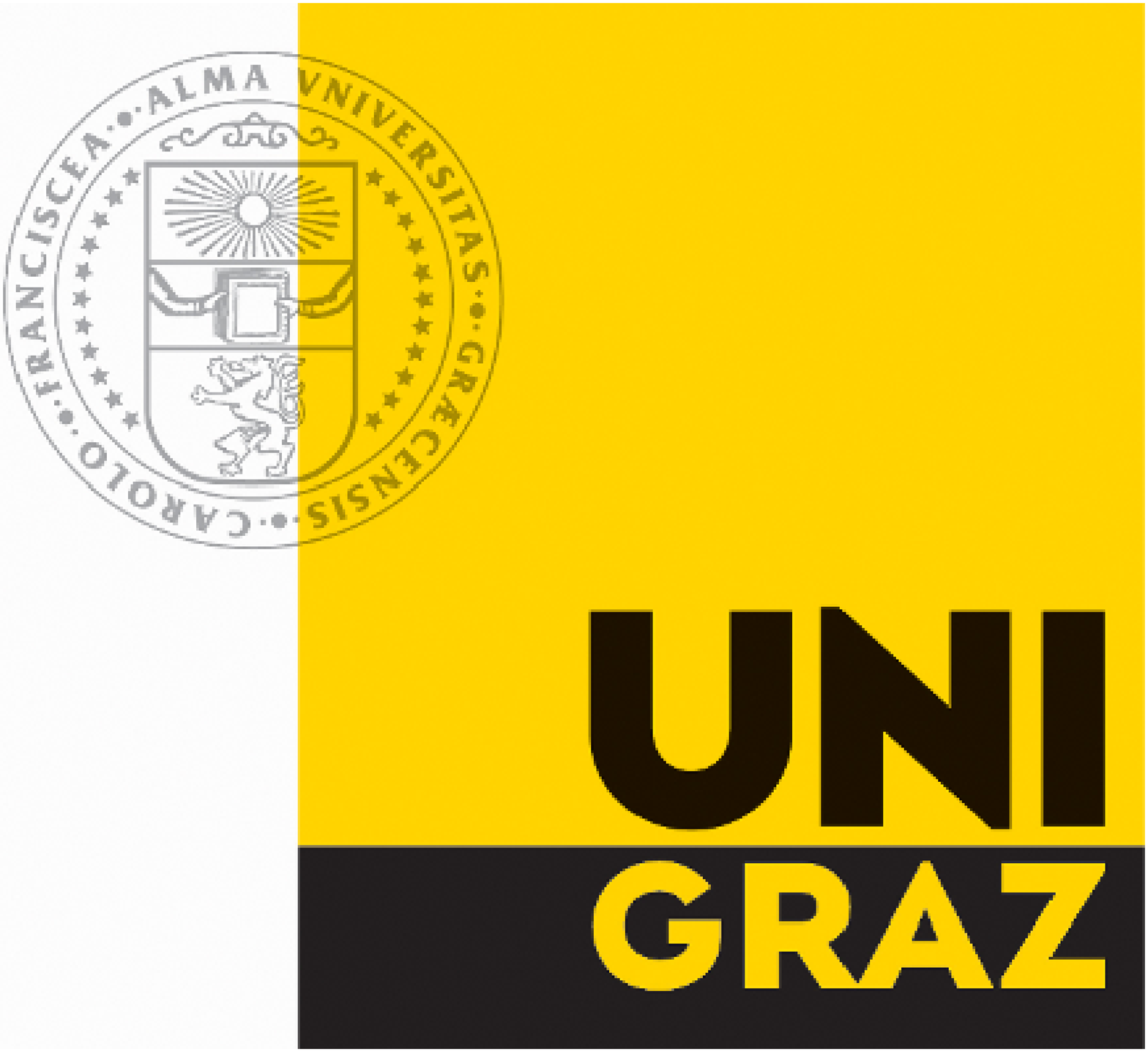}\\
\Large  {\bfseries{Karl-Franzens Universit\"at Graz}}\\
\vspace*{1.5cm}
\Large verfasst am Institut f\"ur Physik\\
\vspace*{0.5cm}
\Large Betreuer:\\
\Large Priv.-Doz.~Mag.~Dr.~Andreas Krassnigg\\
\vspace*{1.5cm}
\Large Graz, 2011
\end{center}
\end{titlepage}

\chapter*{Abstract}
In this thesis, the Dyson-Schwinger - Bethe-Salpeter formalism is investigated and used to study the meson spectrum at zero temperature, as well as the chiral phase transition in finite-temperature QCD.

First, the application of sophisticated matrix algorithms to the numerical solution of both the homogeneous Bethe-Salpeter equation (BSE) and the inhomogeneous vertex BSE is discussed, and the advantages of these methods are described in detail.

Turning to the finite temperature formalism, the rainbow-truncated quark Dyson-Schwinger equation is used to investigate the impact of different forms of the effective interaction on the chiral transition temperature. A strong model dependence and no overall correlation of the value of the transition temperature to the strength of the interaction is found. Within one model, however, such a correlation exists and follows an expected pattern. 

In the context of the BSE at zero temperature, a representation of the inhomogeneous vertex BSE and the quark-antiquark propagator in terms of eigenvalues and eigenvectors of the homogeneous BSE is given. 
Using the rainbow-ladder truncation, this allows to establish a connection between the bound-state poles in the quark-antiquark propagator and the behavior of eigenvalues of the homogeneous BSE, leading to a new extrapolation technique for meson masses. This is used to study the ground- and excited-state meson spectrum for all quark masses from light to bottom, for pseudoscalar, scalar, vector, axialvector and tensor mesons. Good agreement with experiment is found, e.g., for all ground states in the bottomonium system.

In addition, new applications of the inhomogeneous vertex BSE, such as the possibility to calculate on-shell quantities like decay constants, are investigated. Finally, we study the influence of the infrared behavior of the effective interaction on properties of $\pi$ and $\rho$ mesons.

\chapter*{Notes on the eprint-version}
This PhD-thesis has been submitted to the University of Graz in April 2011.  In this eprint-version, the list of references was updated, minor typing-errors corrected, and the readability of the text was improved with respect to the submitted version. In addition, the format has been slightly altered. The original can be retrieved from the library of the University of Graz in electronic form.

\tableofcontents

\mainmatter

\chapter{Introduction}\label{chap:intro}
Within the standard model of elementary particle physics, the strong interaction is described by Quantum Chromodynamics (QCD). This relativistic quantum field theory of quarks and gluons is a gauge theory with gauge group $SU(3)$, such that the elementary degrees of freedom carry a quantum number termed `color' \cite{Fritzsch:1973pi}, that can take three values.

Some the most noteworthy features of QCD are \emph{asymptotic freedom} \cite{Gross:1973id,Politzer:1973fx}, \emph{dynamical breaking of chiral symmetry} \cite{Nambu:1961tp}, and \emph{confinement}. Asymptotic freedom means that at high energies the coupling of the theory becomes small, and perturbation theory is applicable. At low energies, where the coupling is large, QCD phenomenology is governed by the dynamical breaking of chiral symmetry which is responsible for the masses of the light hadrons. Hadrons, which appear in the framework of QCD as bound states of quarks, antiquarks, and gluons, are the only observed particles that interact strongly. This is usually referred to as confinement, and it is one of the most interesting problems in QCD, especially since an exact proof is still missing (for recent reviews, see \cite{Greensite:2011zz,Alkofer:2006fu}).

Due to confinement, the low-energy spectrum of QCD consists solely of bound states. This, together with the large coupling, necessitates the use of non-perturbative methods for any calculation of the physical spectrum of QCD. Two main branches have developed: \emph{Lattice QCD} and \emph{functional approaches}.

In the case of Lattice QCD, a discretization of space-time leads to a regularization of the (Euclidean) path integral \cite{Wilson:1974sk}, such that observables like hadron masses can be calculated more or less directly. While these ab-initio calculations give a good description of the hadron spectrum (see, e.g., \cite{Durr:2008zz}), it is non-trivial to study the properties of the elementary degrees of freedom, quarks and gluons, in this approach. In particular, the theoretically interesting limits of very small and very large momenta are difficult to access.

On the  other hand, one may use a so-called \emph{functional approach} and work directly with the Green functions of the elementary degrees of freedom. They appear as solutions to the Dyson-Schwinger \cite{Dyson:1949bp,Schwinger:1951ex} or Functional Renormalization Group (FRG) \cite{Wilson:1971bg,Stueckelberg:1953dz,Wetterich:1989xg} equations; if all Green functions were known, the theory would be solved. However, the Dyson-Schwinger equations (DSEs) and FRG equations form an infinite tower of equations, and analytic statements about the solutions can only be made in special cases, e.g., in the limit of very small momenta (infrared analysis, see e.g.~\cite{vonSmekal:1997is}).

In this thesis, we use a functional approach to QCD based on Dyson-Schwinger equations, which are perfectly suited to study the properties of quarks and gluons in a nonperturbative framework. However, in order to provide a consistent description of QCD phenomenology, hadronic bound states have to be considered as well. They can be included via the so-called Bethe-Salpeter equation (BSE), which was proposed as a natural extension of the Green function formalism to bound states \cite{Bethe:1951bs,Salpeter:1951sz,Gell-Mann:1951rw}. As outlined in Chap.~\ref{chap:formalism}, the BSE and the DSEs can be derived consistently from the generating functionals of QCD, although it is interesting to note that the application of the BSE to (mesonic) bound states of the strong interaction \cite{Smith:1969az} predates the formulation of QCD. 

Practical calculations, however, require sophisticated numerical methods, and the solution strategies and algorithms used in subsequent chapters are discussed in Chap.~\ref{chap:numerics}. Chap.~\ref{chap:fint} is devoted to an application of the DSE formalism to the restoration of chiral symmetry at finite temperature, and in Chap.~\ref{chap:mesons} we use the same framework to study meson phenomenology. Various aspects of the different versions (homogeneous and inhomogeneous) of the Bethe-Salpeter equation are discussed, which allow a numerical calculation of meson ground states as well as excitations. Furthermore, the influence of the infrared behavior of ghost- and gluon propagators on meson phenomenology is studied. Chap.~\ref{chap:summary} summarizes the results and provides an outlook.

Part of the material contained in this thesis is available in the following articles:\\[6pt]
\begin{bibunit}[phd-doi-new]
\noindent\bibentry{Blank:2010bz}\\[6pt]
\noindent\bibentry{Blank:2010pa}\\[6pt]
\noindent\bibentry{Blank:2010sn}\\[6pt]
\noindent\bibentry{Blank:2010bp}\\[6pt]
\noindent\bibentry{Krassnigg:2010mh}
\end{bibunit}

\chapter{Formalism}\label{chap:formalism}
The aim of this thesis is to study quarks and mesons in the framework of nonperturbative QCD in a functional approach. The tools employed are the (quark) Dyson-Schwinger equation and the Bethe-Salpeter equations for bound states, which --- although originally derived in different settings --- can be treated in a consistent way, as demonstrated in the present chapter.

In the following, we sketch how the Dyson-Schwinger equations (DSEs) for one-particle irreducible Green functions and the Bethe-Salpeter equation (BSE) for the propagator of a quark-antiquark system, via two-particle irreducibility, rigorously follow from the underlying quantum field theory of QCD. On the same footing, starting from the appropriate generating functionals, it is also possible to obtain a consistent truncation scheme of the quark Dyson-Schwinger equation and the BSE, thus allowing for numerical investigations while still preserving the most important symmetries of the theory. In this work, we distinguish three types of Bethe-Salpeter equations: the BSE for the quark-antiquark propagator; the homogeneous BSE; the inhomogeneous Bethe-Salpeter equation for a vertex that connects quark and antiquark to a color-singlet current (vertex BSE). The derivation of the homogeneous BSE and the vertex BSE from the BSE for the quark-antiquark propagator is demonstrated in Secs.~\ref{sec:hom_bse} and \ref{sec:vertex_bse}, respectively. 

\section{One- and two-particle irreducibility}\label{sec:2pi}
This section briefly reviews some of the basic points which are useful for the derivations given later on in this chapter. For a more complete and pedagogical account on generating functionals and the steps leading to one-particle irreducibility and also to Dyson-Schwinger equations, the reader may consult Refs.~\cite{Alkofer:2000wg,Lichtenegger:2010ph}. For an introduction to two-particle irreducibility, see for example Ref.~\cite{Cvitanovic:1983aa}.

In order to introduce the concepts of one- and two-particle irreducibility (often referred to as 1PI and 2PI) in the context of the path-integral quantization of a field theory one has to start with its basic building blocks, the generating functionals. Formulated in Euclidean space, the generating functional of the full Green functions of QCD is
\beq\label{eq:Z[J]}
Z[J]=\mathcal{N}\int \mathcal{D}\phi\: \exp{(-S_{QCD}[\phi]+\phi\dotp J)}\;,
\eeq
where $\phi=\{\bar{\psi},\psi,A_\mu,\bar{c},c\}$ represents all (fermion, gauge boson, and ghost) fields of the theory. We consider the theory in Euclidean space, and the action $S_{QCD}[\phi]$ of QCD is given by
 \begin{equation}\label{eq:Sqcd}
S_{QCD}[\bar{\psi},\psi,A,\bar{c},c]=\int d^4x\: \left(\bar{\psi}(\slashed{\partial}+\ii g\slashed{A}+m)\psi +\frac{1}{4}F_{\mu\nu}F_{\mu\nu}+\mathcal{L}_{gf}\right)\;.
\end{equation}
In contrast to lattice calculations, the functional approach considered here is gauge-fixed, which is reflected in the gauge-fixing Lagrangian $\mathcal{L}_{gf}$. For Landau gauge, it reads
\beq
\mathcal{L}_{gf}=\frac{1}{2\alpha}(\partial_\mu A^i_\mu)^2+\bar{c}^i\left( \partial^2\delta^{ij}+g f^{ijk}\partial_\mu A_\mu^k\right)c^j\;.
\eeq
The gauge field $A_\mu=A^i_\mu t^i$ is an element of the algebra $su(3)$, and the $t^i$ denote the generators of the gauge group $SU(3)$. The superscripts $i,j,k$ denote the color components, and repeated indices are summed over. The field strength tensor $F_{\mu\nu}$ appearing in the action is given by
\beq
F_{\mu\nu}=(\partial_\mu A^i_\nu-\partial_\nu A^i_\mu-g f^{ijk} A^j_\mu A^k_\nu)\,t^i\;.
\eeq

The generating functional or \emph{partition function} $Z[J]$ is a functional of the sources $J=\{J_{\bar{\psi}},J_\psi,J_A,J_{\bar{c}},J_c\}$, and the scalar product $\phi\dotp J$ indicates a sum over all types of fields, as well as summation/integration over all discrete and continuous indices and variables. The full Green functions of QCD follow from $Z[J]$ via functional differentiation, for example a full two point function is given by 
\beq
Z^{(2)}[J]=\frac{\delta^2 Z[J]}{\delta J_i\delta J_k}\;.
\eeq
Note that this expression still depends on the sources $J$; Since here only vacuum Green functions are considered, the sources have to be set to zero at the end of the calculation.
%At the end of the calculation, if one is interested in vacuum Green functions, the sources have to be set to zero. 

To study only the connected Green functions of the theory, the definition of a new generating functional, 
\beq\label{eq:W[J]}
 W[J]=-\Log[Z[J]]\;,
 \eeq
can be used. To get from connected to one-particle irreducible Green functions, one further step has to be taken, namely a Legendre transform
 \beq
 \Gamma[\phi]=W[J]+\phi\dotp J\;.
 \eeq
It defines another generating functional, the 1PI effective action $\Gamma[\phi]$, which is, in contrast to $Z[J]$ and $W[J]$, no longer a functional of the sources $J$, but of the classical fields $\phi$ given by
 \beq
 \phi_i=\frac{\delta W[J]}{\delta J_i}\;.
 \eeq

In a diagrammatic representation, a 1PI diagram is constructed such that it does not become disconnected by cutting one internal line. The 1PI Green functions, however, are not a subset of the connected Green functions. This can easily be seen by the relation between the propagator $G^{(2)}_{ik}=\delta^2 W[J]/(\delta J_i \delta J_k)$ and the 1PI function $\Gamma^{(2)}_{ik}=\delta^2 \Gamma[\phi]/(\delta \phi_i \delta \phi_k)$,
 \beq\label{eq:duality}
 G^{(2)}_{ik} \Gamma^{(2)}_{kj}=\delta_{ij}\;.
 \eeq
Thus, the 1PI function $\Gamma^{(2)}$ is the inverse of the propagator $G^{(2)}$. This is important when constructing a diagram: 1PI functions (which are sometimes referred to as (proper) vertex functions) always have to be connected to each other via propagators, i.e. they do not have external legs.

The sources as well as the classical fields are defined as functions of one space-time point $x$, such that they may be called 'local'. This is well suited to study the properties and interactions of the elementary fields of the theory under investigation. Bound states in the context of a quantum field theory, however, are represented not by elementary fields but by composite fields. 

In order to study them, we follow \cite{Cornwall:1974vz} and introduce a new type of (bi-local) sources $J_{ik}(x,y)$, which depend on two space-time points $x$ and $y$ and are able to connect two fields, as indicated by the indices $i,\,j$. In the presence of these sources the generating functional $Z[J]$ reads
\beq\label{eq:Z[J]2PI}
Z[J]=\mathcal{N}\int \mathcal{D}\phi\: \exp{\left(-S_{QCD}[\phi]+\int d^4x\:d^4y\phi_i(x)J_{ik}(x,y)\phi_k(y)\right)}\;,
\eeq
where we consider the case of vanishing local sources $J_i=0$ for simplicity. The indices $i,k$ label the field type, and repeated indices are summed over. As before, the generating functionals of connected and irreducible Green functions can be defined as
\beq
W_{[2]}[J]=-\Log[Z[J]]\;,\quad \text{and}
\eeq
\beq\label{eq:effectiveaction}
\Gamma_{[2]}[B]=W_{[2]}[J]+ \int d^4x\:d^4y\:B_{ij}(x,y)J_{ji}(y,x)\;,
\eeq
where $B_{ij}(x,y)$ denotes the classical bilocal field corresponding to the source $J_{ji}(y,x)$. $\Gamma[B]$ is referred to as two particle irreducible (2PI) effective action, and functional derivatives of $\Gamma_{[2]}[B]$ are the 2PI Green functions. The functional derivatives of $W_{[2]}[J]$ are in this case not connected, but bi-locally connected.

In strict analogy to Eq.~(\ref{eq:duality}) in the 1PI case, one can show that the inverse of the bi-locally connected two point-pair function, which represents the propagator of a \emph{particle pair}, is a 2PI function (see for example \cite{Cvitanovic:1983aa,McKay:1989rk}).

\section{Dyson-Schwinger and Bethe-Salpeter equations}\label{sec:inhom_bse}
The Dyson-Schwinger equations (DSEs) provide relations among the Green functions of a quantum field theory, and are sometimes referred to as quantum equations of motion. They can be obtained from a master equation, which in turn is derived from the generating functional $Z[J]$ (see e.g.~\cite{Alkofer:2000wg,Itzykson:2005it,Lichtenegger:2010ph}). The master equation in the 1PI case reads
\beq\label{eq:masterGamma}
\frac{\delta \Gamma[\phi]}{\delta \phi_i}-\frac{\delta S_{QCD}}{\delta \phi_i}\left[\phi+\frac{\delta^2W[J]}{\delta J \delta J}\frac{\delta}{\delta\phi}\right]=0\;.
\eeq
In the second term, the expression in the square brackets is the argument of the functional derivative of the action, meaning that at any occurrence of a field $\phi_i$ the corresponding $\phi_i+\frac{\delta^2W[J]}{\delta J_i \delta J_k}\frac{\delta}{\delta\phi_k}$ has to be inserted (summation over repeated indices is implied). Eq.~\eq{masterGamma} is the DSE for a one-point function, therefore to get higher $n$-point functions one has to apply subsequent functional derivatives. Alternatively, the derivation can be done in a diagrammatic way and has been implemented as the Mathematica-package DoDSE, as described in \cite{Alkofer:2008nt}. As already noted in connection with the generating functionals, for vacuum Green functions the classical fields have to be chosen such that the sources vanish, $J=0$. 

\begin{figure}
\centering \includegraphics[width=0.7\textwidth]{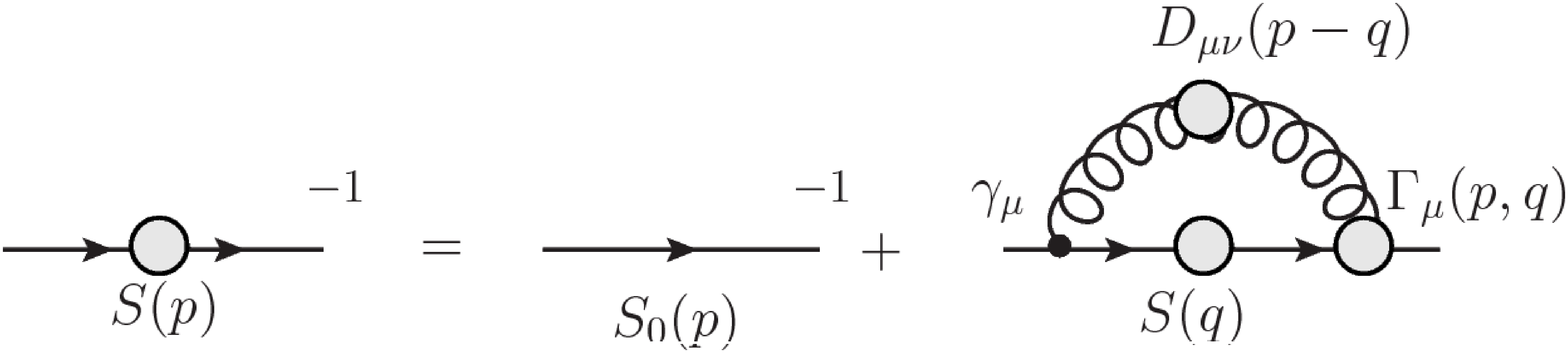}
\caption[The quark DSE \eq{gapfull} in pictorial representation.]{\label{fig:gapfull} The quark DSE \eq{gapfull} in pictorial representation. The blobs denote dressed propagators and vertices.}
\end{figure}

In this work mainly one Dyson-Schwinger equation is used, namely the quark gap equation. Its pictorial representation is given in Fig.~\fig{gapfull}. In Euclidean momentum space, the equation reads \cite{Alkofer:2000wg}
\beq\label{eq:gapfull}
S(p)^{-1}=Z_2\left(\ii \slashed{p}+Z_m m \right) + g^2 Z_{1F}\int_q\: \frac{\lambda^a}{2}\gamma_\mu S(q) \Gamma^a_\nu(p,q) D_{\mu\nu}(p-q)\;.
\eeq
It is a nonlinear, inhomogeneous integral equation for the renormalized dressed quark propagator $S(p)$ which can be represented by two dressing functions $A(p^2)$ and $B(p^2)$, such that
\beq\label{eq:squark}
S(p)^{-1}=\ii \slashed{p} A(p^2)+B(p^2)\;.
\eeq
The first term on the left-hand side of Eq.~\eq{gapfull} is the inverse bare propagator $S_0^{-1}=\ii \slashed{p}+Z_m m$ (multiplied by the renormalization constant $Z_2$) including the bare quark mass $m$ as well as the mass renormalization constant $Z_m$. The second term is usually referred to as (proper) self-energy. It involves the bare quark-gluon vertex $\frac{\lambda^a}{2}\gamma_\mu$ ($\lambda^a$ denotes the Gell-Mann $SU(3)$ color matrices), the full 1PI quark gluon vertex $Z_{1F}\,\Gamma^a_\nu(p,q)$ (with the renormalization constant $Z_{1F}$), the gluon propagator $D_{\mu\nu}(p-q)$ and again the dressed quark propagator. The index $a$ indicates color and is summed over. In the derivation of Eq.~\eq{gapfull}, the 1PI quark-gluon vertex appears as functional derivative of the effective action, $\delta^3\Gamma[\phi]/(\delta \bar{\psi}\,\delta A_\mu\,\delta \psi )$, and the gluon propagator as second functional derivative of $W[J]$, $\delta^2W[J]/(\delta J_{A_\mu}\,\delta J_{A_\nu})$.

In order to obtain a similar equation for the two-particle propagator of, e.~g., the quark-antiquark system, $G^{(2)}_{[2]}:=\delta^2W_{[2]}[J]/(\delta J_{\psi\bar{\psi}} \delta J_{\psi\bar{\psi}})$, one has to separate `interacting' and `non-interacting' parts. Following \cite{Cornwall:1974vz} (cf.~also \cite[App. A]{Cvitanovic:1983aa}), this can be done by writing the 2PI effective action $\Gamma_{[2]}[B]$ as 
\beq\label{eq:GammaInt}
\Gamma_{[2]}[B]=\Tr\left[S_0^{-1} B\right]-\Tr\Log\left[B\right]+\tilde{\Gamma}[B]\;,
\eeq
where $\tilde{\Gamma}[B]$ represents the `interacting contributions' to the 2PI effective action. The vacuum value of the classical field $B_{\psi\bar{\psi}}$, where the corresponding source vanishes, i.e., where $J_{\psi\bar{\psi}}=0$, is given by \cite{Munczek:1994zz}
\beq
B_{\psi\bar{\psi}}=S\;,
\eeq
where $S$ is the exact quark propagator. In this case, the inverse two-particle propagator is
\beq\label{eq:bse1}
(G^{(2)}_{[2]})^{-1}=\left.\frac{\delta^2\Gamma_{[2]}[B]}{\delta B_{\psi\bar{\psi}} \delta B_{\psi\bar{\psi}}}\right|_{ B_{\psi\bar{\psi}}=S}=S^{-1} S^{-1}+\frac{\delta^2\tilde\Gamma[S]}{\delta S \delta S}\;.
\eeq
From this equation, one can define the \emph{interaction kernel}
\beq\label{eq:definek}
K:=-\frac{\delta^2\tilde\Gamma[S]}{\delta S \delta S}\;,
\eeq
such that Eq.~(\ref{eq:bse1}) can be written in the more familiar form
\beq\label{eq:bseGinv}
(G^{(2)}_{[2]})^{-1}=S^{-1} S^{-1}-K\;,
\eeq
or equivalently
\beq\label{eq:bseG}
G^{(2)}_{[2]}=S\,S+S\,S\,K\,G^{(2)}_{[2]}\;,
\eeq
which is the well-known Bethe-Salpeter equation (BSE) for the propagator of the quark-antiquark system.

It should be noted that this construction is in a strict sense only valid if the quark and the antiquark have the same flavor; it is however possible to extend the procedure to the general case.

\section{Truncation and models}\label{sec:truncation}
In principle, the quark gap equation and the BSE given in Eqs.~\eq{gapfull} and \eq{bseG}, respectively, are exact equations. However, in practice the %While the quark gap equation (\ref{eq:gapfull}) as well as the BSE (\ref{eq:bseG}) are exact equations, they can not be solved directly. The 
gap equation is, via the hierarchy of Dyson-Schwinger equations, coupled to infinitely many other equations, starting with the DSE for the 1PI quark-gluon vertex and for the gluon propagator. The BSE contains the so-far unknown interaction kernel $K$. Therefore, one has to resort to truncations of the system to obtain numerical solutions of these equations. There exists, however, a connection between the quark self energy and the interaction kernel, which may be used as a guiding principle and is derived in the following (cf.~\cite{Munczek:1994zz}).

From the form of the 2PI effective action given in Eq.~\eq{GammaInt}, it follows that
\beq
\frac{\delta \Gamma_{[2]}[B]}{\delta B_{\psi\bar{\psi}}}=S_0^{-1}-B_{\psi\bar{\psi}}^{-1}+\frac{\delta \tilde{\Gamma}[B]}{\delta B_{\psi\bar{\psi}}}\;.
\eeq
In the vacuum, where all sources vanish, one obtains from Eq.~\eq{effectiveaction}
\beq
J_{\bar\psi \psi}=\left.\frac{\delta \Gamma_{[2]}[B]}{\delta B_{\psi\bar{\psi}}}\right|_{B_{\psi\bar{\psi}}=S}=0\;,
\eeq
such that
\beq
S^{-1}=S_0^{-1}+\frac{\delta \tilde\Gamma[S]}{\delta{S}}\;.
\eeq
This is an exact equation for the inverse quark propagator, and by comparison with Eq.~\eq{gapfull} it is clear that $\frac{\delta \tilde\Gamma[S]}{\delta{S}}$ gives the self energy. Therefore, the self energy and the interaction kernel are (up to a sign) the first and second functional derivative of $\tilde\Gamma[S]$, which opens a possibility to find a consistent and numerically feasible truncation of both equations. In addition, if $\tilde\Gamma$ is chosen in accordance with the symmetries of the theory 
%(most notably chiral symmetry)
, these are manifest in the solutions of Eqs.~(\ref{eq:gapfull}) and (\ref{eq:bseG}), as discussed in Ref.~\cite{Munczek:1994zz}. This is especially important for the application of QCD to light hadrons, where chiral symmetry is essential.

\begin{figure}
\centering\includegraphics[width=0.2\textwidth]{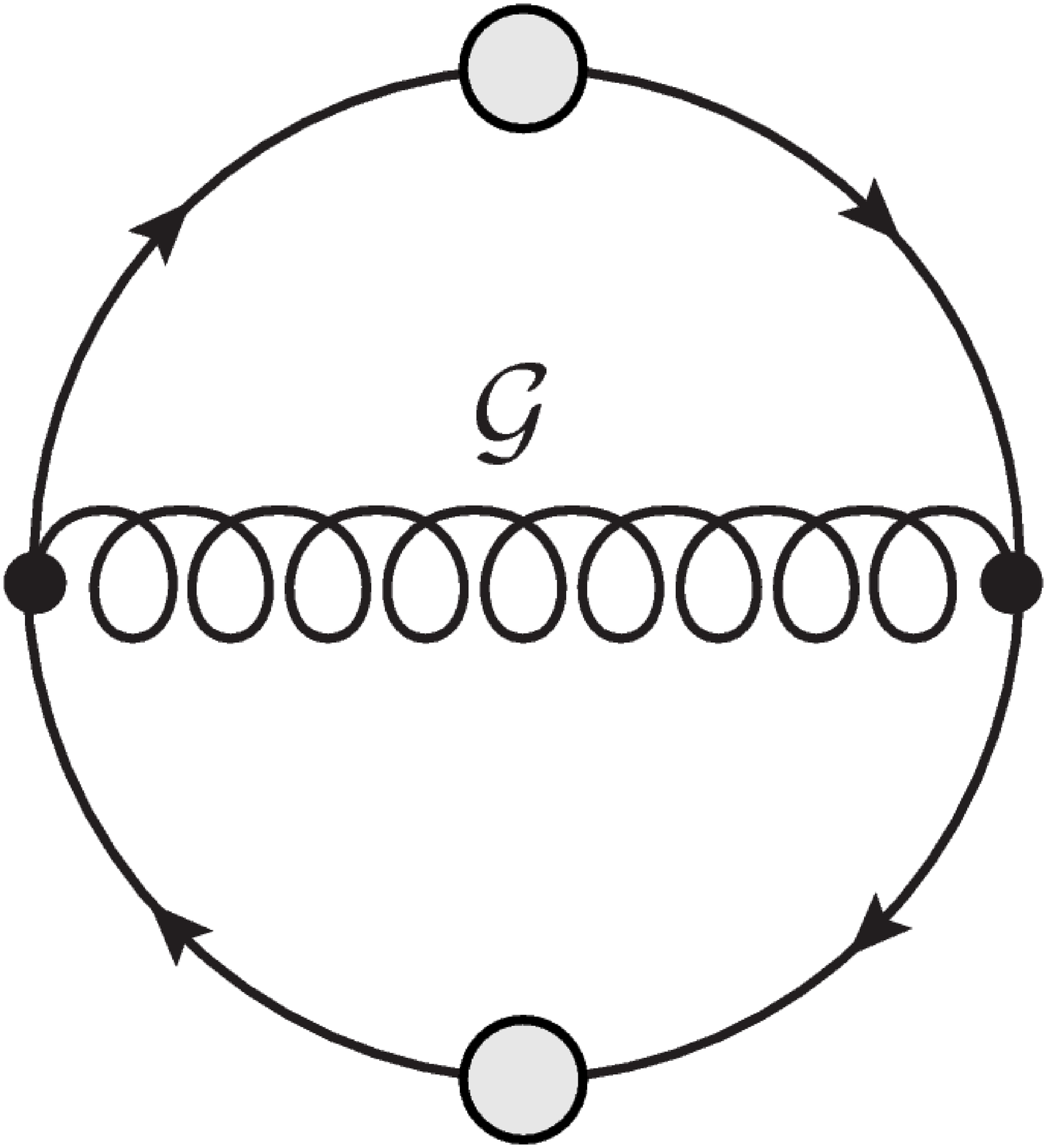}
\caption[Diagrammatic representation of the 2PI effective action leading to the rainbow-ladder truncation.]{\label{fig:gamma2}Diagrammatic representation of $\tilde\Gamma$ which leads to the well-known rainbow-ladder truncation, with the effective interaction $\mathcal G$ (for details see App. \ref{app:models}).}
\end{figure}

A popular choice of $\tilde\Gamma$ is depicted in Fig.~\ref{fig:gamma2}. It corresponds to the rainbow-ladder (RL) truncation of the quark self energy and the interaction kernel $K$ and correctly implements chiral symmetry independent of the functional form of the effective interaction $\mathcal G$. This truncation is used for the majority of calculations presented in this thesis, and the different types of model functions and the corresponding parameters used are given in App.~\ref{app:models}.

\section{Homogeneous BSE and normalization}\label{sec:hom_bse}
The inhomogeneous BSE for the two-particle propagator, Eq.~\eq{bseG}, has poles at those energies where the system exhibits a bound state. The corresponding on-shell condition is
\beq\label{eq:onshell}
P_i^2=-M_i^2\;,
\eeq
where the index $i$ labels the bound state, whose total momentum and mass are given by $P_i$ and $M_i$, respectively.

The propagator $G^{(2)}_{[2]}$ may now be expressed in terms of pole contributions and regular terms (denoted by $R(p,q,P)$) as \cite{Nakanishi:1969ph,Smith:1969az}
\beq\label{eq:polesum}
G^{(2)}_{[2]}(p,q,P)=\sum_i \frac{ \chi(p,P_i) \:\bar\chi(q,-P_i)}{P^2-P_i^2}\:+\:R(p,q,P)\;.
\eeq
$\chi(p,P_i)\equiv S^a(p_+)\Gamma_{[h]}(p,P_i) S^b(p_-)$ is usually referred to as the Bethe-Salpeter wave function consisting of the Bethe-Salpeter amplitude $\Gamma_{[h]}(p,P_i)$ and the dressed (anti)quark propagators $S^{a,b}$. Here, the indices $a,b$ indicate the possibly different flavors of the quark and the antiquark. The charge-conjugate wave function $\bar \chi$ is given by 
\beq\label{eq:cconj}
\bar \chi(p,-P_i)=[C \chi(-p,-P_i) C^{-1}]^t\;,
\eeq
where the charge-conjugation matrix $C=\gamma_2\gamma_4$ and the superscript $t$ denotes the matrix transpose. In general, uppercase Latin letters refer to total momenta ($P$ and $P_i$), lowercase letters to relative momenta of quark and antiquark ($q$, $p$, and $k$), and lowercase letters with subscripts $\pm$ to the momenta of quarks and antiquarks themselves ($p_\pm$, $q_\pm$, and $k_\pm$), which are defined as $p_\pm=p\pm \eta_\pm P$ with the momentum partitioning parameters $\eta_\pm$, $\eta_+ + \eta_-=1$. For a complete review of the kinematics of the Bethe-Salpeter equation and the momenta involved the reader is referred to App.~\ref{app:bsekinematic}, where all necessary definitions can be found.

\begin{figure}
\centering\includegraphics[width=0.576\textwidth]{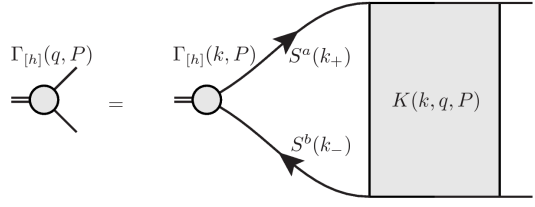}
\caption{\label{fig:hom_bse}The homogeneous BSE, Eq.~(\ref{eq:hom_bse}).}
\end{figure}

Inserting Eq.~\eq{polesum} into Eq.~\eq{bseG} and equating residues at the bound state pole $P^2=P_j^2$ yields
\beq  \label{eq:hom_bse}
\Gamma_{[h]}(q,P_j)=\int_{k}\! K(k,q,P_j) S^a(k_+)\Gamma_{[h]}(k,P_j) S^b(k_-) \;.
\eeq
This is the well-known homogeneous BSE for the on-shell Bethe-Salpeter amplitude (BSA) $\Gamma_{[h]}(p,P_j)$, which is valid only at $P_j^2=-M_j^2$. In pictorial form, it is given in Fig.~\fig{hom_bse}.

Eq.~\eq{hom_bse}, however, only determines the homogeneous BSA up to a constant factor and an additional normalization condition is needed. The \emph{canonical norm} follows from the two-particle propagator, by demanding that the residue at the bound state pole is equal to one, i.e., that the decomposition \eq{polesum} holds exactly. There exist different ways to obtain the normalization condition, see for example \cite{Nakanishi:1969ph} and references therein. Here, we consider one pole term in $G^{(2)}_{[2]}$ with an arbitrary residue $r_i$ (in the neighborhood of the pole, the other terms may be neglected),
\beq\label{eq:resNorm}
G^{(2)}_{[2]}(p,q,P)\:\simeq\: r_i\: \frac{\chi(p,P_i) \:\bar\chi(q,-P_i)}{P^2-P_i^2}\;,
\eeq
or equivalently
\beq
G^{(2)}_{[2]}(p,q,P)\:\simeq\:\frac{r_i}{P_\mu+(P_i)_\mu}\: \frac{\chi(p,P_i) \:\bar\chi(q,-P_i)}{P_\mu-(P_i)_\mu}\;,
\eeq
where the simple pole in the variable $P_\mu$ has been made explicit.

In general, if a function $f(z)$ has a simple pole in $z$ and it can be written as $f(z)=g(z)/h(z)$, then the residue of $f$ at a point $z_0$ is given by $Res(f,z_0)=g(z_0)/h'(z_0)$. This can be used to obtain the residue of $G^{(2)}_{[2]}$, since from Eq. \eq{bseGinv} it follows that
\beq
G^{(2)}_{[2]}= \left[S^{-1} S^{-1}-K\right]^{-1}\;.
\eeq
Therefore, the residues of the pole in $P_\mu$ on both sides of Eq.~\eq{resNorm} are
\begin{multline}
\left[\frac{\partial}{\partial P_\mu}\left( \left[ S^a(q+\eta_+P)S^b(q-\eta_-P)\right]^{-1}\delta(p-q)-K(p,q,P)\right) \right]_{P=P_i}^{-1}\\
=\:\frac{r_i}{2(P_i)_\mu}\: \chi(p,P_i) \:\bar\chi(q,-P_i)\;.
\end{multline}
Reshuffling the terms and taking the trace in a functional sense as well as in Dirac-space, one obtains
\begin{multline}
\frac{2(P_i)_\mu}{r_i}=\Tr\left[\int_q \bar\chi(q,-P_i)\left( \frac{\partial}{\partial P_\mu} \left[S^a(q+\eta_+P)S^b(q-\eta_-P)\right]^{-1}\right) \chi(q,P_i) \right.\\ \left.
-\int_q \int_p\bar\chi(q,-P_i) \left( \frac{\partial}{\partial P_\mu} K(p,q,P)\right) \chi(p,P_i)\right]_{P=P_{i}}\;.
\end{multline}
If $r_i=1$, this is the canonical normalization condition for the Bethe-Salpeter wave function $\chi$ \cite{Nakanishi:1969ph,Smith:1969az}. After expressing $\chi$ in terms of $\Gamma_{[h]}$ and propagators, and working out the derivatives (note that $\bar\Gamma_{[h]}$ is defined analogously to $\bar\chi$), it can be converted to a condition for the BSA (see, e.g., \cite{Maris:1997tm}). As a result, the canonical norm can be defined as 
\begin{multline}\label{eq:canonicalnorm}
\mathcal N = \frac{1}{r_i}=-\frac{(P_i)_\mu}{2 P_i^2} \frac{\partial}{\partial P_\mu}\\
\Tr
\left[ 
\int_q 
S^a(q+\eta_+P)\bar\Gamma_{[h]}(q,-P_i) S^b(q_-) \Gamma_{[h]}(q,P_i) \right. \\
+ \int_q S^a(q_+)\bar\Gamma_{[h]}(q,-P_i) S^b(q-\eta_-P) \Gamma_{[h]}(q,P_i)
\\
\left.+\int_q \int_p
S^a(q_+)\bar\Gamma_{[h]}(q,-P_i) S^b(q_-) K(p,q,P)S^a(p_+)\Gamma_{[h]}(p,P_i) S^b(p_-)
\right]_{P=P_{i}}\;,
\end{multline}
where all dependence on $P$ has been made explicit. Thus, to get a unit residue in Eq.~\eq{resNorm}, $r_i=1$, $\Gamma_{[h]}$ has to be replaced by $\Gamma_{[h]}/\sqrt{\mathcal N}$.

\section{Inhomogeneous vertex BSE}\label{sec:vertex_bse}
The homogeneous BSA represents an on-shell quark-antiquark-meson vertex. In order to obtain off-shell information without having to solve the full inhomogeneous BSE for the two-particle propagator, it is desirable to define a general vertex that connects quark and antiquark to a color singlet current. This quantity inherits the pole structure of the two-particle propagator $G^{(2)}_{[2]}$ and is restricted to one (mesonic) channel, i.e. one set of quantum numbers.

\begin{figure}
\centering\includegraphics[width=0.72\textwidth]{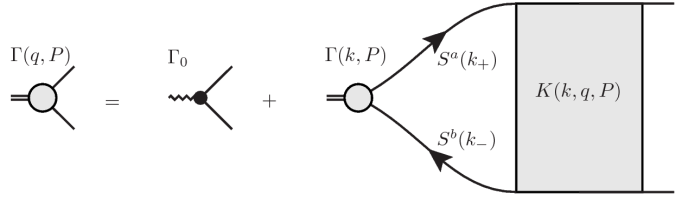}
\caption{\label{fig:inhom_bse} The inhomogeneous vertex BSE, Eq.~(\ref{eq:inhom_bse}).}
\end{figure}

We now derive a Bethe-Salpeter equation for this vertex, and start from the inhomogeneous BSE \eq{bseG} by projecting one side onto a suitable (renormalized) current $\Gamma_0$, such that
\beq
G^{(2)}_{[2]}\Gamma_0=S\,S\,\Gamma_0+S\,S\,K\,G^{(2)}_{[2]}\Gamma_0\;.
\eeq
To obtain the legless structure of a proper vertex analogous to the homogeneous BSA, the external legs on the right are removed, 
\beq\label{eq:Gproj}
S^{-1}S^{-1}G^{(2)}_{[2]}\Gamma_0=\Gamma_0+K\,G^{(2)}_{[2]}\,\Gamma_0\;.
\eeq
We now define the inhomogeneous Bethe-Salpeter amplitude as 
\beq\label{eq:defineG}
\Gamma \equiv S^{-1}S^{-1}G^{(2)}_{[2]}\,\Gamma_0\;,
\eeq
and end up with 
\beq
\Gamma=\Gamma_0+K\,S\,S\,\Gamma\;.
\eeq
In less symbolic notation, with explicit momentum arguments and matrix structure in Dirac space taken into account properly, this equation reads
\beq \label{eq:inhom_bse}
\Gamma(p,P) = \Gamma_0(p,P)+\int_{k} K(p,k,P) S^a(k_+)\Gamma(k,P) S^b(k_-) \;,
\eeq
which is the inhomogeneous vertex Bethe-Salpeter equation (see, e.g., \cite{Maris:1997hd,Bhagwat:2007rj}). In pictorial form, it is given in Fig.~\fig{inhom_bse}.

Note that this definition of the inhomogeneous BSA is consistent with the form of the vertex BSE used in Ref.~\cite{Maris:1997hd}, which can be obtained by introducing the renormalized fully amputated quark-antiquark scattering matrix $M$ as
 \beq
 M=K+K\,S\,S\,K+\ldots=S^{-1}S^{-1}G^{(2)}_{[2]}S^{-1}S^{-1}-S^{-1}S^{-1}\;, 
 \eeq
and inserting this definition into Eq.~\eq{Gproj}. The result is
\beq
\Gamma=\Gamma_0+M\,S\,S\,\Gamma_0\;,
\eeq
which is equivalent to \cite[Eq.~(19)]{Maris:1997hd}.

% A different representation of Eq.~\eq{inhom_bse} can be obtained by introducing the renormalized fully amputated quark-antiquark scattering matrix $M$ as \cite{Maris:1997hd}
% \beq
% M=K+K\,S\,S\,K+\ldots=S^{-1}S^{-1}G^{(2)}_{[2]}S^{-1}S^{-1}-S^{-1}S^{-1}\;.
% \eeq
% With this, Eq.~\eq{Gproj} can be written as (cf. \cite{Maris:1997hd})
% \beq
% S^{-1}S^{-1}G^{(2)}_{[2]}\,\Gamma_0=\Gamma=\Gamma_0+M\,S\,S\,\Gamma_0\;.
% \eeq
%The vertex BSE in this form is also given in Ref.~\cite{Maris:1997hd}, which shows the consistency of the definition Eq.~\eq{defineG}.

\section{Finite temperature}\label{sec:fintform}
So far, only the vacuum, i.e.~zero temperature and density, has been considered. The functional formalism used here, however, is also well suited to study QCD at finite temperature. An extensive discussion of thermal quantum field theory is presented in Ref.~\cite{Kapusta:1989tk}; Here, only the basic concepts and details of immediate importance for the practical calculations are given. 

One starts form the observation that the partition function as known from statistical mechanics, 
\beq
Z=\Tr\left[\exp(-\beta H)\right]\;,
\eeq
with the Hamiltonian $H$ and the inverse temperature $\beta=1/T$, can be written as a path integral. The time variable, however, becomes complex, and instead of `real time' $t$ one has to use `Euclidean time' $\tau=\ii\, t$, with $\tau \in [0,\beta]$.

Thus, the formalism developed before is applicable, but time is compact. As a consequence, one has to introduce boundary conditions in time, such that bosonic fields become periodic and fermionic fields become antiperiodic. If the equations are transformed to momentum space, it is easy to see that the energies, which in Euclidean notation are the fourth component of the 4-momentum, are discrete. These \emph{Matsubara frequencies} are given by
\beq\label{eq:omega_boson}
\Omega_n = 2 n \pi T \quad\textrm{for bosons,}
\eeq
\beq\label{eq:omega_fermion}
\omega_n = (2n+1)\pi T\quad\textrm{for fermions.}
\eeq
Therefore, the integral over the energy has to be replaced by a sum over the Matsubara frequencies.

Due to the differences in the treatment of the fourth momentum component compared to the other three, the equations no longer appear in covariant form. Physically speaking, the system has been coupled to an external heat bath, and the direction of Euclidean time corresponds to the rest frame of the heat bath, defined by the four vector $u_\mu$. 
%If this additional vector is taken into account, a covariant notation can be used to investigate the implications of the finite temperature formalism on the structure of Lorentz and Dirac tensors appearing in the equations (cf.~\cite{Rusnak:1995ex,Ayala:2001mb}). 
If this additional vector is taken into account, it is still possible to use a covariant notation to investigate the implications of the finite temperature formalism on the structure of Lorentz and Dirac tensors appearing in the equations (cf.~\cite{Rusnak:1995ex,Ayala:2001mb}).

In this thesis, the main object under investigation at finite temperature is the quark propagator, whose inverse at zero temperature is given in Eq.~\eq{squark}. In that case, it contains the tensor structures $\slashed{p}$ and $\mathbf 1$, which are the only Lorentz scalars that can be constructed from the four-momenta $\gamma_\mu$ and $p_\mu$. At finite temperature, with the additional four-vector $u_\mu$, the available constructions are $\slashed{p}$ and $(p\dotp u)\slashed{u}$ for the `vector' part of the propagator, as well as $\mathbf{1}$ and $(p\dotp u)\, \slashed{p}\,\slashed{u}$ for the `scalar' and a possible `tensor' part, respectively. Note that the factors of $p\dotp u$ are chosen for later convenience. This leads to the general form \cite{Roberts:2000aa}
%which are the only available constructions with the four-momenta $\gamma_\mu$ and $p_\mu$ . 
%At finite temperature, with the additional four-vector $u_\mu$, one can construct $\slashed{p}$, $p\dotp u\:\slashed{u}$, $\mathbf{1}$, and $p\dotp u\: \slashed{p}\:\slashed{u}$. Note that the factors of $p\dotp u$ are chosen for later convenience. This leads to the general form \cite{Roberts:2000aa}
\begin{multline}\label{eq:sfint}
S(p,u)^{-1}=\ii \slashed{p}\: A(p^2,u^2,p\dotp u)+\ii\,p\dotp u\: \slashed{u}\: C(p^2,u^2,p\dotp u)+ \\
\mathbf{1}\: B(p^2,u^2,p\dotp u)+p\dotp u\:\slashed{p}\:\slashed{u}\: D(p^2,u^2,p\dotp u)
\end{multline}
with the four scalar dressing functions $A,\;B,\;C,\;D$. Choosing
\beq
u_\mu = (0,0,0,1)\quad \text{and}\quad p_\mu=(\vec{p},\omega_n)
\eeq
gives the standard form of the quark propagator at finite temperature,
\begin{multline}\label{eq:sfint1}
S(\vec{p},\omega_n)=\ii \vec{\slashed{p}}\: A(\vec{p}^2,\omega_n)+\ii \gamma_4 \omega_n C(\vec{p}^2,\omega_n)\\
+\mathbf{1} B(\vec{p}^2,\omega_n)+\vec{\slashed{p}}\:\gamma_4\omega_n D(\vec{p}^2,\omega_n)\;.
\end{multline}
% As pointed out in Ref.~\cite{Rusnak:1995ex}, the dressing function $D$ does not contribute in a quantum field theory at zero temperature but nonvanishing chemical potential, since the corresponding tensor structure has the wrong transformation properties under time reversal. At finite temperature in equilibrium, time is removed from the theory, and its place is taken by inverse temperature. However, the argument still holds: the theory and its vacuum are invariant under translations in inverse temperature (which resembles imaginary time), as demanded by the Matsubara formalism described above. This causes invariance under ``inverse temperature reversal'', and therefore $D(\vec{p}^2,\omega_n)\equiv0$.\footnote{This argument emerged from a discussion with M.~Mitter.}
The dressing function $D$, however, is power-law suppressed in the UV \cite{Roberts:2000aa} and does not contribute in all cases investigated here. At zero temperature but nonvanishing chemical potential it vanishes exactly since the corresponding tensor structure has the wrong transformation properties under time reversal \cite{Rusnak:1995ex}.

Similar to the quark propagator, the gluon propagator acquires additional structures as well, such that it reads \cite{Kapusta:1989tk,Roberts:2000aa}
\begin{eqnarray}
D_{\mu\nu}(\vec{k},\Omega) &=&
P_{\mu\nu}^L(\vec{k},\Omega)\;G(\vec{k},\Omega) + P_{\mu\nu}^T(\vec{k},\Omega)\; F(\vec{k},\Omega) ,\label{eq:gluonprop}\\
P_{\mu\nu}^T(\vec{k},\Omega)&=&
\left\{\begin{matrix}0,& \mu\;\mbox{and/or}\;\nu=4\\
\delta_{ij}-\frac{k_i k_j}{\vec{k}^2},&\mu,\nu=i,j=1,2,3\end{matrix}\right.,\\
P_{\mu\nu}^L(\vec{k},\Omega)&=&\delta_{\mu\nu}-\frac{k_\mu k_\nu}{\vec{k}^2+\Omega^2}-P_{\mu\nu}^T,
\end{eqnarray}
where $G(\vec{k},\Omega)$ and $F(\vec{k},\Omega)$ denote its two dressing functions.

In order to calculate the quark propagator the corresponding DSE has to be converted to the finite temperature formalism as well. Therefore, the Matsubara sum and three-dimensional integral have to be inserted for the usual four-dimensional integration ($\int_q\rightarrow T \sum_{l=-\infty}^\infty \int_{\vec{q}}$). Together with the correct form of the quark propagator, Eq.~\eq{sfint1}, the quark DSE at finite temperature can be written as (see, e.g.~\cite{Roberts:2000aa})
\begin{multline}\label{eq:gapfint}
1/Z_1^A\,\ii \vec{\slashed{p}}\: A(\vec{p}^2,\omega_n)+1/Z_1\,\left(\mathbf{1} B(\vec{p}^2,\omega_n)+  \ii \gamma_4 \omega_n C(\vec{p}^2,\omega_n)\right) =\\
Z_2^A/Z_1^A\, \ii \vec{\slashed{p}}+Z_2/Z_1\,\left(\ii \gamma_4 \omega_n+m\right)+\\
T \sum_{l=-\infty}^\infty \int_{\vec{q}}
g^2 D_{\mu\nu}(\vec{k},\Omega) \gamma_\nu \frac{\lambda^a}{2} S(\vec{q},\omega_l)\Gamma^a_\mu(\vec{p},\omega_n;\vec{q},\omega_l)\;,
\end{multline}
where $\vec{k}=\vec{p}-\vec{q}$ and $\Omega=\omega_n-\omega_l$ are the gluon three-momentum and Matsubara frequency, respectively, and $Z_1$, $Z_1^A$, $Z_2$, $Z_2^A$ are the necessary renormalization constants \cite{Roberts:2000aa}. 
They are obtained from the renormalization condition that for a renormalization
scale $\zeta$, and $\vec{p}\,^2+\omega_0^2=\zeta^2$,
\begin{equation}
S^{-1}(\vec{p},\omega_0) =i\vec{\gamma}\cdot\vec{p} + i\gamma_4\omega_0 + m(\zeta), 
\end{equation}
which is the finite temperature analog of the renormalization condition at zero temperature, Eq.~\eq{rencond}.

\chapter{Numerical techniques}\label{chap:numerics}
In the present chapter, we give an account of the numerical methods employed in this thesis to solve the gap equation at zero temperature and the homogeneous as well as the inhomogeneous vertex BSE, whose derivation was discussed in Chap.~\ref{chap:formalism}. 

Our basic point is that the homogeneous and vertex BSE in a fully numerical setup correspond to matrix equations, an eigenvalue equation in the homogeneous and a simple linear system in the inhomogeneous case. From this point of view, it is natural to apply well-known and efficient algorithms that are able to solve such problems for general complex matrices, as also advocated in \cite{Blank:2010bp}. Even though in this work we only solve the BSEs in rainbow-ladder truncation, the methods discussed here are applicable in a more general sense: in more involved truncations as well as in systems involving more than two constituents, like baryons which are described by a covariant three-body equation, cf.~\cite{Eichmann:2009qa,Eichmann:2009zx}.

As a prerequisite, we discuss the solution of the quark propagator for real and complex arguments including the renormalization procedure used here. In Sec.~\ref{sec:bsenum}, following the arguments given in \cite{Blank:2010bp}, we turn to a detailed description of the numerical representation of the BSE \emph{kernel matrix}, which is the basic building block of the homogeneous as well as of the vertex BSE, and discuss the application of the implicitly restarted Arnoldi factorization to the homogeneous and of the Bi-Conjugate Gradients stabilized (BiCGstab) algorithm to the vertex BSE. 

To illustrate their efficiency, we provide in Sec.~\ref{sec:illustration} a detailed description of the application of our methods to the case of pseudoscalar quantum numbers and a comparison to more conventional approaches. In Sec.~\ref{sec:quantnum}, the comparison is extended to other quantum numbers, and in Sec.~\ref{sec:hominhomcompare} we directly compare the homogeneous and inhomogeneous vertex BSE in terms of numerical efficiency.

\section{Quark propagator and renormalization}\label{sec:renormalization}
As already stated before, our main tool to investigate the properties of quarks is the quark DSE, which reads
$$
S(p)^{-1}=Z_2\left(\ii \slashed{p}+Z_m m \right) + g^2 Z_{1F}\int_q\: \frac{\lambda^a}{2}\gamma_\mu S(q) \Gamma^a_\nu(p,q) D_{\mu\nu}(p-q)\;.
$$
This equation is exact, but we can not solve it directly since the gluon propagator and the quark gluon vertex needed as input are not known a priori. Therefore, we rely in most of our calculations on the so-called \emph{rainbow-ladder} (RL) truncation, cf.~App.~\ref{app:models}. In this truncation (and using $Z_2\,Z_m\,m\equiv Z_4\,m(\mu^2)$), the gap equation becomes
\beq\label{eq:gapRL}
S(p)^{-1}=Z_2\,\ii \slashed{p}+Z_4\, m(\mu^2) + \int_q \frac{\lambda^a}{2}\gamma_\mu\, S(q)\, \frac{\lambda^a}{2}\gamma_\nu \frac{\mg\left((p-q)^2\right)}{(p-q)^2}\, T_{\mu\nu}(p-q)\;, 
\eeq
with $T_{\mu\nu}(k)=\delta_{\mu\nu}-\frac{k_\mu k_\nu}{k^2}$ denoting the transverse projector and $\mg(k^2)$ the effective interaction, which absorbs in this setting the dressing function of the gluon propagator, any momentum dependence of the quark-gluon vertex, the coupling $g^2$, as well as the quark-gluon-vertex renormalization constant $Z_{1F}$. Following \cite{Maris:1997tm}, we denote by $m(\mu^2)$ the renormalized quark mass at scale $\mu$.

In order to solve Eq.~\eq{gapRL}, i.e.~to obtain the two dressing functions $A(p^2)$ and $B(p^2)$ defined in Eq.~\eq{squark}, it is projected on the two tensor structures $T_1=\ii \slashed{p}$ and $T_2=\mathbf 1$ by using that $\Tr[T_i\cdot T_j]=0$ if $i\neq j$. In total, 
\begin{subequations}\label{eq:gapab}
\begin{multline}
A(p^2) = Z_2+\frac{4}{3\,p^2}\int_q \frac{\mg\left((p-q)^2\right)}{(p-q)^2}\frac{A(q^2)}{q^2 A^2(q^2)+B^2(q^2)}\\
\times\left( p\dotp q+2\,\frac{\left((p-q)\dotp p\right)\left((p-q)\dotp q\right)}{(p-q)^2}\right)
\end{multline}
\beq
B(p^2)=Z_4\,m(\mu^2)+4\,\int_q \frac{\mg\left((p-q)^2\right)}{(p-q)^2}\frac{B(q^2)}{q^2 A^2(q^2)+B^2(q^2)}  \;.
\eeq
\end{subequations}
The most straightforward way to solve these coupled, nonlinear, inhomogeneous integral equations numerically is direct iteration. First, we choose a parametrization of the momenta (cf.~App.~\ref{app:integration}), then we apply the so-called Nystr\"om or quadrature method (cf.~\cite[Chap.~4]{Delves:1985aa}), which amounts to replacing an integral by a sum over suitable quadrature weights and points and neglecting the error term in order to discretize the integration variables. Choosing the integration points to discretize the momentum dependence on the left-hand side as well allows to iterate Eqs.~\eq{gapab}.

Depending on the form of the effective coupling, the occurring integrals need to be regularized. This is implemented in a translationally-invariant form via a Pauli-Villars cutoff applied to the gluon momentum \cite{Pauli:1949zm}. Subsequent renormalization then gives the renormalized dressing functions and the values of the renormalization constants $Z_2$ and $Z_4$.

Following \cite{Maris:1997tm}, we use a subtractive renormalization scheme and fix $Z_2$ and $Z_4$ by demanding that at a renormalization scale $\mu$ in the perturbative domain the propagator becomes bare,
\beq\label{eq:rencond}
A(\mu^2)=1\;\;\textrm{and}\;\;B(\mu^2)=m(\mu^2)\;.
\eeq
Writing the inverse propagator as
\beq
S^{-1}(p^2)=Z_2\ii \slashed{p}+Z_4 m(\mu^2)+\ii \slashed{p}(A'(p^2)-1)+(B'(p^2)-m(\mu^2))\;,
\eeq
with the regularized but unrenormalized dressing functions (defined via Eqs.~\eq{gapab} if $Z_2=Z_4=1$) denoted by $A'(p^2)$ and $B'(p^2)$, one obtains for nonzero quark mass the renormalization conditions
\begin{subequations}\label{eq:ren_ab}
\begin{eqnarray}
A(p^2)&=&1+A'(p^2)-A'(\mu^2)  \\
B(p^2)&=&m(\mu^2)+B'(p^2)-B'(\mu^2)\;.
\end{eqnarray}
\end{subequations}
The renormalization constants are then given by
\begin{subequations}\label{eq:z2z4}
\begin{eqnarray}
Z_2&=&2-A'(\mu^2)  \\
Z_4&=&2-\frac{B'(\mu^2)}{m(\mu^2)}\;.
\end{eqnarray}
\end{subequations}
The chiral limit solutions are obtained for $m(\mu^2)=0$, and in this case 
\beq
B(p^2)=B'(p^2)\;.
\eeq
For our numerical calculations we follow Refs.~\cite{Maris:1997tm,Maris:1999nt} and choose $\mu=19\,\mathrm{GeV}$. 

\section{Quark propagator for complex momenta}\label{sec:gap_complex}
In order to use the quark propagator as an input to a BS equation in Euclidean space it is not sufficient to know the form of the dressing functions for real arguments. It follows from the kinematics of the BSE that their arguments $k_{\pm}$ in general are complex vectors. As explained in App.~\ref{app:bsekinematic}, the dressing functions $A(p^2)$ and $B(p^2)$ constituting the propagator have to be calculated on a parabolic region in the complex $p^2$-plane which is defined by the total momentum of the quark-antiquark system. From Eqs.~\eq{gapab}, it is clear that the quark propagator for any complex momentum $p$ can be obtained by just inserting the corresponding value and performing the integration over the (real) momentum $q$, once the solution on the real axis is known. 

This method is straightforward, but some problems are encountered. First, the (numerical) integral over the loop momentum $q$ only converges at reasonable rates as long as $p$ is close to the real axis. Second, to compute the integral the effective interaction $\mathcal G$ has to be evaluated for complex arguments, although it is in most cases designed to (effectively) reproduce the behavior of the running coupling of QCD, $\alpha_s(p^2)$, only on the real axis (cf.~\cite{Maris:1997tm}).

A different approach has been proposed in \cite{Fischer:2005en}. It is based on a shift in the integration variable of Eq.~\eq{gapRL}, such that the external momentum appears in the quark and not in the gluon momentum and therefore $\mg(p^2)$ is only needed for real arguments. Iterating Eqs.~\eq{gapab} in this setup requires the knowledge of $A(p^2)$ and $B(p^2)$ not only on the real axis, but inside the whole parabola.

If, however, $A(p^2)$ and $B(p^2)$ are assumed to be holomorphic in the region of interest, and one bears in mind that a holomorphic function is completely specified on a region by its values on a closed, surrounding contour, one can iterate on such a contour (ideally, one chooses a parabola which is cut at some point in the UV, cf.~Fig.~\ref{fig:parabola}), and use e.g.~the Cauchy integral formula to compute the values inside. The values outside the parabola in the UV needed for the iteration can be obtained by the straightforward analytic continuation of a fit of the solution on the real axis. 

The only remaining problem is the numerical stability of the Cauchy integral formula if points close to the parabola need to be calculated. Here, we apply the method proposed in \cite{Ioakimidis:1991io}, which was first used in this context in \cite{Krassnigg:2008gd}. It is based on the Cauchy theorem, and allows to extract the values of a holomorphic function $f(z)$  for any point $z_0$ inside a closed contour by numerical integration, such that $f(z_0)$ is given by
\beq
f(z_0)=  \left(\sum_{j=1}^n\frac{w[z_j] f(z_j)}{z_j - z_0}\right)\left/\left(\sum_{j=1}^n \frac{w[z_j]}{z_j - z_0}\right)\right.\;,
\eeq
where the $w[z_j]$ denote the quadrature weights corresponding to the points $z_j$ on the contour. The sums run over all contour points $j=1,\ldots,n$. Similar formulae exist for all derivatives of $f(z_0)$. In this work, in order to calculate the canonical norm of the homogeneous BSA, Eq.~\eq{canonicalnorm}, we need only the first derivative of the dressing functions $A(p^2)$ and $B(p^2)$, which can be calculated using
\beq
f'(z_0)=\left(\sum_{j=1}^n\frac{w[z_j] f(z_j)}{(z_j - z_0)^2}-f(z_0) \sum_{j=1}^n\frac{w[z_j]}{(z_j - z_0)^2}\right)\left/\left(\sum_{j=1}^n \frac{w[z_j]}{z_j - z_0}\right)\right.\;.
\eeq

Therefore, a numerical representation of the quark propagator in the complex plane is given by the values of $A(p^2)$ and $B(p^2)$ on points on a closed parabola (including the points themselves and the corresponding weights) plus a fit to the dressing functions on the real axis in the UV outside the parabola. Note that the evaluation of the propagator given in this form is numerically very efficient, since no interpolation is needed.

\section{Bethe-Salpeter equations}\label{sec:bsenum}
Once the quark propagator is known in the complex plane, one can proceed to the numerical solution of the Bethe-Salpeter equations (\ref{eq:inhom_bse}) and (\ref{eq:hom_bse}), as discussed in detail in Ref.~\cite{Blank:2010bp}.

The first step towards a numerical representation of these equations is the analysis of the Lorentz and Dirac structure of the respective amplitudes. This structure is a result of the particular representation of the symmetry properties of the state under consideration under the Lorentz group, including the state's parity and spin. Therefore, the (in)homogeneous BSA $\Gamma(q,P)$ is decomposed into Lorentz-covariant parts $T_i(q,P)$ and Lorentz-invariant parts $F^i(P^2,q^2,q\cdot P)$,
respectively, as
\beq\label{eq:decomp}
\Gamma(q,P) = \sum_{i=1}^N T_i(q,P) \;F^i(P^2,q^2,q\dotp P)\;,
\eeq
where the number of terms $N$ as well as the tensor structure of the $T_i$ and $\Gamma$ depend on the quantum
numbers of the state (for the explicit construction, see App.~\ref{app:basis}). The $T_i(q,P)$, which carry the Lorentz and Dirac structure, are usually referred to as \emph{covariants}, whereas we call the Lorentz and Dirac scalar quantities $F^i(P^2,q^2,q\dotp P)$ the \emph{components} of the amplitude $\Gamma(q,P)$. The $T_i(q,P)$ represent a basis for the BSA and one is to some extent free to choose the details thereof.

For numerical convenience, the basis elements are constructed such that they are orthonormal with respect to the (generalized) scalar product
 \beq \label{eq:covariants_norm}
 \mathrm{Tr}\left(T_i\cdot T_j\right)=\delta_{i,j}\;,
 \eeq
where any occurring Lorentz indices are understood to be summed over. The details of the construction for mesons of any parity and spin discussed in this thesis are given in App.~\ref{app:basis}. Using the decomposition (\ref{eq:decomp}), the BSEs (\ref{eq:inhom_bse}) and (\ref{eq:hom_bse}) can be rewritten as coupled integral equations of the components depending on the scalar products of the momenta via the corresponding projections on the basis $T_i$.

Note, however, that the structure of the covariants $T_i$ does not determine all quantum numbers of the quark-antiquark system. For constituents of equal masses, the homogeneous amplitude is an eigenstate of the operation of charge conjugation defined in Eq.~\eq{cconj}, thus defining the quantum number of charge-conjugation parity (C-parity), $C=\pm 1$. This determines the symmetry properties of the components with respect to $q\dotp P$ (cf.~App.~\ref{app:cparity}), such that in this case a restriction of the symmetry of the components allows to select either $C=+1$ or $C=-1$. Here, however, we consider a general dependence on $q\dotp P$, such that the methods described in this section are also applicable in the case of constituents of unequal masses, where C-parity is not well-defined.

We proceed by considering the integrand in the homogeneous and the inhomogeneous vertex BSE, which is in both cases given by
\beq  \label{eq:int_term}
K(k,q,P) \;S^a(q,P)\;\Gamma (q,P)\; S^b (q,P)\;.
\eeq
The amplitude $\Gamma (q,P)$ is expanded in the chosen Dirac basis $T_j (q,P)$ and the result is projected on $T_i (k,P)$. Doing so, one obtains a matrix structure in the space of covariants, and Eq.~(\ref{eq:int_term}) can be written as a matrix-vector multiplication in this space involving the BSE \emph{kernel matrix} $\mathcal{K}^i_j(k,q,P)$,
\begin{multline}\label{eq:kernelmatrix}
\mathcal{K}^i_j(k,q,P) F^j(P^2,q^2,q\dotp P)=\\
\mathrm{Tr}\left[T_i(k,P)\; K(k,q,P)\;S^a(q,P) T_j(q,P)\; S^b(q,P)\right]  F^j(P^2,q^2,q\dotp P)\;,
\end{multline}
where the sum over the repeated index $j$ is implied.

The index $j$ of the components $F^j(P^2,q^2,q\dotp P)$ can thus be viewed as a vector index, which has to be contracted with the corresponding index of the kernel matrix $\mathcal{K}^i_j(k,q,P)$.

The next step is to make the dependence on the continuous momentum variables $P^2$, $q^2$, and $q\dotp P$ numerically accessible. As in the case of the gap equation (cf.~Sec.~\ref{sec:renormalization}), we apply the quadrature method to discretize the integrals and use the same set of points also for the momentum dependence on the left-hand side. The homogeneous and the inhomogeneous vertex Bethe-Salpeter equations can then be written as matrix equations in the covariants and the discretized momenta and read 
\beq
F^{i,\mathcal{P}}_{[h]} = \mathcal{K}_{j,\mathcal{Q}}^{i,\mathcal{P}} F^{j,\mathcal{Q}}_{[h]}\label{eq:hom_index}
\eeq
in the homogeneous case, and
\beq
F^{i,\mathcal{P}}=F_0^{i,\mathcal{P}}+ \mathcal{K}_{j,\mathcal{Q}}^{i,\mathcal{P}} F^{j,\mathcal{Q}}
\eeq
in the inhomogeneous case. The indices $i,\;j$ label the components, the multi-indices $\mathcal{P},\;\mathcal{Q}$ stand for all discretized momentum variables (summation over repeated indices is implied). The BSE kernel matrix $\mathcal{K}_{j,\mathcal{Q}}^{i,\mathcal{P}}$ is the same in both equations, and subsumes the interaction kernel, the dressed propagators of the constituents, the Dirac- and Lorentz structure, and the discretized integrations. It is applied to a vector $F_{(h)}^{i,\mathcal{P}}$ or $F^{i,\mathcal{P}}$, representing the homogeneous or inhomogeneous BSA.

\subsection{Homogeneous BSE}\label{sec:hommethod}
\begin{figure}
\begin{center}
\includegraphics[width=0.6\columnwidth,clip=true]{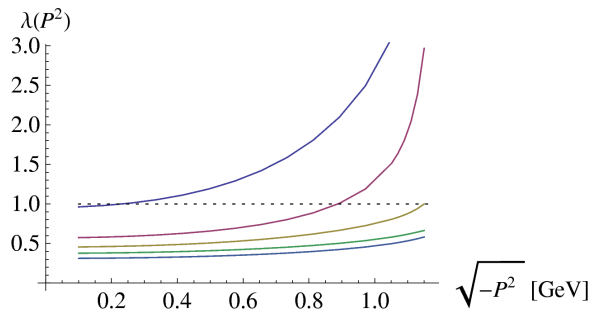}
\caption[The five largest eigenvalues of the homogeneous BSE plotted over $\sqrt{-P^2}$.]{\label{fig:hom_solution}The five largest eigenvalues of the homogeneous BSE plotted over $\sqrt{-P^2}$.
If $\lambda=1$ (indicated the dotted line), the bound state mass $M$ is given by $M=\sqrt{-P^2}$.
The ground-state (leftmost intersection) solution vector has positive C-parity (pion), the second
has negative (exotic) and the third again has positive C-parity (excited pion).}
\end{center}
\end{figure}

With the results of the preceding section, the homogeneous BSE, given in Eq.~(\ref{eq:hom_index}) in index notation, can be written as
\beq \label{eq:hom_bse_matvec}
\vec{F}_{[h]}=\mathcal{K}\cdot \vec{F}_{[h]}\;
\eeq
using matrix-vector notation. As already mentioned in Sec.~\ref{sec:hom_bse}, this equation is only valid at the on-shell points of the bound states in the respective channel, i.e.~at certain values of the total momentum squared $P^2=-M_n^2$, where $n=0,1,2,\ldots$ numbers the ground- and all excited states in the channel. To find such a value of $P^2$, one investigates the spectrum of $\mathcal{K}$ as a function of $P^2$, since Eq.~(\ref{eq:hom_bse_matvec}) corresponds to an eigenvalue equation (with the dependence on $P^2$ made explicit)
\beq \label{eq:hom_bse_matvec_lam}
\lambda(P^2)\vec{F}_{[h]}(P^2)=\mathcal{K}(P^2)\cdot \vec{F}_{[h]}(P^2)\;,
\eeq
where the eigenvalue $\lambda(P^2)=1$. In other words, to numerically approach a solution of the equation, a part of the result has to be already known, namely the values $M_n^2$, or --- more precisely --- the mass of the state one is looking for. The way out is a self-consistency argument, where the eigenvalue spectrum is plotted as a function of $P^2$ and those points with $\lambda_n(P^2)=1$ are identified: the largest eigenvalue determines the ground state, the smaller ones in succession the excitations of the system (cf.~Fig.~\ref{fig:hom_solution}).

A great variety of algorithms is available to numerically tackle these kinds of problems, and the most commonly used is a simple iterative method. Similar to the other algorithms discussed in this section, it relies on the multiplication of the matrix $\mathcal{K}$ on a vector and can successively be applied to find also excited states by projecting on states already obtained, see e.g., Ref.~\cite{Krassnigg:2003wy}. This simple method, however, is not able to resolve pairs of complex conjugate eigenvalues, which may occur in the BSE, cf.~\cite{Ahlig:1998qf}. In addition its convergence properties are not favorable, as demonstrated in Sec.~\ref{sec:efficiency}.

These difficulties are overcome by the use of more advanced algorithms. For this purpose, we utilize the implicitly restarted Arnoldi factorization \cite{Sorensen:1996aa} (implemented as the MPI-based ARPACK library), which is also frequently applied in lattice QCD studies, e.g.~\cite{Joergler:2007sh}.

\subsection{Inhomogeneous vertex BSE}
\label{sec:inhom_methods}
In the most compact notation, the inhomogeneous BSE \eq{inhom_bse} can be written as
\beq \label{eq:inhom_bse_matvec}
\vec{F}(P^2)=\vec{F_0}(P^2)+\mathcal{K}(P^2)\cdot \vec{F}(P^2)
\eeq
where the matrix $\mathcal K(P^2)$ is identical to the one in Eq.~(\ref{eq:hom_bse_matvec}), and the vector $\vec{F_0}$ is given by the decomposition of $\Gamma_0$ according to Eq.~(\ref{eq:decomp}), $\Gamma_0=\sum_i T_i F_0^i$ together with 
the discretization of a possible momentum dependence.

Again, the simplest method to treat this problem is a direct iteration. Mathematically, this corresponds to the representation of the solution by a von Neumann series (cf.~\cite[Chap.~4]{Delves:1985aa}), which can be shown to converge as long as the norm of the operator $\mathcal K$ is smaller than one, $\|\mathcal K \| < 1$. For matrices, this norm can be related to the largest eigenvalue, such that for $P^2>-M_0^2$, the iteration converges. When $P^2$ approaches the ground state position $-M^2_0$ from above, the number of iterations necessarily grows, and no convergence is obtained if $P^2\leq -M_0^2$, as demonstrated in Sec.~\ref{sec:efficiency}.

However, a solution is possible for any $P^2$ if one rewrites Eq.~(\ref{eq:inhom_bse_matvec}) as
\beq \label{eq:inhom_sol}
\vec{F} = (\mathbf 1 - \mathcal K)^{-1}\cdot \vec{F}_0 \;,
\eeq
i.e., $\vec{F}$ is given by the inhomogeneous term $\vec{F}_0$ multiplied by the matrix inverse of $(\mathbf 1 - \mathcal K)$. $\vec{F}$ can then be computed by e.g.~inverting the matrix exactly, which has been successfully used to resolve bound-state poles in the inhomogeneous BSA, as shown in \cite{Bhagwat:2007rj}. On the downside, the direct inversion of a matrix is computationally expensive, and it is not straightforward to parallelize the procedure.

A better approach is to view Eq.~(\ref{eq:inhom_sol}) as a linear system whose solution is to be found. Equations of this type are very common and several algorithms have been developed for their solution. In particular, if the matrix $(\mathbf 1 - \mathcal K)$ is big, Eq.~(\ref{eq:inhom_sol}) is a typical application for the so-called Conjugate Gradient (CG) algorithms. Many types of these iterative Krylov-space methods are available. In the case of the BSEs considered here, the matrices involved are neither hermitian nor symmetric, such that a good choice is the well-known Bi-Conjugate-Gradients stabilized (BiCGstab) algorithm \cite{vanderVorst:1992aa}, which is widely used for example in lattice QCD (cf.~\cite[Chap.~6.2]{Gattringer:2010zz}, where also the algorithm is described in detail). This is the method we used to obtain the results presented in Sec.~\ref{sec:inhomresults}.

\begin{figure}
\begin{center}
\includegraphics[width=0.65\columnwidth,clip=true]{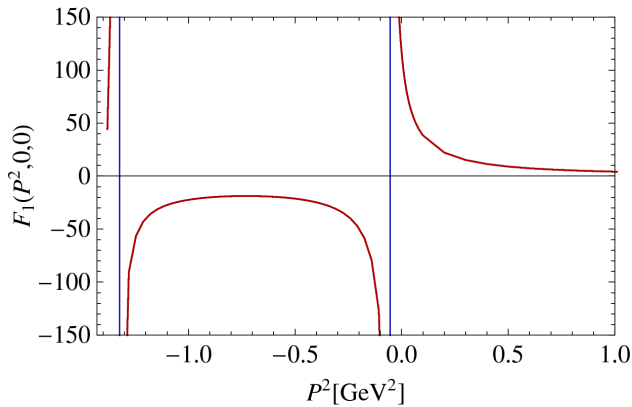}
\caption[Component $F_1(P^2,0,0)$ of the inhomogeneous pseudoscalar amplitude.]{Component $F_1(P^2,0,0)$ of the inhomogeneous pseudoscalar amplitude calculated using BiCGstab
vs. the square of the total momentum $P^2$. The vertical lines mark the pole positions, corresponding
to the pion ground- and first excited state ($J^{PC}=0^{+-}$).\label{fig:inhom_solutions}}
\end{center}
\end{figure}

\section{Illustration: the pseudoscalar BSEs}\label{sec:illustration}
As an illustration, we discuss in detail the application of the algorithms presented above to solve the homogeneous and inhomogeneous BSEs for pseudoscalar quantum numbers and compare their efficiency in terms of the number of matrix-vector multiplications needed to achieve a specified accuracy. We employ the rainbow-ladder truncation, i.e.~the
rainbow approximation in the quark DSE together with a ladder truncation of the corresponding quark-antiquark BSE. As effective interaction we use the model proposed by Maris and Tandy \cite{Maris:1999nt} with parameter $\omega=0.4$ and light quarks (for details concerning the truncation and the effective interaction, see App.~\ref{app:models}).

\subsection{Kernel setup}
Choosing the rest frame of the quark-antiquark system, and applying the parametrization and discretization 
described in App.~\ref{app:integration}, the kernel matrix Eq.~(\ref{eq:kernelmatrix}) in our setup becomes (we use the orthonormal pseudoscalar covariants constructed in App.~\ref{app:basis})
\begin{multline}
\mathcal K ^{i,r,s}_{j,l,m}(P) =  -\frac{4}{3 (2 \pi)^3} w[q_l^2] w[z_m] \int_{-1}^1 \!\! dy\; 
\frac{\mathcal G\left((p-q)^2\right)}{(p-q)^2}T^{\mu\nu}(p-q) \\
\times\mathrm{Tr}\left[ T_i(p,P) \gamma_\mu S(q_+)T_j(q,P) S(q_-) \gamma_\nu \right]\;,
\end{multline}
where $w[q_l^2]$, $w[z_m]$ denote the quadrature weights ($T^{\mu\nu}$ represents the transverse projector) and the replacements $p^2\rightarrow p_r^2$, $z_p\rightarrow z_s$, $q^2\rightarrow q_l^2$, $z\rightarrow z_m$ have been made in all occurring momenta to implement the discretization. Therefore, the indices $i;j$ label the components and $r,s;l,m$ the momentum space points. For these illustrative calculations, we use $N_q=32$ and $N_z=24$, such that $\mathcal K$ has the 
dimensions $(32,24,4)\times(32,24,4)$.

\subsection{Numerical efficiency}\label{sec:efficiency}
To compare the efficiency of our algorithm of choice to solve the homogeneous BSE, the implicitly restarted Arnoldi factorization, to the standard method we compute the one to four largest eigenvalues of $\mathcal K$ and compare the convergence in terms of the number of iterations needed to obtain an accuracy of the eigenvector of $\epsilon = 10^{-8}$, at a (typical) value of  $P^2=-M_0^2=0.0527\,\mathrm{GeV}^2$. For the simple iteration, our stopping criterion demands that the absolute change in any element of the eigenvector from one iteration step to the next does not exceed the desired accuracy $\epsilon$, while the vector is normalized to one with respect to the standard scalar product of $\mathbf C^n$, in order to be comparable to the ARPACK-library which works with the same scalar product.

\begin{figure}
\includegraphics[width=0.49\textwidth]{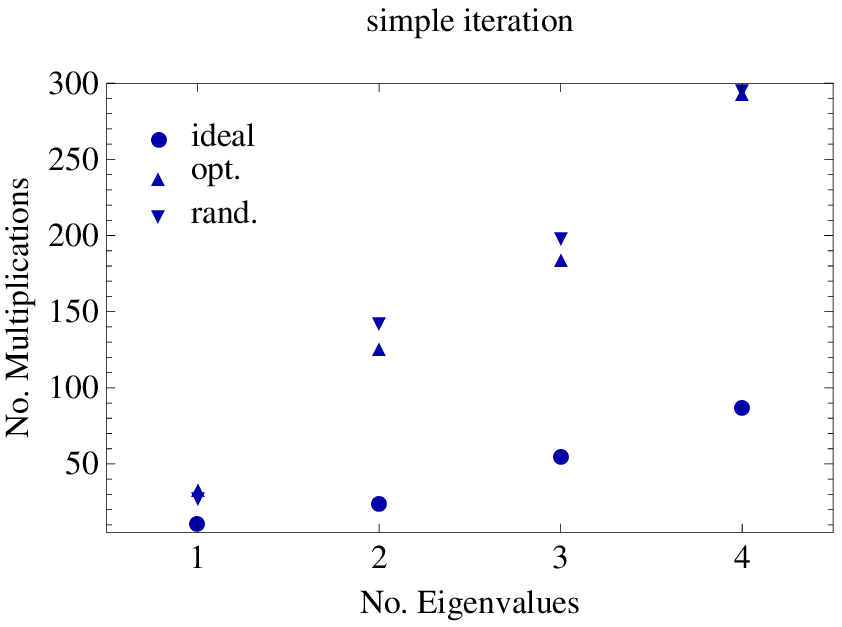}\hspace{0.015\textwidth}\includegraphics[width=0.49\textwidth]{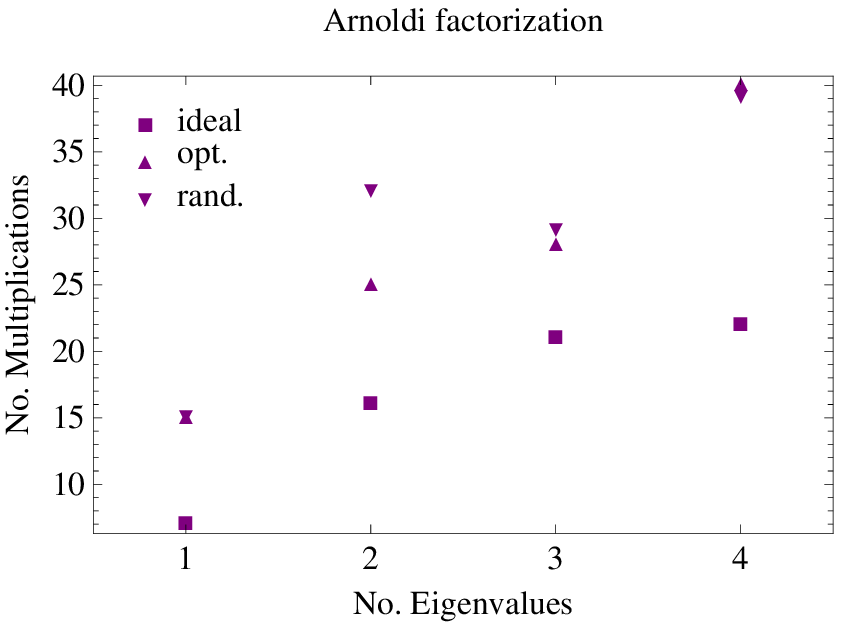}
\caption[The number of matrix-vector multiplications needed for convergence of the simple iteration and the Arnoldi factorization for different initial conditions.]{\label{fig:hom_initial} The number of matrix-vector multiplications needed for convergence of the
simple iteration (left panel) and the Arnoldi factorization (right panel), plotted against the number of eigenvalues computed at a (typical) fixed value of $P^2=-M_0^2$ for random (``rand''), optimized (``opt''), and ideal initial conditions.}
\end{figure}

We compare three different initial conditions for both iterative procedures. First, we use the default settings provided by ARPACK, which are based on a random vector (cf.~the ARPACK users guide \cite{Lehoucq:1998aa}) opposed to a truly random initial vector $F_{\textrm{in}}^j$ of complex numbers (where real and imaginary part are uniformly distributed in the interval $[0,1]$) for the simple iteration. This choice is denoted by ``rand''. Second, we use optimized initial conditions which anticipate the UV behavior of the amplitudes while not imposing further symmetries by choosing
\beq
F_{\textrm{in}}^j(P^2,q^2,z)=\frac{(1+z)(1+\ii)}{1+q^2}\;,
\eeq
which we denote by ``opt''. Third, we consider ``ideal'' initial conditions by utilizing the eigenvectors obtained at a nearby value of $P^2$, here $P^2=-0.053361\,\mathrm{GeV}^2$, as $F_{\textrm{in}}^j$. In ARPACK, if more than one eigenvalue is sought, the initial vector is set to the sum of the previously computed eigenvectors.

\begin{figure}
\begin{center}
\includegraphics[width=0.65\columnwidth,clip=true]{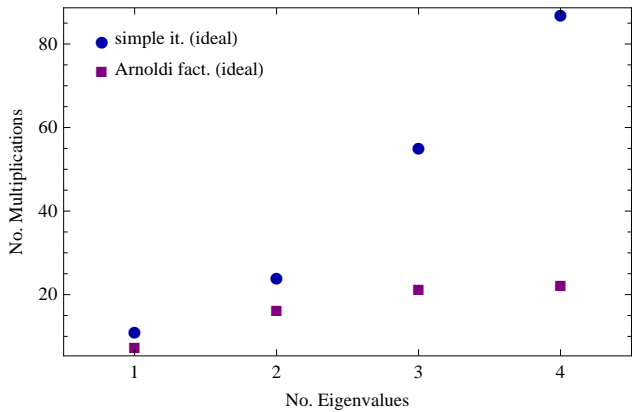}
\caption[Direct comparison of the number of matrix-vector multiplications needed for convergence of the
simple iteration and the Arnoldi factorization.]{\label{fig:hom_compare} The number of matrix-vector multiplications needed for convergence of the
simple iteration (circles) and the Arnoldi factorization (squares), plotted against the number of eigenvalues computed at
a (typical) fixed value of $P^2=-M_0^2$, for ideal initial conditions.}
\end{center}
\end{figure}

The results are given in Figs.~\fig{hom_initial} and \fig{hom_compare}, where the sensitivity of both algorithms to the initial conditions and the comparison of the efficiency in the ideal case are shown. From Fig.~\fig{hom_initial} it is clear that the Arnoldi factorization is less sensitive to a change in initial conditions than the simple iteration, and in general is more efficient. Even in the ideal case (cf.~Fig.~\fig{hom_compare}) for the first eigenvalue the advanced algorithm is $36\%$ more efficient ($7$ iterations compared to $11$), and becomes even more advantageous for an increasing number of eigenvalues.

A further interesting observation from Fig.~\ref{fig:hom_initial} is that the Arnoldi factorization for random initial conditions was more efficient for three eigenvalues than when only two were requested. This is most likely due to a ``clustering'' of eigenvalues number two and three for the algorithm, an effect which appears for eigenvalues close together
and is also related to the eigenvectors. In this particular case, eigenvectors two and three have opposite $C$-parity
or $z$-symmetry (cf.~App.~\ref{app:cparity}), which may make them more easily distinguishable for the algorithm and more easy to obtain as a result. The ARPACK library is very efficient at evaluating all eigenvalues in such a cluster, while convergence is slower if one asks for only one or a few of the eigenvalues in the cluster (see also \cite{Lehoucq:1998aa}).

To solve the inhomogeneous vertex BSE (\ref{eq:inhom_bse}) we apply the direct iteration (summation of the von Neumann series) and the inversion using the BiCGstab algorithm, in the setup described above for pseudoscalar quantum numbers. In this case not only the structure of the amplitude, but also the structure of the inhomogeneous term $\Gamma_0$ determine the quantum numbers of the solution. Following \cite{Maris:1997hd}, a possible choice for pseudoscalars is
\beq \label{eq:gamma0}
\Gamma_0=Z_4\gamma_5\;,
\eeq
with $Z_4$ the renormalization constant from Eq.~\eq{z2z4}. With this choice (pseudoscalar, positive C-parity), as described in App.~\ref{app:cparity} no poles corresponding to negative C-parity appear in the solution, as can be seen from Fig.~\ref{fig:inhom_solutions}. The curve shown in this figure has been obtained with the BiCGstab algorithm, because as described in Sec.~\ref{sec:inhom_methods} the direct iteration fails to converge if $P^2\leq-M_0^2$. 

\begin{figure}
\begin{center}
\includegraphics[width=0.65\columnwidth,clip=true]{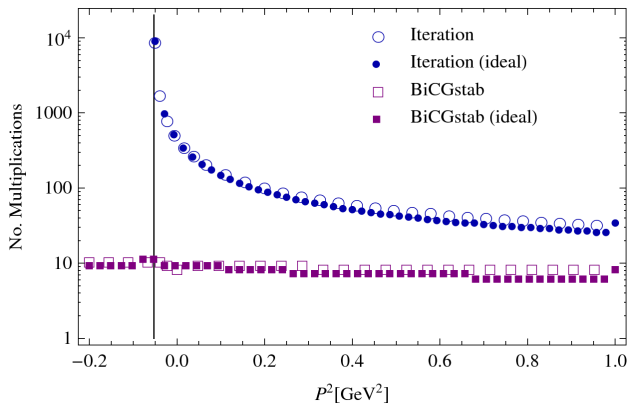}
\caption[The number of matrix-vector multiplications needed for convergence
of the iterative solution of the inhomogeneous vertex BSE and the BiCGstab algorithm.]{\label{fig:inhom_compare} The number of matrix-vector multiplications needed for convergence
of the iterative solution of Eq.~(\ref{eq:inhom_bse}) (circles) and the BiCGstab algorithm (squares), plotted on a logarithmic scale against the square of the total momentum $P^2$ of the amplitude, for ``ideal'' initial conditions (full symbols) as well as for $F_{\textrm{in}}^j=0$ (open symbols). The vertical line indicates the position of the ground state $P^2=-M_0^2$ of the system. Note that the straightforward iteration does not converge for $P^2\leq-M_0^2$.}
\end{center}
\end{figure}

This is demonstrated in Fig.~\ref{fig:inhom_compare}, where the number of matrix-vector multiplications needed for convergence is plotted against $P^2$ for both methods. Here, we compare two choices of initial conditions: ``ideal'' starting values analogous to those discussed for the homogeneous equation, where the result of the calculation for the preceding value of $P^2$ (starting from $P^2=1\,\textrm{GeV}^2$ in this case) is taken as initial guess, and the standard choice $F_{\textrm{in}}=0$. For both algorithms, the inversion was calculated to an accuracy of $\epsilon = 10^{-8}$, such that 
$$\|\vec{F}_0-(\mathbf 1 - \mathcal K)\cdot F_\textrm{fin}\|\leq \epsilon\;,$$
where $ F_\textrm{fin}$ denotes the numerical result of the inversion, and the norm is a maximum-norm, which corresponds to the maximal absolute value of any element of the vector. 

It is clear that, independent of the initial condition, the number of matrix-vector multiplications needed for the direct iteration diverges as $P^2$ approaches $-M_0^2$ (note that Fig.~\ref{fig:inhom_compare} uses a logarithmic scale on the vertical axis). The inversion with BiCGstab, however, converges for all $P^2$ with nearly the same speed, needing approximately 10 matrix-vector multiplications. 

A further observation from Fig.~\ref{fig:inhom_compare} is that very close to the pole the standard initial conditions lead to faster convergence
(by 1 iteration) than the ideal ones. This is most likely connected to the sign change in the components due to the pole, which is not taken into
account in our definition of ideal initial conditions. In addition, these initial conditions also strongly depend on the step size.

\section{Other quantum numbers}\label{sec:quantnum}
In order to check whether the results described above are limited to the special case of pseudoscalar quantum numbers, we discuss the application of the algorithms presented in this work to states with different parity and spin.

In the homogeneous case, we consider scalar, vector, axialvector and tensor quantum numbers and use the same model and parameters as above in the pseudoscalar case. The bound states in these channels have been investigated thouroughly in \cite{Krassnigg:2009zh} and recently \cite{Krassnigg:2010mh}, where all details concerning the parameter dependence and a comparison to experiment are given. 

For the different quantum numbers, we compute the first eigenvalue (which on-shell corresponds to the ground state) for the four values of $J^{PC}=0^{++}\,,\;1^{--}\,,\;1^{+-}\,,\;2^{++}$ at a typical value of $P^2$ employing the Arnoldi factorization as well as the simple iteration, for both optimal and ideal initial conditions, as explained above. Note that the number of momentum-space points used for the integration is the same in all cases, while the number of covariants differs. For pseudoscalars and scalars, the basis consists of four covariants, while for the other quantum numbers eight covariants have to be used (see, e.g., \cite{Krassnigg:2010mh}), which increases the size of the BSE kernel matrix.

The results, collected in Tab.~\ref{tab:quantnumhom}, show that in each case the Arnoldi factorization was more efficient. Indeed, the advanced algorithm seems to be even more advantageous when applied to more complicated systems, which advocates its use also in studies of e.g.~baryons.

\begin{table}
\caption{Number of matrix-vector multiplications needed to achieve convergence of the Arnoldi factorization (Arn.) and the simple iteration (Iter.) when applied to the homogeneous BSE for the indicated quantum numbers $J^{PC}$ at typical values of the total momentum squared $P^2$, for ``optimized'' and  ``ideal'' initial conditions (with starting values computed at $-(\sqrt{-P^2}-0.01)^2$ in each case). \label{tab:quantnumhom}}
\begin{center}
\begin{tabular}{cc|cc|cc}
 \multicolumn{2}{c}{Initial conditions:} & \multicolumn{2}{c}{optimal} &\multicolumn{2}{c}{ideal} \\
$J^{PC}$    & $P^2$[$\textrm{GeV}^2$]  & Arn.&Iter.&Arn.&Iter.\\
\hline
$0^{++}$  & -0.509796 & 15 & 40   &  10 &  15 \\
$1^{--}$  & -0.599695 & 15 & 74   &  10 &  31  \\
$1^{+-}$  & -0.739772 & 20 & 103  &  10 &  15  \\
$2^{++}$  & -1.254400 & 20 & 63   &  15 &  46  \\
\end{tabular}
\end{center}
\end{table}

\section{Comparison of homogeneous and vertex BSE}\label{sec:hominhomcompare}
As shown in Sec.~\ref{sec:inhomresults}, it is possible to calculate observables like meson masses and decay constants by using only the inhomogeneous vertex BSE instead of the homogeneous BSE. Therefore, it is interesting to compare the numerical efficiency of the solution methods for both equations directly, in order to see if the use of the vertex BSE provides a numerical advantage.

To this end, we compare the matrix vector multiplications needed to solve the inhomogeneous BSE using the BiCGstab algorithm for a wide range of total momentum-squared to the solution of the eigenvalue problem for the homogeneous equation (using the Arnoldi factorization) in the same range for one and three eigenvalues. This calculation allows to obtain the masses of the pseudoscalar ground state and the first excitation for positive $C$-parity in our setup.

\begin{figure}
\begin{center}
\includegraphics[width=0.7\columnwidth,clip=true]{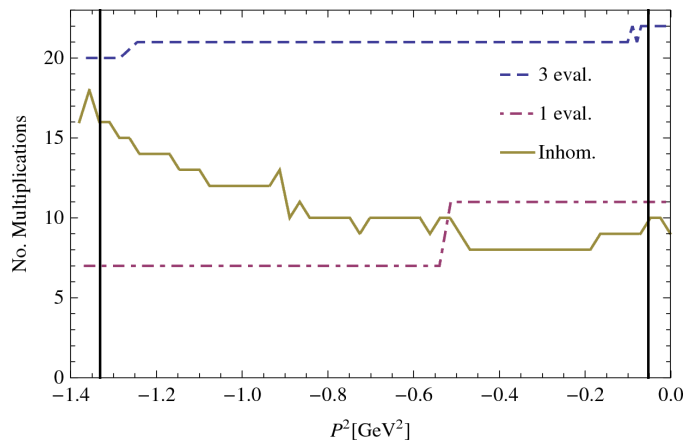}
\caption[Comparison of the efficiency of the solution of the homogeneous vs.~the inhomogeneous BSE in the pseudoscalar case.]{\label{fig:hom_inhom_compare} The number of matrix-vector multiplications needed for convergence
of the inhomogeneous BSE using BiCGstab (solid line) compared to the calculations of one eigenvalue (dash-dotted line) and three eigenvalues
(dashed line) with the Arnoldi factorization. The vertical lines mark the ground and excited state in the $0^{-+}$ channel at
$P_0^2=-0.0527\textrm{GeV}^2$ and $P_1^2=-1.3315\textrm{GeV}^2$.}
\end{center}
\end{figure}

The results of the comparison are given in Fig.~\ref{fig:hom_inhom_compare}. Even though ideal initial conditions were used in all cases, the solution of the inhomogeneous BSE took fewer multiplications than the eigenvalue calculations in the region of the on-shell points of the ground state (corresponding to the first eigenvalue) and the first excitation (corresponding to the third eigenvalue), which are indicated by the vertical lines in the figure. This shows that in this case it is more efficient in terms of matrix-vector multiplications to use the inhomogeneous BSE to calculate the mass spectrum, especially the excited state.

Demanding an even closer numerical correspondence of homogeneous and vertex BSE, we have also tried to extract the entire homogeneous ground state amplitude from the inhomogeneous BSA by fitting the residues in $P^2$ at each point of our relative-momentum-squared grid in a rather naive approach. While this appears possible in principle, the limiting factors in this case are the accuracies of the fit as well as the determination 
of the residues at the pole of the inhomogeneous vertex amplitude under investigation. In our pseudoscalar example, we were able to obtain the leading component of the homogeneous amplitude (which corresponds to the covariant $T_1$ in Eq.~(\ref{eq:tpsN})) with an accuracy of $2.6\%$ when
compared to the solution of the corresponding homogeneous BSE. The other components showed somewhat larger deviations such that this procedure 
cannot be performed without resorting to more involved fitting methods. Thus we conclude that, while obtaining the homogeneous BSA from the
corresponding inhomogeneous vertex BSE is not impossible, with comparable numerical effort a determination directly from the homogeneous BSE is much more accurate.

\begin{figure}
\begin{center}
\includegraphics[width=0.7\columnwidth,clip=true]{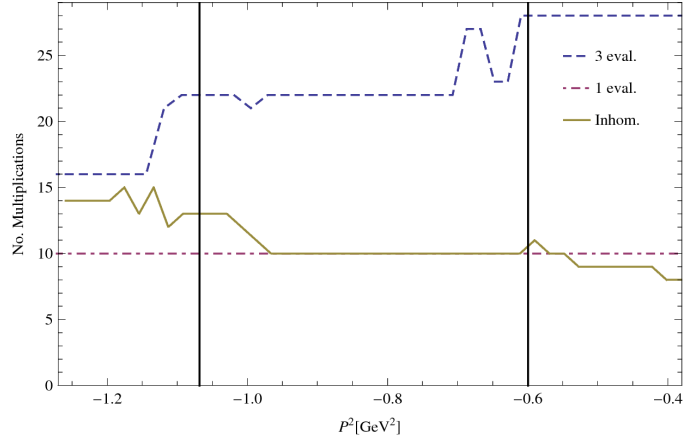}
\caption[Comparison of the efficiency of the solution of the homogeneous vs.~the inhomogeneous BSE for $J^{PC}=1^{--}$.]{\label{fig:hom_inhom_compareVE} The number of matrix-vector multiplications needed for convergence
of the inhomogeneous BSE using BiCGstab (solid line) compared to the calculations of one eigenvalue (dash-dotted line) and three eigenvalues
(dashed line) with the Arnoldi factorization. The vertical lines mark the ground and excited state in the $1^{--}$ channel at
$P_0^2=-0.5997\textrm{GeV}^2$ and $P_1^2=-1.0682\textrm{GeV}^2$.}
\end{center}
\end{figure}

The results for the $1^{--}$ channel are presented in Fig.~\ref{fig:hom_inhom_compareVE}. For the excited state,
the advantages of the inhomogeneous equation are even more pronounced than for the pseudoscalar channel, whereas for the ground state the two
equations are almost equivalent in terms of efficiency.

\chapter{The chiral phase transition}\label{chap:fint}
The framework of Dyson-Schwinger and Bethe-Salpeter equations provides a consistent approach to QCD phenomenology, which is very successful in describing not only the properties of the fundamental degrees of freedom, the quarks and gluons, but also mesons and baryons in the vacuum. In this setup,  various aspects of chiral symmetry and its dynamical breaking have been studied some time ago via the DSE of the quark propagator (see, e.g.~\cite{Maris:2003vk} and references therein). For the calculation of bound states, sophisticated studies in different truncations have been performed (see e.g.~\cite{Krassnigg:2009zh,Eichmann:2007nn,Eichmann:2009qa,Fischer:2009jm,Fischer:2008wy}). 

%wuwuwuwuwuwuwuwuwuwuwuwuwuwuwuwuwuwuwuwuwuwuwuwuwuwuwuwuwuwuwuwuwuwuwuwuwuwuwuwuwuwu
As discussed in Sec.~\ref{sec:fintform}, it is straightforward to extend the formalism to finite temperature and chemical potential, and several studies consider the restoration of chiral symmetry at finite temperature and the behavior of meson masses at and beyond the phase transition at zero chemical potential \cite{Maris:2000ig,Blaschke:2000gd,Blaschke:2000zm,Blaschke:2006ss,Horvatic:2007wu,Horvatic:2007qs,Alkofer:1989vr}. However, the situation is complicated by the additional variables introduced via the Matsubara formalism, such that most of these studies resort to simplifications that are already overcome in the zero temperature case. In addition, certain aspects of the phase diagram \cite{Maris:1997eg,Blaschke:1999ab,Nickel:2006vf,Nickel:2006kc,Chen:2008zr,Klahn:2009mb} including the deconfinement transition \cite{Fischer:2009wc,Fischer:2009gk,Fischer:2010fx} have also been studied in this formalism.

In the present chapter, which is based on Ref.~\cite{Blank:2010bz}, we focus on quarks at finite temperature and study the chiral phase transition via the properties of the quark propagator in rainbow truncation. It was shown some time ago that certain classes of models in the QCD gap equation in this truncation yield a second-order phase transition with mean-field critical exponents \cite{Holl:1998qs,Maris:1999bj}. While the main emphasis of that investigation was the critical behavior and universality class of the phase transition, it also provided the corresponding values for $T_c$, which ranged between $120$ and $174$ MeV among the model interactions investigated. In somewhat different setups, using a separable interaction kernel in the gap equation \cite{Blaschke:2000gd,Blaschke:2006ss,Blaschke:2007ce,Horvatic:2007wu} or neglecting retardation effects \cite{Alkofer:1989vr}, one obtains values in the range between $110$ and $146$ MeV.

The pointwise behavior of all these interactions as functions of the gluon momentum is rather different, although in all cases it was determined by adjusting the relevant parameters to meson phenomenology at zero temperature. Therefore, it is interesting to ask whether there is any simple relationship between the momentum dependence of the interaction and the value of $T_c$. Such a relationship could e.g.~be analogous to that found recently in studies of meson properties using a particular form of model interaction with a one-parameter setup. There the value of the free parameter determines an effective range of the intermediate- and low-momentum parts of the interaction. While ground-state properties of pseudoscalar and vector mesons were unaffected by variations in the model parameter, masses of excitations of any kind (orbital or radial) showed a strong and systematic dependence on the model parameter \cite{Holl:2004fr,Holl:2004un,Holl:2005vu,Krassnigg:2008gd,Krassnigg:2009zh} thus identifying excited-state properties as prime targets to study the particular details of the effective interaction in the DSE formalism, in particular in the nonperturbative regime (cf.~also Sec.~\ref{sec:grexc}). In an analogous fashion, the goal here is to identify $T_c$ as a quantity with the same capability.

\section{Modeling at finite temperature}\label{sec:fintmodel}
The QCD gap equation at finite temperature, Eq.~\eq{gapfint}, retains the same form it has at zero temperature (cf.~Fig.~\ref{fig:gapfull}), such that for a solution the quark-gluon vertex as well as the gluon propagator are needed. Here, however, we apply the rainbow-truncation, where, as stated in App.~\ref{app:models}, the gluon propagator and quark-gluon vertex are taken as bare and multiplied by an effective interaction in order to mimic the behavior of the full system. At finite temperature, the effective interaction is modified such that it follows the structure of the gluon propagator, Eq.~\eq{gluonprop}, which has two dressing functions. Defining $s := \vec{k}\,^2 + \Omega^2 + m_g^2$ with  the Debeye mass $m_g$ \cite{Bender:1996bm}, the effective dressing functions are given by
\begin{eqnarray}
G(\vec{k},\Omega)&=&\frac{\mg(s)}{s} \\
F(\vec{k},\Omega)&=&\left.\frac{\mg(s)}{s}\right|_{m_g=0}\;.
\end{eqnarray}
In the following, we investigate the impact of different forms of the effective interaction $\mg$ on the chiral phase transition. We consider the models proposed by Munczek and Nemirovsky (MN) \cite{Munczek:1983dx}, Maris and Roberts (MR) \cite{Maris:1997tm}, Maris and Tandy (MT) \cite{Maris:1999nt}, and Alkofer, Watson, and Weigel (AWW) \cite{Alkofer:2002bp}. The different functional forms of these effective interactions and the corresponding parameter sets are collected in App.~\ref{app:models}. 

\section{Scaling analysis and critical exponents}
\begin{figure}
\begin{center}
\includegraphics[width=0.7\columnwidth,clip=true]{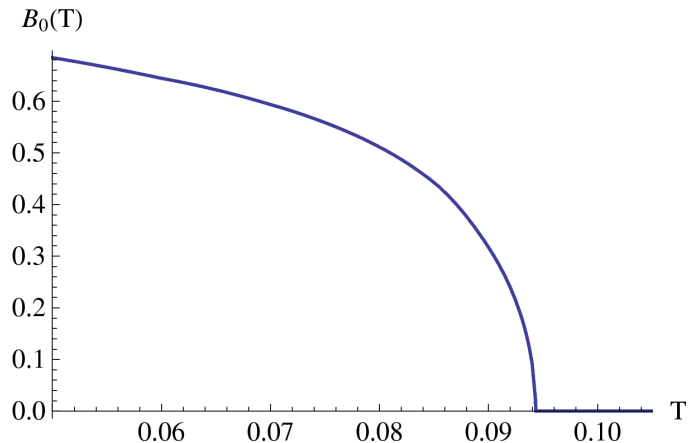}
\end{center}
\caption[The order parameter $B_0$ vs. temperature $T$.]{The order parameter $B_0$ vs. temperature $T$, shown exemplarily
for model MT2.\label{fig:orderpar}}
\end{figure}

With all ingredients of the QCD gap equation at hand one can obtain a solution numerically and study the chiral phase transition temperature $T_c$ as well as the nature of the transition. The corresponding order parameter is the chiral condensate; Here, however, we use the equivalent order parameter $B_0:=B(0,\omega_0)$, since it is easier to access and can be calculated more accurately. A typical behavior of the order parameter for a chiral-limit solution is shown in Fig.~\ref{fig:orderpar} and indicates a second order phase transition. Indeed, it has been found previously that the rainbow-truncated quark DSE generally yields a second order chiral phase transition with mean-field critical exponents \cite{Holl:1998qs,Maris:1999bj}. Using the scaling analysis described below, we have confirmed this behavior for all interactions considered here and thus enlarged the set of interaction types it had been shown for by MT and AWW. 

In the case of a second order phase transition, the order parameter $B_0(t,h)$ obeys the scaling laws
\begin{eqnarray}
B_0(t,h) &\propto& \left.(-t)^{\beta}\right|_{h=0}\;\;\;t\rightarrow0^-\quad \text{and}\label{eq:scaling1}\\
B_0(t,h) &\propto& \left.h^{1/\delta}\right|_{t=0}\;\;\;h\rightarrow0^+\;.\label{eq:scaling2}
\end{eqnarray}
Here, $t = \frac{T}{T_c}-1$ is the reduced temperature and $h=\frac{m}{T}$ the reduced mass, a measure for the explicit breaking of chiral symmetry by a non-vanishing current quark mass $m$. $\beta$ and $\delta$ are the critical exponents of the phase transition.

\begin{figure}
\begin{center}
\includegraphics[width=0.495\textwidth]{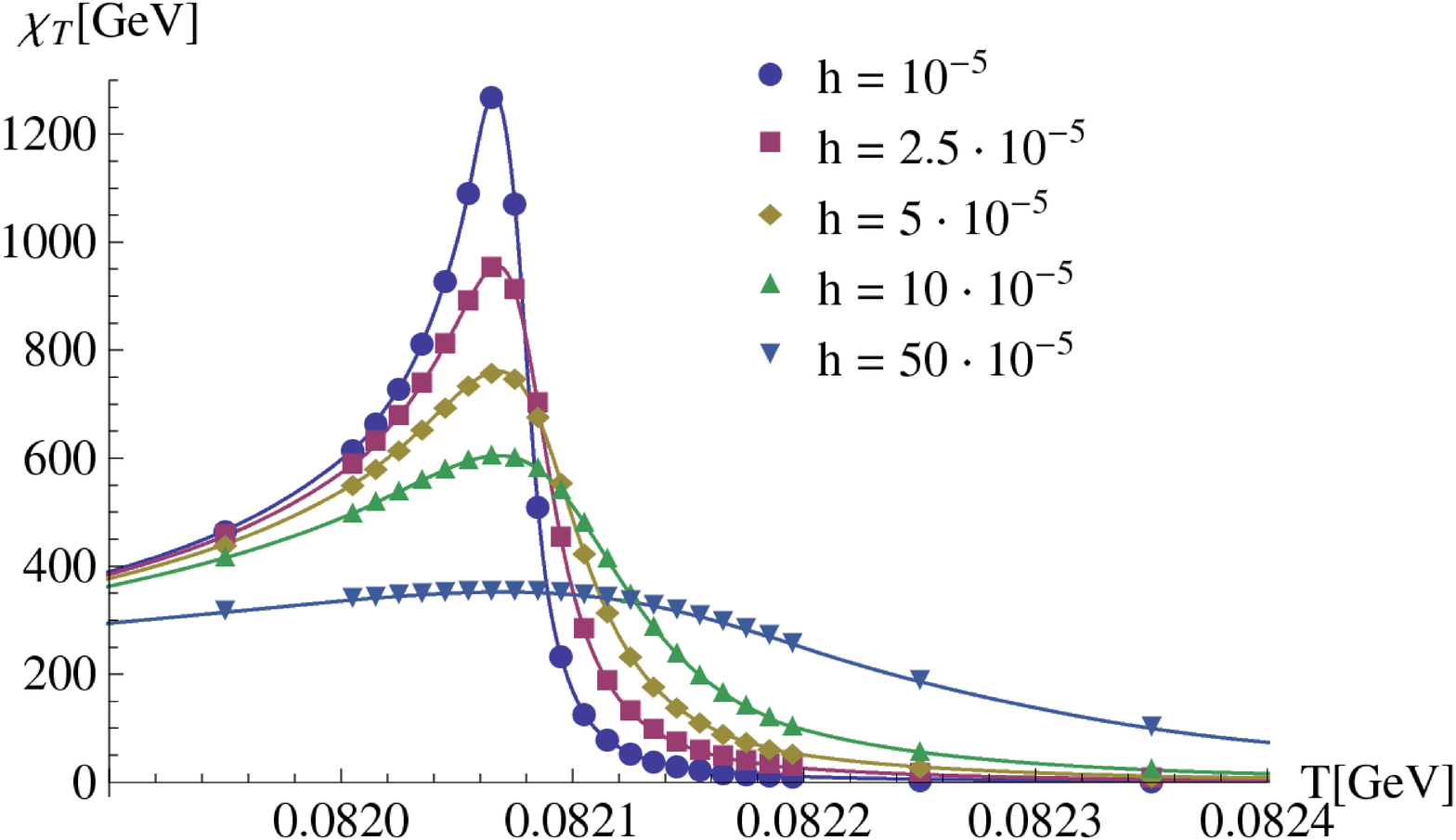}
\includegraphics[width=0.495\textwidth]{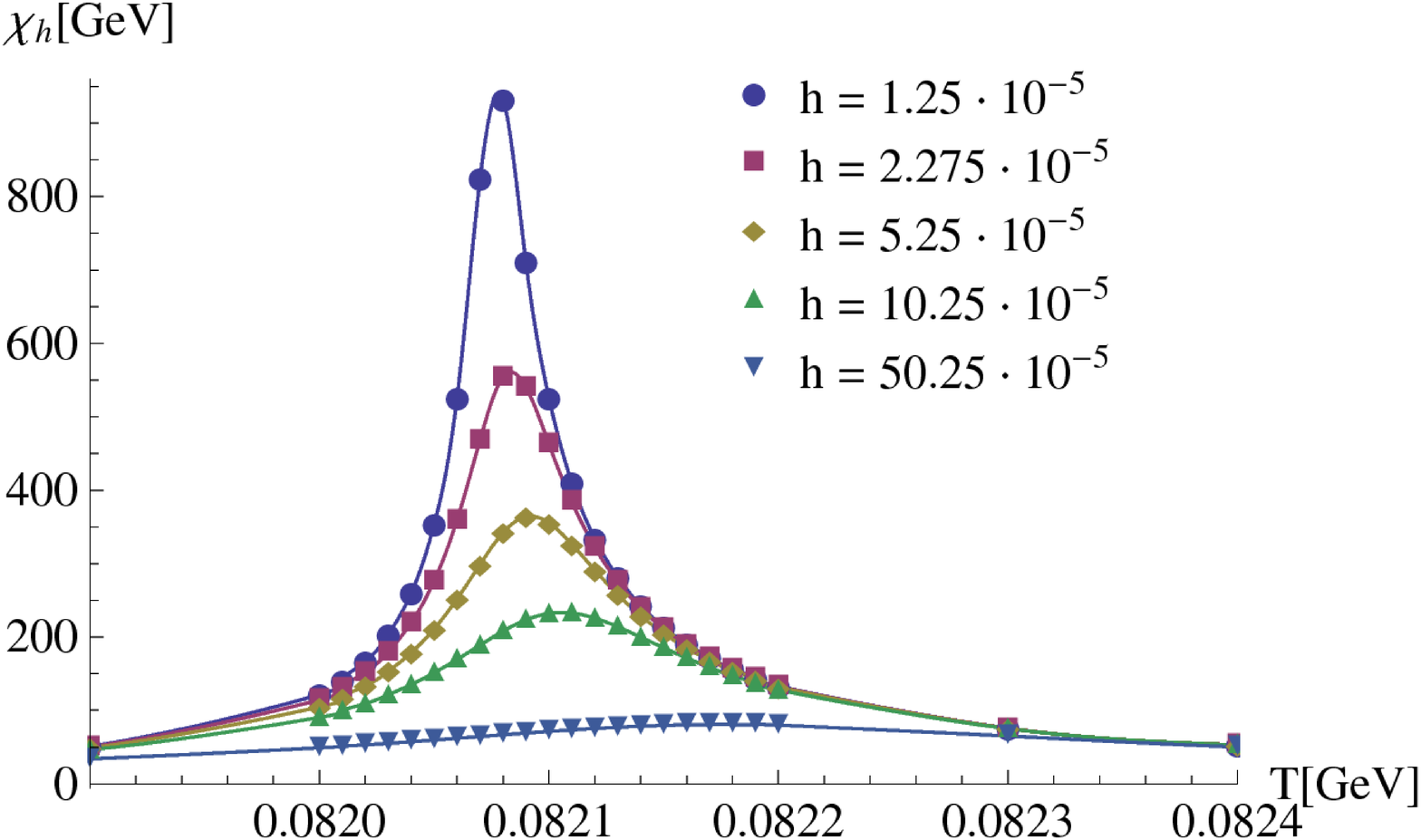}
\end{center}
\caption[Chiral susceptibilities $\chi_T(T)$ and $\chi_h(T)$ for the MT1 model
versus temperature $T$ for various values of $h$.]{Chiral susceptibilities $\chi_T(T)$ (left) and $\chi_h(T)$ (right) for the MT1 model
plotted versus temperature $T$ for various values of $h$. The curves between the data points were obtained by
interpolation using cubic splines.\label{fig:pseudocritical}}
\end{figure}

Although it is straightforward to use Eqs.~(\ref{eq:scaling1})-(\ref{eq:scaling2}) to obtain $\beta$ and $\delta$, this procedure requires to solve the gap equation at $h=0$ or $T=T_c$. This is numerically difficult, such that a direct evaluation does not allow fits to extract $T_c$ with the necessary precision and reliably observe scaling. Therefore we exploit further scaling relations and use chiral susceptibilities for our analysis. They are defined by
\begin{eqnarray}
 \chi_T &:=& \left.-T_c \frac{\partial B_0(T,h)}{\partial T}\right|_{h\;\text{fixed}}\;,\quad\text{and}\\
\chi_h &:=& \left.\frac{\partial B_0(T,h)}{\partial h}\right|_{T\;\text{fixed}}\;.
\end{eqnarray} 
The maxima of these quantities for nonvanishing $h$ are referred to as pseudocritical points, $\chi_T^{pc}$ and $\chi_h^{pc}$, respectively. The corresponding pseudocritical temperatures are denoted by $T_T^{pc}$ and $T_h^{pc}$. They are obtained as the maxima of the chiral susceptibilities with respect to temperature, as depicted in Fig.~\ref{fig:pseudocritical}.

\begin{figure}
\begin{center}
\includegraphics[width=0.7\columnwidth,angle=0,clip=true]{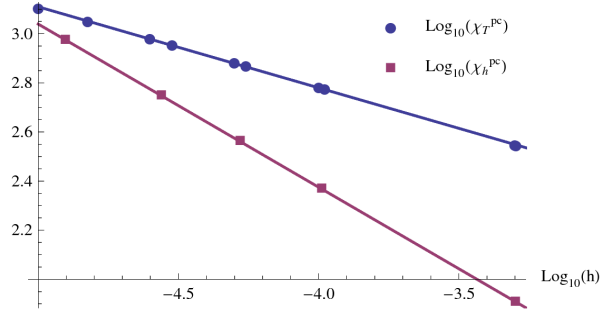}
\end{center}
\caption[Peak heights of the chiral
susceptibilities $\chi_T^{pc}$ and $\chi_h^{pc}$ plotted versus the reduced mass $h$.]{Peak heights of the chiral
susceptibilities $\chi_T^{pc}$ and $\chi_h^{pc}$ plotted versus the reduced mass $h$ on log-log
scale, for the MT1 model. The lines are linear fits through the calculated points which are used
to obtain the critical exponents according to Eqs.~(\ref{eq:scaling3}) - (\ref{eq:scaling4}).\label{fig:scaling}}
\end{figure}

The pseudocritical points $\chi_T^{pc}$ and $\chi_h^{pc}$ also obey scaling laws. Their behavior for
$T\sim T_c$ and $h\sim 0$ is described by
\begin{eqnarray}
 \chi_T^{pc} &\propto& h^{-1+1/\delta}\;,\label{eq:scaling3}\\
\chi_h^{pc} &\propto& h^{\frac{1}{\beta\delta}\left(1-\beta\right)}\;.\label{eq:scaling4}
\end{eqnarray}
Following \cite{Blaschke:1998mp} we use these scaling relations to obtain $\beta$, $\delta$, and
$T_c$ for all models and parameter sets given in Tab.~\ref{tab:models}, as illustrated
in Fig.~\ref{fig:scaling}.

\section{Chiral transition temperature}
As already mentioned above, all interactions yield a second order phase transition with mean-field critical exponents, regardless of the strength of the interaction, its range, or its pointwise behavior in any particular momentum range. In particular, the presence of the $\delta$-function term does not make a difference in this respect. Another general observation is the fact that the function $D(\vec{p}\,^2,\omega_k)$ in Eq.~(\ref{eq:sfint}) is identically zero for all interactions in rainbow truncation.

The values obtained for the chiral transition temperature $T_c$, however, are rather different among the various interactions. In the following we will argue how this feature can be exploited to discern various forms within a given truncation. The results for $T_c$ are summarized in Tab.~\ref{tab:results} for the four interactions on a range of model parameters well-used in meson phenomenology.

\begin{table}
\caption{Results for the transition temperature $T_c$ rounded to MeV for the interactions
and parameter sets defined in
Tab.~\ref{tab:models}.\label{tab:results}}
\begin{center}
\begin{tabular}{l r r r r }
$\omega$ [GeV] & N/A & 0.3 & 0.4 & 0.5  \\\hline\hline
Model& MN & & &  \\
$T_c$& 169& & &  \\\hline
Model& & AWW1& AWW2& AWW3 \\
$T_c$& &  82 &  94 & 101  \\\hline
Model& &  MT1&  MT2&  MT3\\
$T_c$& &  82 &  94 &  96  \\\hline
Model& &  MR1&  MR2&  MR3\\
$T_c$& & 120 & 133 & 144\\\hline\hline
\end{tabular}
\end{center}
\end{table}

The MN interaction has to be treated somewhat separately, since the only free parameter in this case is $D$ whose choice completely determines $\mathcal{G}$. In terms of plain numbers it is interesting to see that MN yields the highest value for $T_c$, followed by MR. AWW and MT give the smallest numbers.

\begin{figure}
\subfigure[$T_c$ vs.~$\omega$]{\includegraphics[height=0.24\columnwidth,angle=270,clip=true]{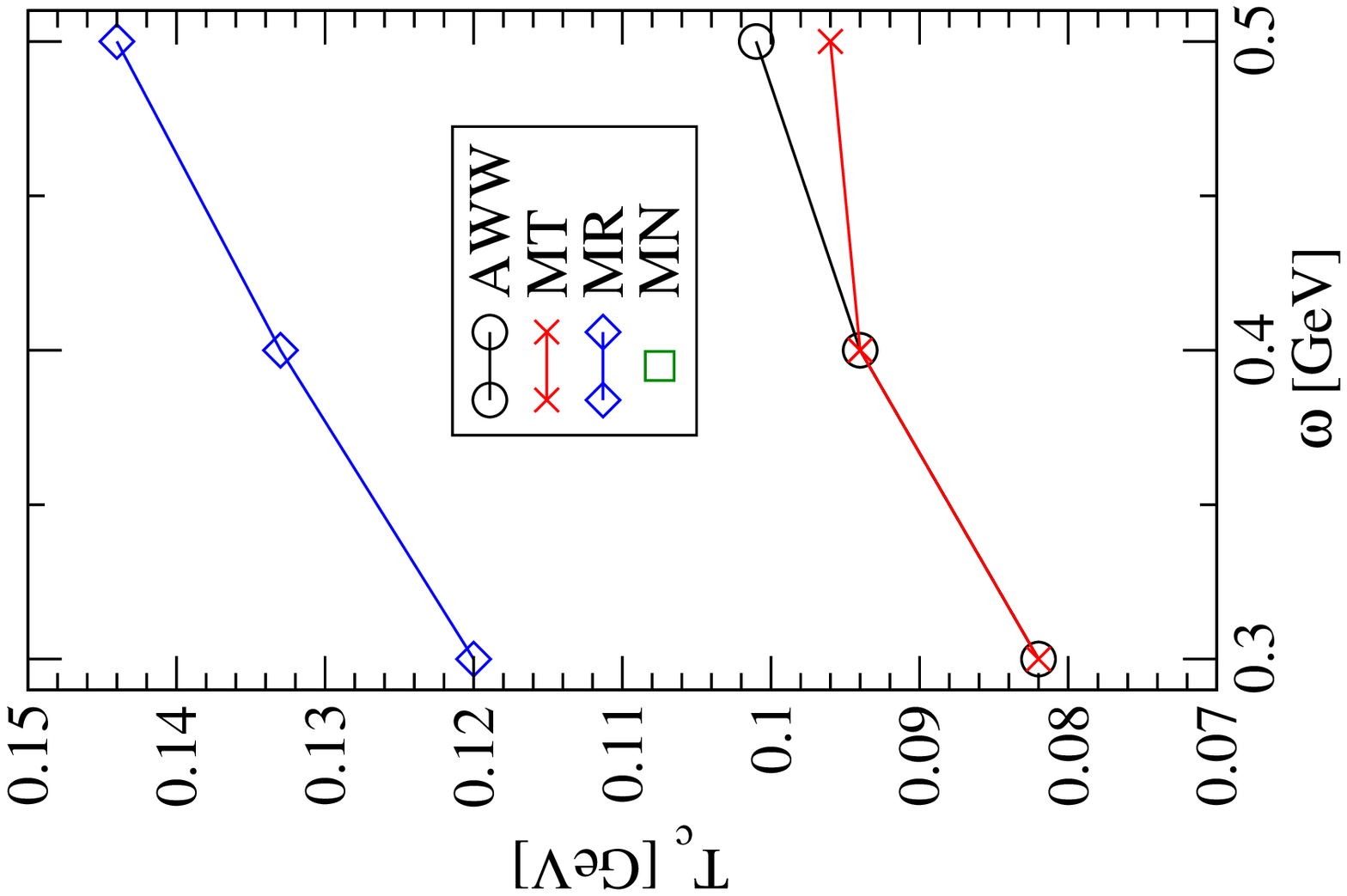}}
\subfigure[$T_c$ vs.~$\omega\,D$]{\includegraphics[height=0.24\columnwidth,angle=270,clip=true]{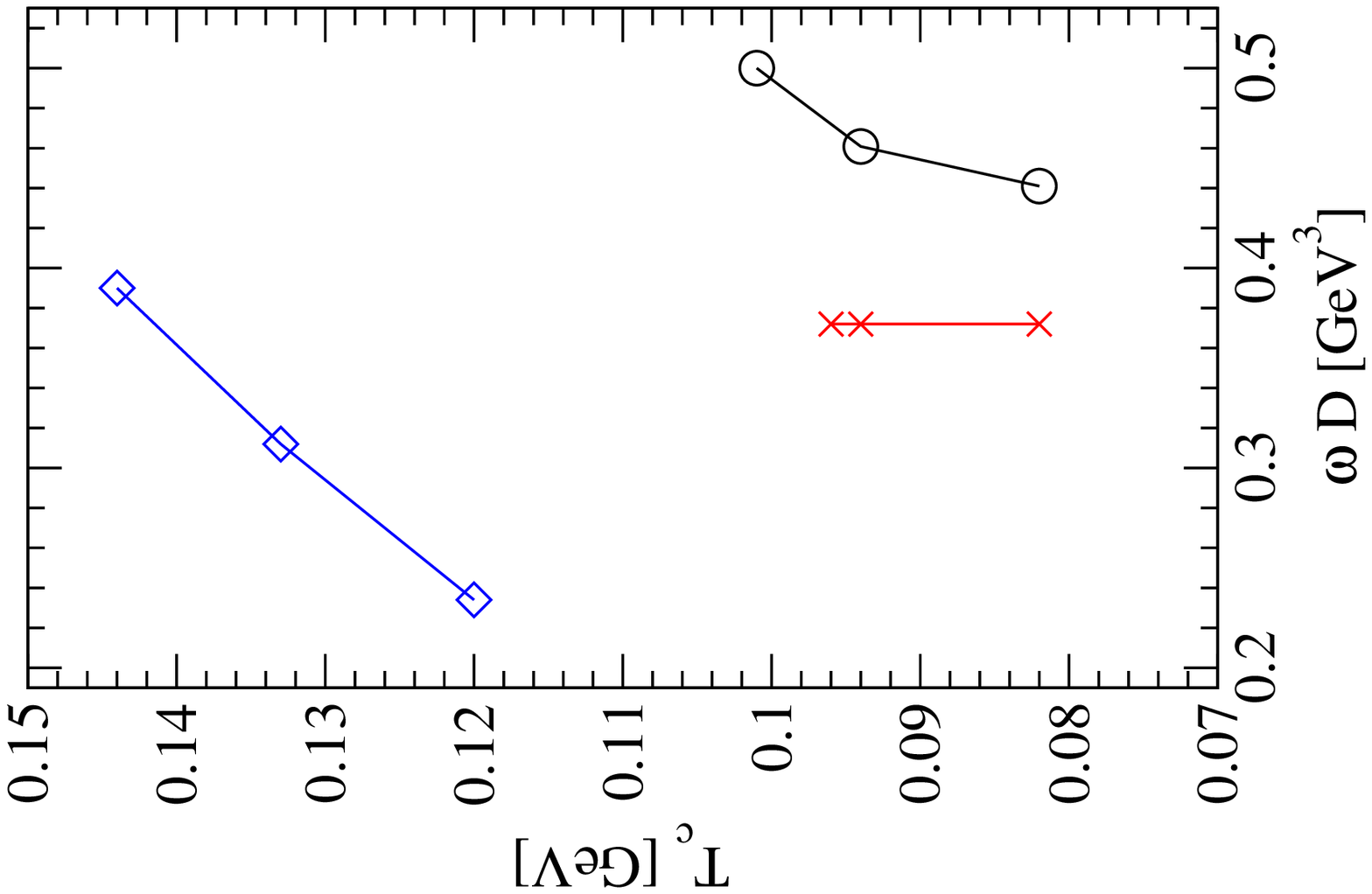}}
\subfigure[$T_c$ vs.~$D$]{\includegraphics[height=0.24\columnwidth,angle=270,clip=true]{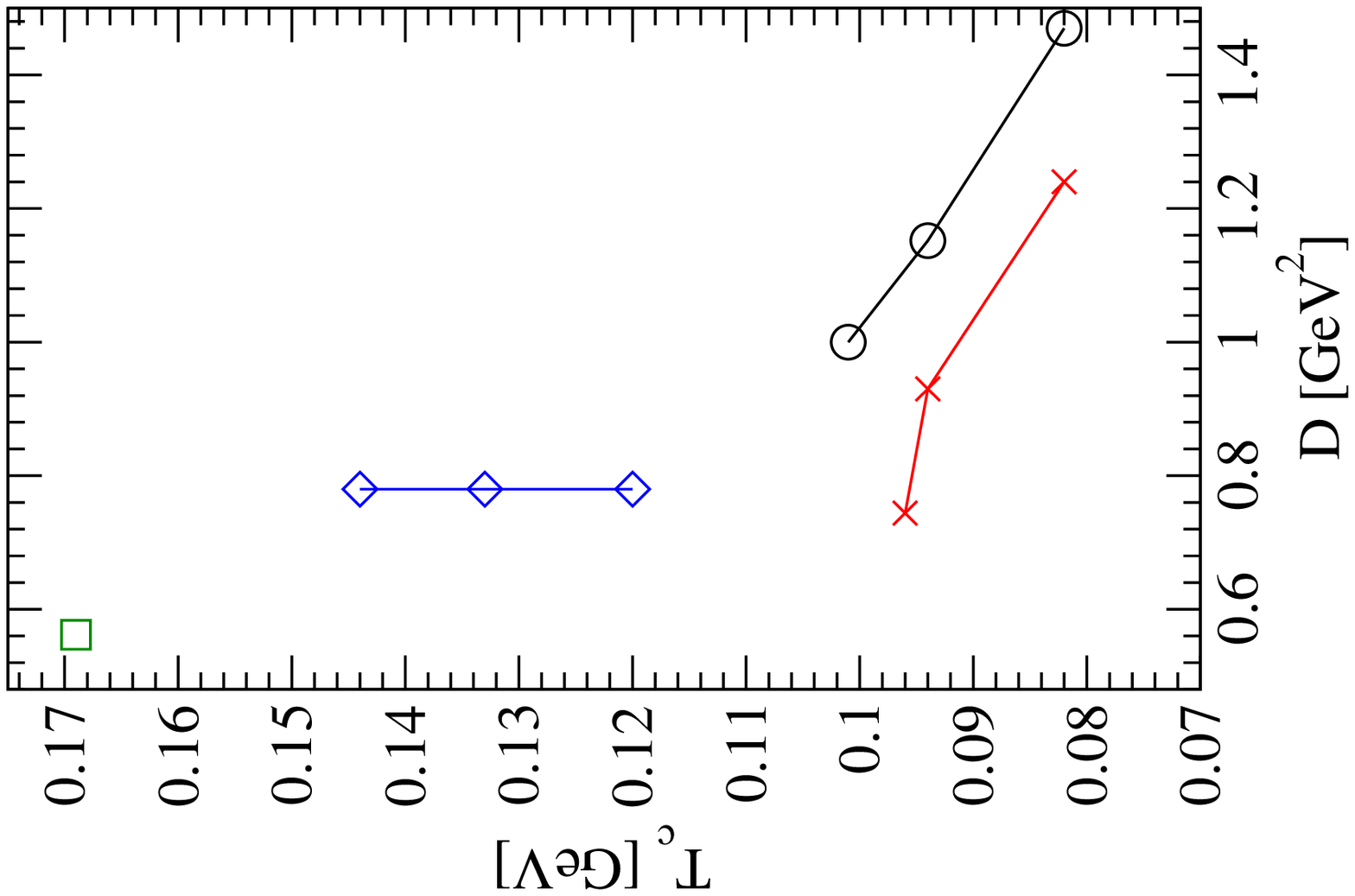}}
\subfigure[\protect\parbox{0.15\columnwidth}{$T_c$ vs.~chiral condensate}]{\includegraphics[height=0.24\columnwidth,angle=270,clip=true]{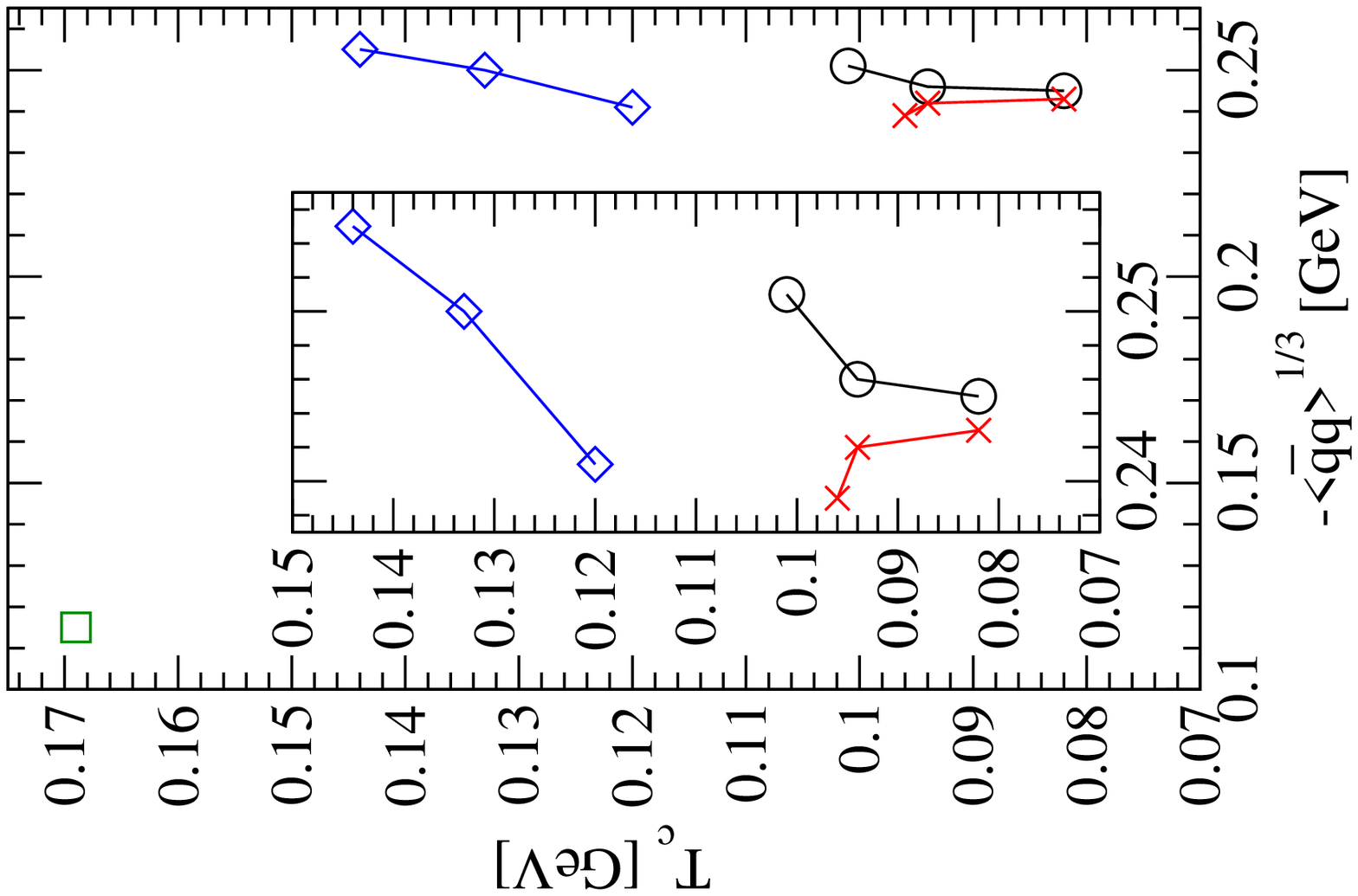}}
\caption[Plot of the chiral transition temperature for all model with respect to the different model parameters.]{The different colors and symbols on the lines denote the
corresponding interactions, namely: black (circles) -- AWW; red (X) -- MT; and blue (diamonds) -- MR.
The result for MN (single green square) is only shown in panels (c) and (d), since it does not
have a value of $\omega$ associated with it. The insert in panel (d) enlarges the region of interest
for the $\omega$-dependent interactions.\label{fig:results}}
\end{figure}

For a given interaction, $T_c$ increases with $\omega$, which is illustrated in panel (a) of Fig.~\ref{fig:results}. A straight-forward interpretation of this observation goes back to $\omega$ representing an inverse effective range of the interaction. In this picture, $T_c$ is higher for an interaction, for which this range is smaller, i.e. the major strength in the interaction comes from a region defined by the larger momentum scale $\omega$. However, this effect is obviously minor compared to the differences with regard to the form of the interaction, in particular the appearance of the $\delta$-function term.

To further illustrate the details of the differences between the various forms for the interaction as well as to highlight their characteristic features, we plot $T_c$ as a function of three more specific properties of each model, namely $\omega\,D$ in panel (b), $D$ in panel (c), and the chiral condensate in panel (d). While the attempt to quantitatively correlate $T_c$ globally (i.e.~across the different interactions) to any of these fails, a clear structure is visible. Throughout all panels of Fig.~\ref{fig:results} the AWW and MT results are rather close together, but clearly separate from MR, which again is clearly separate from MN.

Both $\omega\,D$ and $D$ investigated in the two middle panels have been interpreted as an ``integrated strength'' of the interaction. In fact, the latter is proportional to the integral over $d^4q$ of the interaction at zero temperature for all interactions investigated here, if one leaves out the UV part $\mathcal F_{UV}$ in Eqs.~(\ref{eq:mt}) and (\ref{eq:mr}). Fig.~\ref{fig:results} in panel (c) clearly shows that there is no overall simple dependence of $T_c$ on $D$. In the MT interaction, a constant value for $\omega\,D$ leads to (almost) unchanged masses and decay constants for ground-state pseudoscalar and vector mesons, for which reason $\omega\,D$ can be termed ``integrated strength'' in this case instead of $D$. Again, Fig.~\ref{fig:results} in panel (b) indicates no simple dependence of $T_c$ on $\omega\,D$. In both cases, the absence of a simple relation is exemplified already by the fact that three points with the same $\omega\,D$, the characteristic feature for MT, in Fig.~\ref{fig:results} (b) and three points with the same $D$, the characteristic feature for MR, in Fig.~\ref{fig:results} (c) each correspond to three different values of $T_c$.

Finally, in Fig.~\ref{fig:results} (d) we plot $T_c$ as a function of an order parameter of chiral symmetry breaking, the chiral condensate, evaluated at $T=0$. In a very simple picture, one could argue that, given a certain form of the dependence of the order parameter on the temperature (see, e.g.~Fig.~\ref{fig:orderpar}), an enlargement of the value of the order parameter at $T=0$ would naturally lead to an increase in $T_c$. However, panel (d) clearly shows that the interactions used in our setup are not that simple.

Overall, these observations lead to the important possibility to establish distinct ranges of values for $T_c$ accessible for each type of interaction, which has the potential to rule out certain forms in a given truncation. We note here that various effects need to be taken into account beyond the simple setup presently used; we will attempt to quantify a certain part of the truncation effect below. At this point, however, we can still try to attribute the clear differences apparent from all panels in Fig.~\ref{fig:results} to the structure of the interaction. The object of interest in this respect is the appearance and relative strength of the $\delta$-function term in $\mathcal{G}$, Eqs.~(\ref{eq:mn}) -- (\ref{eq:mr}): In the AWW and MT interactions, Eqs.~(\ref{eq:aww}) and (\ref{eq:mt}), there is no such term. In the MR interaction, Eq.~(\ref{eq:mr}), it carries half the strength of the coupling in the sense of the integral $\int d^4q \mathcal{G}\sim D$ described above. In the MN interaction, Eq.~(\ref{eq:mn}), it is the only term and obviously represents all of the interaction's strength. From this, one can say that a less pronounced $\delta$-function term in the interaction will lead to lower values of $T_c$.

\section{Conclusions}
We have investigated the impact of various forms of an effective quark-gluon interaction in the rainbow-truncated
quark DSE in QCD at finite temperature on the critical temperature of the chiral phase transition.
The established mean-field behavior of this transition in the present truncation is confirmed throughout.
Regarding non-universal aspects of the transition, it is apparent that there is no simple overall relationship between
the distribution of the infrared strength of the interaction and the critical temperature, although 1) a trend
can be extracted identifying the part of the interaction strength carried by a $\delta$-function term in some
of the models as a qualitatively distinctive parameter, and 2) within the various models simple relations
of $T_c$ and relevant model parameters do exist. 

\section{Corrections to rainbow truncation\label{brt}}
While the rainbow truncation of the gap equation provides a simple setup for computations,
one has to keep in mind that the necessary corrections in any given circumstance may
change its results considerably. Therefore we discuss here some of the most important effects
from which a change, in particular to the critical temperature, is to be expected.

In \cite{Blank:2010bz} it was argued that from a phenomenological point of view one can try to quantify the contribution of the rainbow-truncated gap equation to the full (untruncated) result. Such an attempt has been made for meson properties at zero temperature in \cite{Eichmann:2008ae} and has been extended to the baryon sector and exemplified for
several hadronic observables in \cite{Nicmorus:2008vb}. The idea is to change the values of the available
model parameters such that the resulting hadron properties, e.g.~the rho-meson mass, are deliberately
overestimated by the rainbow-ladder truncation result. In this way one expects corrections beyond
this truncation to bring it to the experimental value.

While we do not want to discuss the particular
assumptions made in \cite{Eichmann:2008ae,Nicmorus:2008vb}, the effect described there is relevant
for our present study, since we start with model parameters fixed to meson phenomenology at zero
temperature. To get an estimate of the effects from a change of parameters at $T=0$ on $T_c$, we
simply adapt the two parameter sets referred to as ``A'' (which corresponds to our MT2 parameter set) and
``B'' (which has increased strength in the interaction to overestimate the rho-meson mass as discussed above)
in Ref.~\cite{Nicmorus:2008vb} and add $T_c$ to the Tab.~I given there. In this table, a systematic ratio
of $\approx 0.74$ is found for value in A divided by value in B for several hadron properties (e.g.~$m_\rho$,
$f_\pi$, $m_N$, $m_\Delta$, etc.). Our values for $T_c$ in this respect are $94$ MeV for set A and
$129$ MeV for set B, which produces a ratio of $0.73$ and thus fits perfectly into the picture.

For the present discussion here the relevant point is that parameter changes
inferred from the phenomenological estimation of corrections expected beyond rainbow-ladder truncation
are in agreement with investigations at zero temperature. In particular, one obtains an increase of $T_c$ by
about one third, depending on the details of the corrections assumed.

\chapter{Mesons}\label{chap:mesons}
Strongly interacting particles, the so-called hadrons, can be divided into two classes, baryons and mesons, which can be distinguished since the number of baryons is a conserved quantum number. Thus, a meson is characterized as a hadron with baryon number zero. 

Mesons, in their simplest form, may be considered as bound states of a quark (baryon number $\frac{1}{3}$) and an antiquark (baryon number $-\frac{1}{3}$), within the \emph{constituent quark model}. These types of models, which are very successful in describing the hadron spectrum, provide a consistent classification scheme for mesonic states according to their flavor content via group theoretical arguments (see e.g., the review on the quark model in the Review of Particle Physics \cite{Nakamura:2010zzi}). Nevertheless, the calculation of the meson spectrum in terms of constituent quarks has certain drawbacks. Especially the masses of the low-lying pseudoscalar meson octet ($\pi$, K, $\eta$ mesons) are not easily described by constituents with masses of approximately $300\,$MeV. 

The strong interaction, however, is well described by the relativistic quantum field theory of QCD, which was introduced already in Chap.~\ref{chap:intro}. The elementary degrees of freedom of QCD, quarks and gluons, allow the classification of hadrons in the same way as the quark model. In addition, the properties of the light pseudoscalar mesons can be understood from a theoretical point of view: Their comparatively light masses are explained by their appearance as (pseudo) Goldstone bosons of the dynamically broken chiral symmetry of massless QCD.

Thus, in order to provide a consistent description of mesons within the standard model of particle physics, it is necessary to treat them as composite states in the framework of QCD. Non-perturbative techniques are essential in this context, and e.g.~lattice-regularized QCD is very successful in describing the hadron spectrum for ground states (see, e.g.~\cite{Durr:2008zz,Aoki:2009ix} and references therein) and excitations (see, e.g.~\cite{Engel:2010my,Dudek:2009qf} and references therein). A different approach, as outlined in Chap.~\ref{chap:formalism}, is provided by the DSE/BSE formalism, which we employ in the following to study mesons as relativistic bound states of QCD.

In this context, mesonic states have already been investigated extensively using various approximations (see, e.g., \cite{Krassnigg:2009zh,Fischer:2009jm,Fischer:2008wy} and references therein). For the most part the homogeneous BSE has been used, which provides good descriptions not only for the ground-state spectrum for various quantum numbers, but also for further mesonic properties such as decay constants, charge radii and electromagnetic form factors \cite{Bhagwat:2006pu,Holl:2005vu,Maris:1999bh}, as well as hadronic decays \cite{Jarecke:2002xd,Mader:2010ma,Mader:2011zf}. Some investigations employ the inhomogeneous vertex BSE as well, but mainly as a description for the quark-photon vertex in form factor calculations, or to calculate correlations at finite temperature \cite{Maris:2000ig}.

However, the homogeneous BSE and the inhomogeneous vertex BSE, as discussed in Sec.~\ref{sec:vertex_bse}, are intimately related. In Sec.~\ref{sec:spectralbse}, we further explore these relations and use them to gain insight into the structure of the Bethe-Salpeter equations. This allows to investigate in detail the consequences of the rainbow-ladder truncation, which leads (as shown in Sec.~\ref{sec:extrapolate}) to a new extrapolation technique to calculate the masses of bound states beyond the usually accessible region. It is applied in Sec.~\ref{sec:spectroscopy} to calculate the ground- and excited-state meson spectrum.

The understanding of the structure of the BSEs also gives rise to new applications of the vertex BSE (Sec.~\ref{sec:inhomresults}). We show in Sec.~\ref{sec:inhomonshell} that bound-state properties such as decay constants can be calculated solely from the vertex BSE, and in Sec.~\ref{sec:inhomremovepole} that it is possible to remove bound-state poles from the solution of the vertex BSE while keeping all non-resonant contributions.

In addition, the DSE/BSE approach links the fundamental Green functions of QCD directly to hadronic observables. In the rainbow-ladder truncation, this link is provided by the effective interaction, which can be related to the running coupling and thus the ghost- and gluon propagator dressing functions. The infrared behavior of these dressing functions is currently debated, and in Sec.~\ref{sec:bseir} we investigate the influence of the different available forms on the properties of bound states, especially the $\pi$ and $\rho$ mesons.

\newpage
\section{Spectral representation of the BSEs}\label{sec:spectralbse}
In the numerical approach discussed in Sec.~\ref{sec:hommethod}, the homogeneous BSE is treated as an eigenvalue equation via the representation of the BSE kernel and the propagators as matrices. This allows to obtain (numerical) eigenvalues and eigenvectors, which on-shell describe bound states and their properties. However, from a mathematical point of view each diagonalizable matrix is completely specified by its eigenvalues and eigenvectors. Therefore, even if they are not physical, off-shell eigenvalues and eigenvectors of the homogeneous BSE can be used to gain insights into the system under investigation. In the present section, we explore the theoretical consequences of a representation of the homogeneous BSE via its eigensystem, which reveals further deep relations between the homogeneous BSE, the quark-antiquark propagator, and the vertex BSE.

\subsection{Notation}
For this study of the  structure of the BSEs, we adopt an unusual notation. We denote a Bethe-Salpeter amplitude and its charge-conjugate by \beq
\Gamma(p,P)\equiv \grp\;\textrm{and}\;\;\bar{\Gamma}(p,-P)\equiv \glp\;,
\eeq
respectively, reminiscent of the bra-ket notation commonly used in quantum mechanics. A pair of dressed quark propagators is represented by the operator $\dip$, such that 
\beq
\dip\equiv S^a(p_+)\otimes S^b(p_-)\;.
\eeq
Applying $\dip$ on an amplitude gives a Bethe-Salpeter wave function,
\beq
\dip\grp=S^a(p_+)\Gamma(p,P)S^b(p_-)\equiv\xrp \;.
\eeq
Analogously, the charge-conjugate wave function is given by
\beq
\xlp\equiv\bar{\chi}(p,-P)=S^b(p_-)\bar{\Gamma}(p,-P)S^a(p_+)\;.
\eeq
Note that the flavor indices $a$ and $b$ labeling the propagators are absorbed into the symbols used for the operator $\dip$, the wave function $\xlp$, and the amplitude $\grp$, in order not to overcomplicate the notation.

An `inner' product of a wave function with an amplitude corresponds to
\beq\label{eq:innerprod}
\left\langle\chi(p,P)|\Gamma(p,P)\right\rangle\equiv \Tr\left[\int_p S^b(p_-)\bar{\Gamma}(p,-P)S^a(p_+) \Gamma(p,P) \right]\;,
\eeq
while the `inner' product of two amplitudes or two wave functions is not well-defined. The homogeneous BSE \eq{hom_bse} is constructed by applying the quark-antiquark scattering kernel, represented by the operator $K(p,q,P)$, to a wave function $\diq\grq$,
\beq
\grp=K(p,q,P)\diq\grq\;.
\eeq
Here, the argument $q$, which appears in the operator $K$ as well as in the amplitude $\grq$, is integrated over. This is similar to the matrix notation used in Sec.~\ref{sec:bsenum}, where the loop integral was represented as the multiplication of the BSE kernel matrix $\mathcal K$ on a vector representing the amplitude.

\subsection{Homogeneous BSE}
Using the notation defined above, the homogeneous BSE \eq{hom_bse} for a state (or excitation) $i$ can be written as
\beq\label{eq:hom_bse_spectral}
\left|\Gamma_i(p,P_i)\right\rangle=K(p,q,P_i)D(q,P_i)\left|\Gamma_i(q,P_i)\right\rangle\;.
\eeq
As mentioned already in Sec.~\ref{sec:hommethod}, this is an eigenvalue equation for the operator $K\,D$ with eigenvalue $\lambda_i(P_i^2)=1$. The eigenvector is the homogeneous BSA. To find a numerical solution of Eq.~\eq{hom_bse_spectral}, it is usually generalized to arbitrary $P^2$,
\beq\label{eq:bseright}
\lambda_i(P^2)\girp=K(p,q,P)\diq\girq\;,
\eeq
where $\girp$ is a right-eigenvector of $K\,D$ to the eigenvalue $\lambda_i(P^2)$.

Although these eigenvectors can (in general) not be straight-forwardly interpreted with respect to physical states, they can be used to obtain insight into the structure of the Bethe-Salpeter equations. It was demonstrated in Sec.~\ref{sec:bsenum} that the linear operator $K\,D$ (the BSE kernel matrix $\mathcal K$ in the numerical setup) can be approximated to arbitrary precision by finite matrices, and thus has a discrete spectrum. Consequently, it can be expressed as a spectral sum via its eigenvalues and (left- and right-\nolinebreak)\linebreak[2] eigenvectors. The right-eigenvectors, as stated above, are the solutions of Eq.~\eq{bseright}. To find an equation for the left-eigenvectors, we consider the charge-conjugate of Eq.~\eq{bseright},
\beq\label{eq:bsebar}
\gilp\lambda_i(P^2)=\gilq\diq K(q,p,P)\;.
\eeq
On-shell, this equation gives the charge-conjugate BSA $\bar{\Gamma}_i(p,-P_i)$, but it is not an eigenvalue equation for $K\,D$. If, however, Eq.~\eq{bsebar} is multiplied from the right by \dip,
\beq
\gilp\dip \lambda_i(P^2)=\gilq\diq K(q,p,P) \dip\;,
\eeq
we see that the left-eigenvector of $K\,D$ to the eigenvalue $\lambda_i(P^2)$ is 
\beq \xilp=\gilp\dip\;.\eeq

With all these ingredients, the spectral representation of the operator $K\,D$ can be expressed as a sum over the `outer' product of all right- and left-eigenvectors multiplied by the corresponding eigenvalues,
\beq\label{eq:spectralKD}
K(p,q,P)\diq=\sum_i\lambda_i(P^2)\girp\xilq\;.
\eeq
Here, the left- and right-eigenvectors are mutually orthogonal and normalized such that
\beq\label{eq:spectralnorm}
\xg=\delta_{j,i}\;.
\eeq

\subsection{Quark-antiquark propagator and vertex BSE}
The operator $K\,D$ is the basic building block of all Bethe-Salpeter equations, and with the results of the preceding section, the quark-antiquark propagator as well as the inhomogeneous BSA can be expressed as a sum over eigenvalues and eigenvectors of the BSE kernel matrix.

We first consider the BSE for the quark-antiquark propagator. From Eq.~\eq{bseGinv}, it follows that 
\beq
G^{(2)}_{[2]}(p,q,P)=D(p,P)\left(\mathbf1-K(p,q,P)D(q,P)\right)^{-1}\;. 
\eeq
Inserting the spectral representation Eq.~\eq{spectralKD} into the above equation leads to
\begin{multline}\label{eq:spectralG}
G^{(2)}_{[2]}(p,q,P)=D(p,P)\sum_i\frac{1}{1-\lambda_i(P^2)}\girp\xilq\\
=\sum_i\frac{1}{1-\lambda_i(P^2)}\xirp\xilq\;.
\end{multline}
If all possible channels are considered, this is an exact representation of $G^{(2)}_{[2]}(p,q,P)$ which shows the intimate connection of the eigenvalues and eigenvectors of the BSE kernel matrix to the physical quark-antiquark propagator, when they are calculated off-shell.

The inhomogeneous vertex BSE \eq{inhom_bse} can be written as
\beq
\grp=\left|\Gamma_0\right\rangle+K(p,q,P)D(q,P)\grq\;,
\eeq
with the inhomogeneous term represented by the ket-vector $\left|\Gamma_0\right\rangle$. Therefore, the inhomogeneous BSA is given by
\beq
\grp=\left(\mathbf1-K(p,q,P)D(q,P)\right)^{-1}\left|\Gamma_0\right\rangle\;,
\eeq
as also discussed in Sec.~\ref{sec:inhom_methods}. If the spectral representation Eq.~\eq{spectralKD} is inserted, we obtain
\beq\label{eq:spectralinhom}
\grp=\sum_i\frac{1}{1-\lambda_i(P^2)}\girp\xilq\Gamma_0\rangle\;,
\eeq
which corresponds to a spectral representation of the inhomogeneous BSA. 

\subsection{Canonical Norm}
For convenience, the eigenvectors considered here are normalized according to the condition \eq{spectralnorm}. If, however, the on-shell amplitudes are used to compute physical observables other than the mass, they have to be normalized canonically. The appropriate normalization condition, derived in Sec.~\ref{sec:hom_bse}, is based on the physical decomposition of the quark-antiquark propagator into bound state poles, which in the notation employed in this chapter reads
\beq
G^{(2)}_{[2]}(p,q,P)=\sum_i \frac{1}{\mathcal{N}_i}\: \frac{\xirpi\xilqi}{P^2-P_i^2}+R(p,q,P)\;,
\eeq
where we denote the canonical norm of state $i$ by $\mathcal{N}_i$. In the vicinity of one pole $i$, the other pole- and regular terms become negligible, and inserting Eq.~\eq{spectralG} we see that
\beq
\frac{1}{1-\lambda_i(P^2)}\xirp\xilq=\frac{1}{\mathcal{N}_i}\: \frac{\xirpi\xilqi}{P^2-P_i^2}\;.
\eeq
In summary, the canonical norm is given by
\beq
\mathcal{N}_i=\lim_{P^2\rightarrow P_i^2}\left( \frac{1-\lambda_i(P^2)}{P^2-P_i^2}\right)\;.
\eeq
Using the rules of de l'H\^{o}spital, it can be expressed as
\beq\label{eq:canonical_norm_eval}
\mathcal{N}_i=\lim_{P^2\rightarrow P_i^2}\left(-\frac{d}{d(P^2)}\lambda_i(P^2)\right)=-\lambda'_i(P_i^2)\;,
\eeq
such that the canonical norm of an amplitude normalized according to Eq.~\eq{spectralnorm} is given by the slope of the eigenvalue curve at the on-shell point. With different notation, this form of the normalization condition has already been derived in \cite{Nakanishi:1965zza}, and its validity independent of the truncation was demonstrated in \cite{Fischer:2009jm}.

\newpage
\section{Extrapolation of meson masses}\label{sec:grexc}
%introduction to section
In most studies to date the BSE approach was used to investigate the ground-state spectrum. While a study of excitations is possible in principle \cite{Holl:2004fr,Holl:2004un,Krassnigg:2006ps,Krassnigg:2008gd}, two main problems appear: First, the analytic structure of the quark propagator limits the range accessible for a direct calculation, and either sophisticated methods for analytic continuation (e.g., \cite{Bhagwat:2002tx}) or extrapolation techniques (e.g., \cite{Bhagwat:2007rj}) need to be applied. Second, one can not exclude the appearance of so-called spurious or unphysical solutions (see, e.g., \cite{Ahlig:1998qf} for a collection of arguments), such that the identification of the physical excitations in a channel is unclear, especially for states of ``exotic'' C-parity.

However, in order to attempt a solution to the second problem, the properties of the states in question need to be known. This in turn is only possible if the first problem has already been solved. In addition, a study of mesons of higher spins involves calculation  of bound states of higher masses, such that even for ground states extrapolation techniques are needed.

Based on the results obtained in Sec.~\ref{sec:spectralbse}, we therefore develop a new method for extrapolating the eigenvalues of the homogeneous BSE beyond the limits set by the complex conjugate poles in the quark propagator, which is subsequently applied to the excitations of the pseudoscalar, scalar, vector, axialvector and tensor channel. 

\subsection{Consequences of the truncation}
So far, we have considered the general case, and no truncation has been applied. From this point of view, we can now investigate the consequences of assumptions used in a numerical setup. 

In the following, we consider a truncation constructed such that the quark-antiquark scattering kernel is independent of the total momentum,
\beq\label{eq:KnoP}
K(p,q,P)\equiv K(p,q)\;.
\eeq
The rainbow-ladder truncation is a prominent example where the above condition holds, which entails that the only total-momentum dependence in the homogeneous BSE \eq{hom_bse_spectral} lies in the two propagators, combined into the operator $D(p,P)$ in the notation employed here. 

From Eq.~\eq{spectralKD}, the interaction kernel can be written as a sum over (right-)eigenvectors of $KD$ by amputating the propagators on the right-hand side, such that
\beq\label{eq:knop2}
K(p,q)=\sum_i\lambda_i(P^2)\girp\gilq\;.
\eeq
Since $K$ is independent of $P$, we can express the interaction kernel by using eigenvectors and eigenvalues at any value of $P$, which will be used in the following to establish a connection between the poles in the quark-antiquark propagator and the eigenvalues $\lambda_i(P^2)$.

\subsection{Behavior of eigenvalues}
As explained in Sec.~\ref{sec:hommethod}, the on-shell points defining the masses of the bound states in the channel under consideration are obtained by investigating the behavior of the eigenvalues with respect to the total momentum-squared, $\lambda(P^2)$. If $\lambda(P^2)=1$, the system exhibits a bound state. In addition, from Eq.~\eq{canonical_norm_eval} we see that the canonical norm can be calculated from the derivative of this eigenvalue curve, again at the point where $\lambda(P^2)=1$. Therefore, it is interesting to investigate the consequences of Eq.~\eq{KnoP} on the behavior of these curves.

An eigenvalue curve which crosses one is directly related to a pole in the quark-antiquark propagator $G_{[2]}^{(2)}(p,q,P)$. Focusing on one pole $i$, $G_{[2]}^{(2)}(p,q,P)$ can be expressed as the pole term plus regular terms denoted by $R_i(p,q,P)$, which contain the non-resonant as well as the other pole contributions. We have
\beq
G_{[2]}^{(2)}(p,q,P)=\frac{1}{\mathcal{N}_i}\: \frac{\xirpi\xilqi}{P^2-P_i^2}+R_i(p,q,P)\;.
\eeq
As above, we assume spectral normalization and denote the canonical norm by $\mathcal{N}_i$. On the other hand, via Eq.~\eq{spectralG}, $G_{[2]}^{(2)}$ is given by 
\beq
G_{[2]}^{(2)}(p,q,P)=\sum_i\frac{\xirp\xilq}{1-\lambda_i(P^2)}\;,
\eeq
and it follows that 
\beq\label{eq:rest}
R_i(p,q,P)=\sum_{j\neq i}\frac{\xx{j}{P}{p}{q}}{1-\lambda_j(P^2)} +\tilde{R}_i\;,
\eeq
where 
\beq\label{eq:rtilde}
\tilde{R}_i\equiv\frac{\xx{i}{P}{p}{q}}{1-\lambda_i(P^2)}-\frac{1}{\mathcal{N}_i}\: \frac{\xx{i}{P_i}{p}{q}}{P^2-P_i^2}\;. 
\eeq

In order to link the physical pole in the quark-antiquark propagator to the eigenvalue curve, we consider
\beq\label{eq:trkg}
\Tr[K\,G]=\Tr\left[K(p,q)\frac{1}{\mathcal{N}_i}\: \frac{\xx{i}{P_i}{q}{p}}{P^2-P_i^2}+K(p,q) R_i(q,p,P)\right]\;,
\eeq
where the trace is taken in the functional sense as well as in Dirac (and flavor) space, analogous to Eq.~\eq{innerprod}. In a diagrammatic language, this operation corresponds to joining the two ends of one diagram.

In the first term of Eq.~\eq{trkg}, we can make use of the generality of Eq.~\eq{knop2} with respect to the choice of $P$, which allows to express the kernel via eigenvectors at $P=P_i$. In the second term, Eq.~\eq{knop2} is used directly, leading to
\begin{multline}
\Tr[K\,G]= \Tr\left[ \sum_j\lambda_j(P_i^2) \gga{j}{P_i}{p}{q} \frac{1}{\mathcal{N}_i}\: \frac{\xx{i}{P_i}{q}{p}}{P^2-P_i^2}\right.\\
\left. + \sum_j\lambda_j(P^2) \gga{j}{P}{p}{q} R_i(q,p,P) \right]\;.
\end{multline}
Evaluating the trace and using the orthogonality relation \eq{spectralnorm} gives
\begin{multline}\label{eq:trkg1}
\Tr[K\,G]= \frac{1}{\mathcal{N}_i\left(P^2-P_i^2\right)} + \sum_j\lambda_j(P^2) \gal{j}{P}{q} R_i(q,p,P)\gar{j}{P}{p}\;.
\end{multline}
Using Eq.~\eq{rest}, the second term in Eq.~\eq{trkg1} becomes
\begin{multline}
\sum_j\lambda_j(P^2) \gal{j}{P}{q} R_i(q,p,P)\gar{j}{P}{p}=\\
\sum_j\lambda_j(P^2) \gal{j}{P}{q} \sum_{k\neq i}\frac{\xx{k}{P}{p}{q}}{1-\lambda_k(P^2)} \gar{j}{P}{p}\\
+\sum_j\lambda_j(P^2) \gal{j}{P}{q} \tilde{R}_i(q,p,P)\gar{j}{P}{p}\;,
\end{multline}
which by use of Eq.~\eq{spectralnorm} reduces to 
\beq\label{eq:trkg2}
\sum_{j\neq i}\frac{\lambda_j(P^2)}{1-\lambda_j(P^2)}+\sum_j\lambda_j(P^2) \gal{j}{P}{q} \tilde{R}_i(q,p,P)\gar{j}{P}{p}\;.
\eeq
However, if Eq.~\eq{spectralG} is used directly to evaluate $\Tr[K\,G]$, we obtain
\beq\label{eq:kgspec}
\Tr[K\,G]=\sum_{j}\frac{\lambda_j(P^2)}{1-\lambda_j(P^2)}\;.
\eeq
Upon combining Eqs.~\eq{trkg1}, \eq{trkg2}, and \eq{kgspec}, we arrive at
\beq\label{eq:lambdaFull}
\frac{\lambda_i(P^2)}{1-\lambda_i(P^2)}=\frac{1}{\mathcal{N}_i\left(P^2-P_i^2\right)}+\sum_j\lambda_j(P^2) \gal{j}{P}{q} \tilde{R}_i(q,p,P)\gar{j}{P}{p}\;.
\eeq
The singular part of the operator $\tilde{R}_i$, defined in Eq.~\eq{rtilde}, vanishes by construction near $P=P_i$, such that the pole part dominates the behavior of $\lambda_i(P^2)$. If we assume this to hold also further away from the pole, the second term in Eq.~\eq{lambdaFull} can be neglected and the behavior of the eigenvalue with respect to $P^2$ is determined via
\beq\label{eq:lambdaLin}
\frac{\lambda_i(P^2)}{1-\lambda_i(P^2)}=\frac{1}{\mathcal{N}_i\left(P^2-P_i^2\right)}\;.
\eeq
If this holds in a realistic setup, it can be used to extrapolate masses of bound states beyond the region that is usually accessible, as discussed in the next section.

\subsection{Numerical results}\label{sec:extrapolate}
The methods advocated in Chap.~\ref{chap:numerics} allow the reliable calculation of the biggest eigenvalues of the BSE kernel matrix, such that the behavior of $\lambda_i(P^2)$ for at least $i=1,\ldots,5$ can be investigated on numerically safe grounds.

\begin{figure}
\centering\includegraphics[width=0.6\columnwidth]{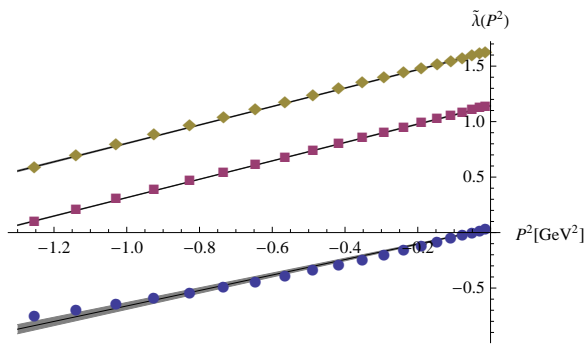} 
\caption[$\tilde{\lambda}(P^2)$ for quantum numbers $0^{-+}$ and light quarks.]{$\tilde{\lambda}(P^2)$ for quantum numbers $0^{-+}$ and light quarks. The lines joining the data points represent a linear fit, the gray bands are the corresponding statistical error bands.\label{fig:psfit}}
\end{figure}

From Eq.~\eq{lambdaLin} it follows that 
\beq\label{eq:lambdati}
 \tilde{\lambda}_i(P^2)\equiv\frac{1}{\lambda_i(P^2)}-1=\mathcal{N}_i\left(P^2-P_i^2\right)\;,
\eeq
which is linear in $P^2$ and goes through zero at $P^2=P_i^2$, i.e.~on-shell, which makes it suitable for extrapolations, provided that the second term in Eq.~\eq{lambdaFull} is indeed negligible.

For our investigation we use the MT model, for parameter sets MT1, MT2, and MT3 (cf.~App.~\ref{app:models}), for light, strange, charm, and bottom quarks with current quark masses as discussed in \cite{Krassnigg:2009zh}. Starting with pseudoscalar quantum numbers, the behavior of $\tilde{\lambda}(P^2)$ is studied for eigenvalues one to three with positive C-parity for light quarks. The result is shown in Fig.~\ref{fig:psfit}, where the calculated values of $\tilde{\lambda}(P^2)$ are plotted together with a linear fit and the corresponding statistical error bands. As can be seen, the deviation from the linear behavior is small, even away from the on-shell point which is given via $\tilde{\lambda}(P_i^2)=0$. 

\begin{figure}
\centering\includegraphics[width=0.6\columnwidth]{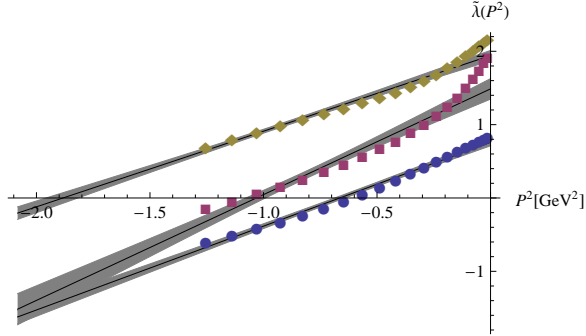} 
\caption[$\tilde{\lambda}(P^2)$ for quantum numbers $1^{--}$ and light quarks.]{$\tilde{\lambda}(P^2)$ for quantum numbers $1^{--}$ and light quarks, analogous to Fig.~\ref{fig:psfit}.\label{fig:vefit}} 
\end{figure}

In Fig.~\ref{fig:vefit} the same situation is given for quantum numbers $1^{--}$. In this case, differences from the linear approximation occur at small $P^2$, however in the relevant area around $\tilde{\lambda}(P^2)=0$ the linear fit represents the data points with good accuracy. In order to use all available data while still putting more emphasis on the region close to the on-shell point, each data point was weighted by $1/|\tilde{\lambda}(P^2)|$. 

It should be noted that an inspection of all curves investigated in the following shows that Figs.~\ref{fig:psfit} and \ref{fig:vefit} represent extreme cases, and the deviations from the linear behavior do in general not exceed the ones shown in Fig.~\ref{fig:vefit}. 

\begin{figure}
\vspace*{-0.5cm}
\subfigure[$J^{PC}=0^{-+}$]{\parbox{\columnwidth}{\includegraphics[width=\columnwidth]{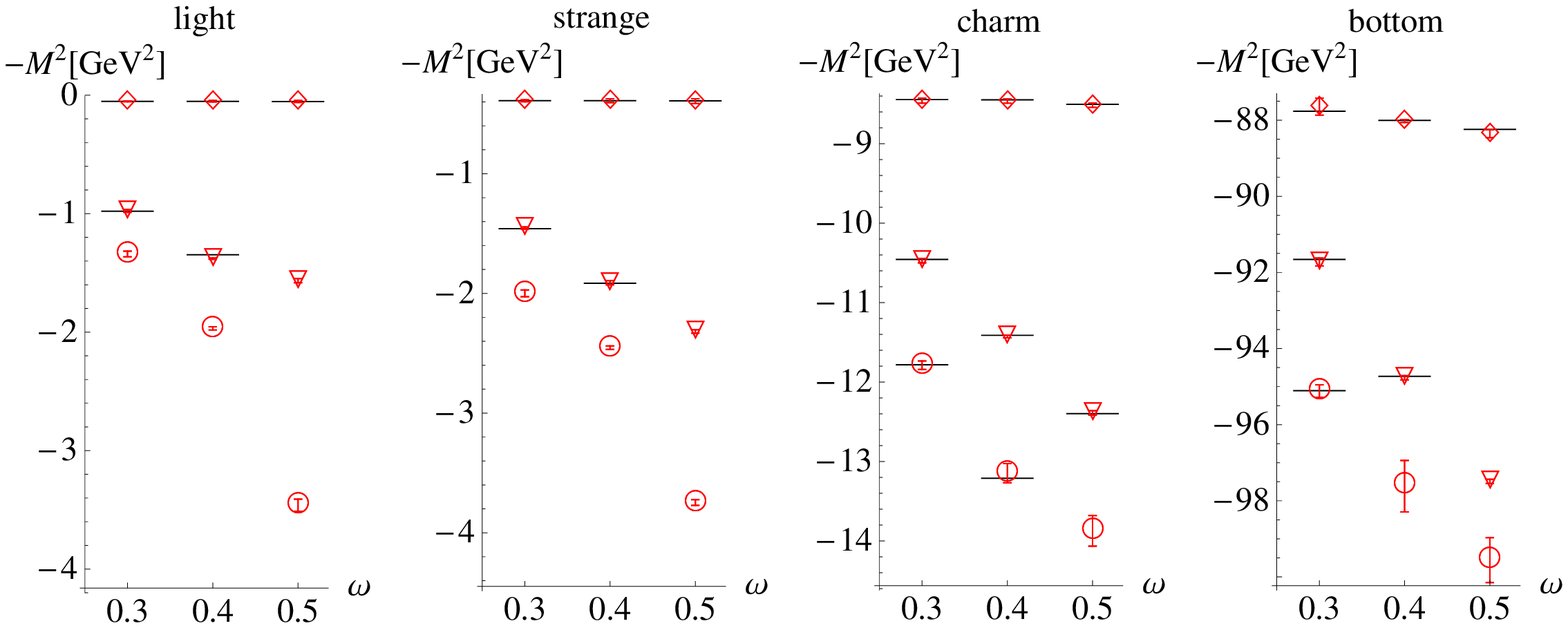}\\ \vspace*{-1cm}}}\vspace*{-0.7cm}
\subfigure[$J^{PC}=0^{++}$]{\parbox{\columnwidth}{\includegraphics[width=\columnwidth]{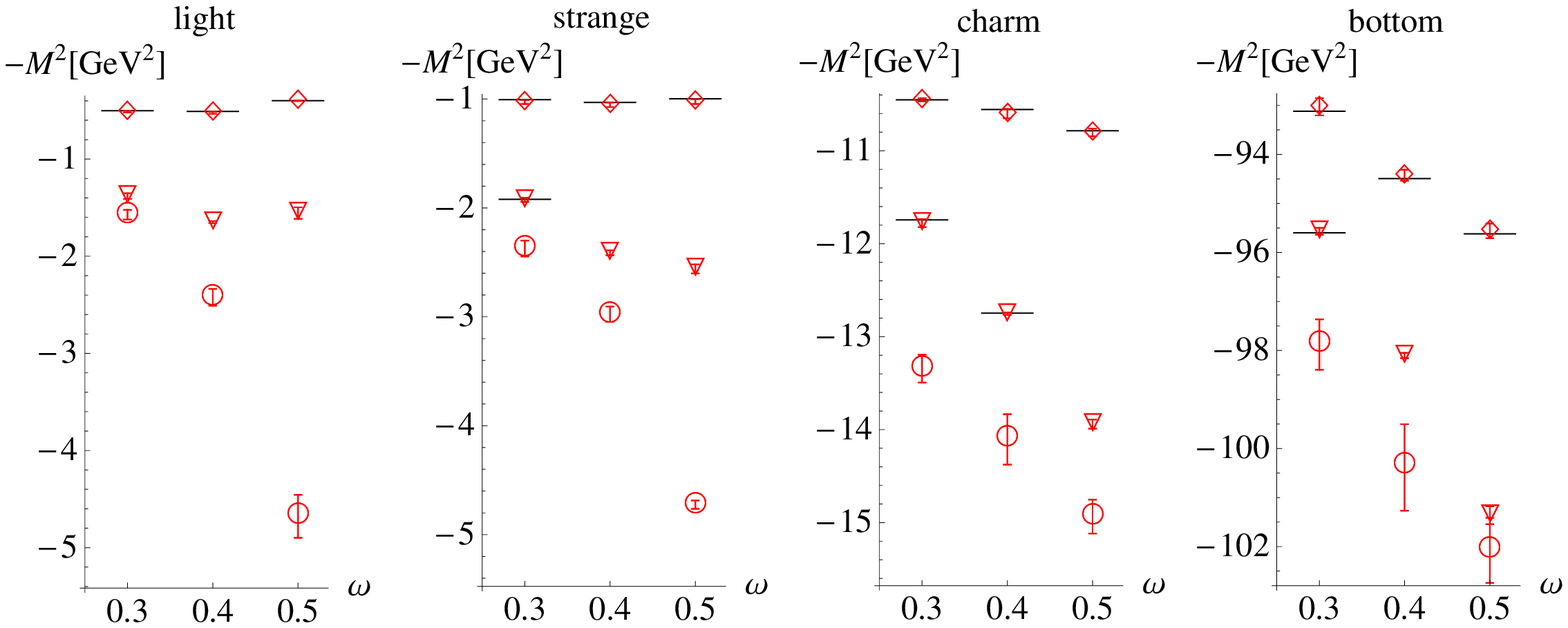}\\ \vspace*{-1cm}}}\vspace*{-0.7cm}
\subfigure[$J^{PC}=1^{--}$]{\parbox{\columnwidth}{\includegraphics[width=\columnwidth]{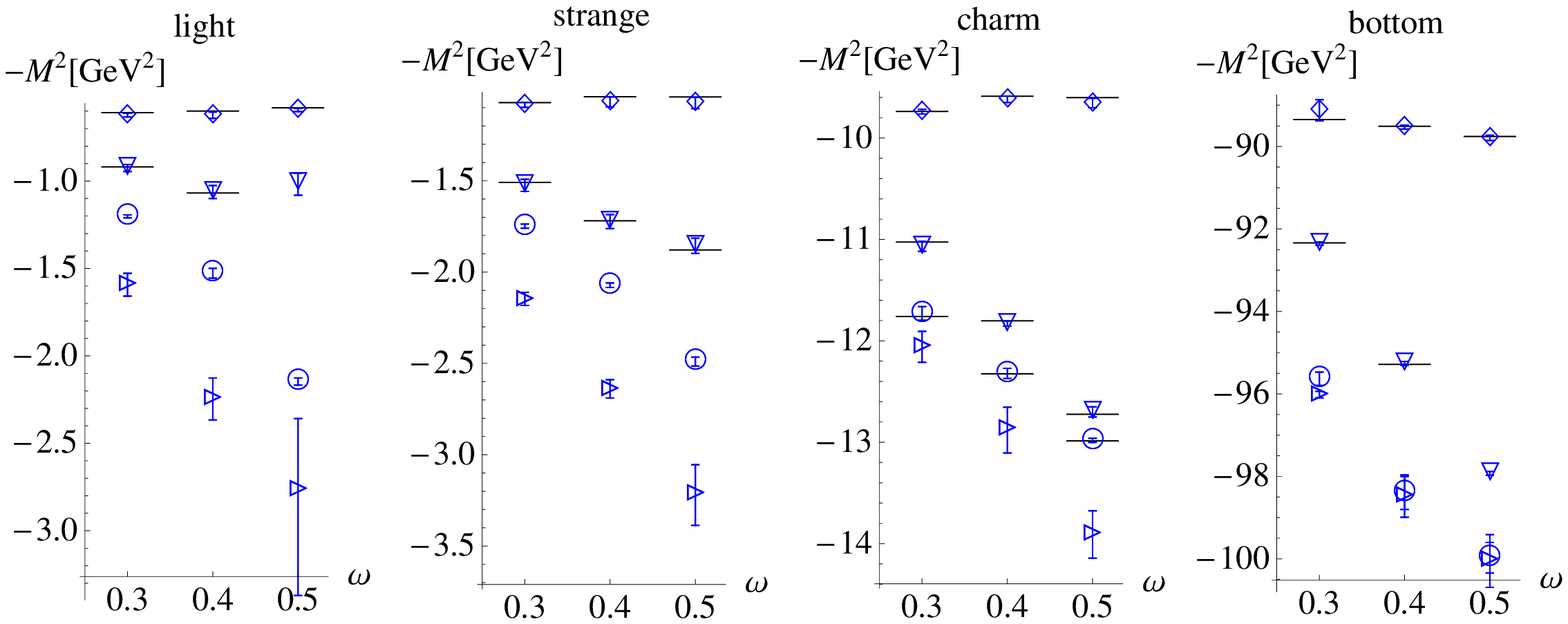}\\ \vspace*{-1cm}}}
\caption[$-M^2$ vs.~$\omega$ for the ground state, first and second excitation for $J^{PC}=0^{-+}$ and $0^{++}$ . For $J^{PC}=1^{--}$, also the third excitation is shown.]{$-M^2$ vs.~$\omega$ for the ground state, first and second excitation for $J^{PC}=0^{-+}$ (panel (a)) and $0^{++}$ (panel (b)). In panel (c), for $J^{PC}=1^{--}$, also the third excitation is shown. The black lines correspond to the results of the direct calculation where available, the symbols are the results of the interpolation/extrapolation. The errorbars correspond to the $95\%$ confidence interval of the linear fit.\label{fig:int1}}
\end{figure}

\begin{figure}
\vspace*{-0.5cm}
\subfigure[$J^{PC}=2^{++}$]{\parbox{\columnwidth}{\includegraphics[width=\columnwidth]{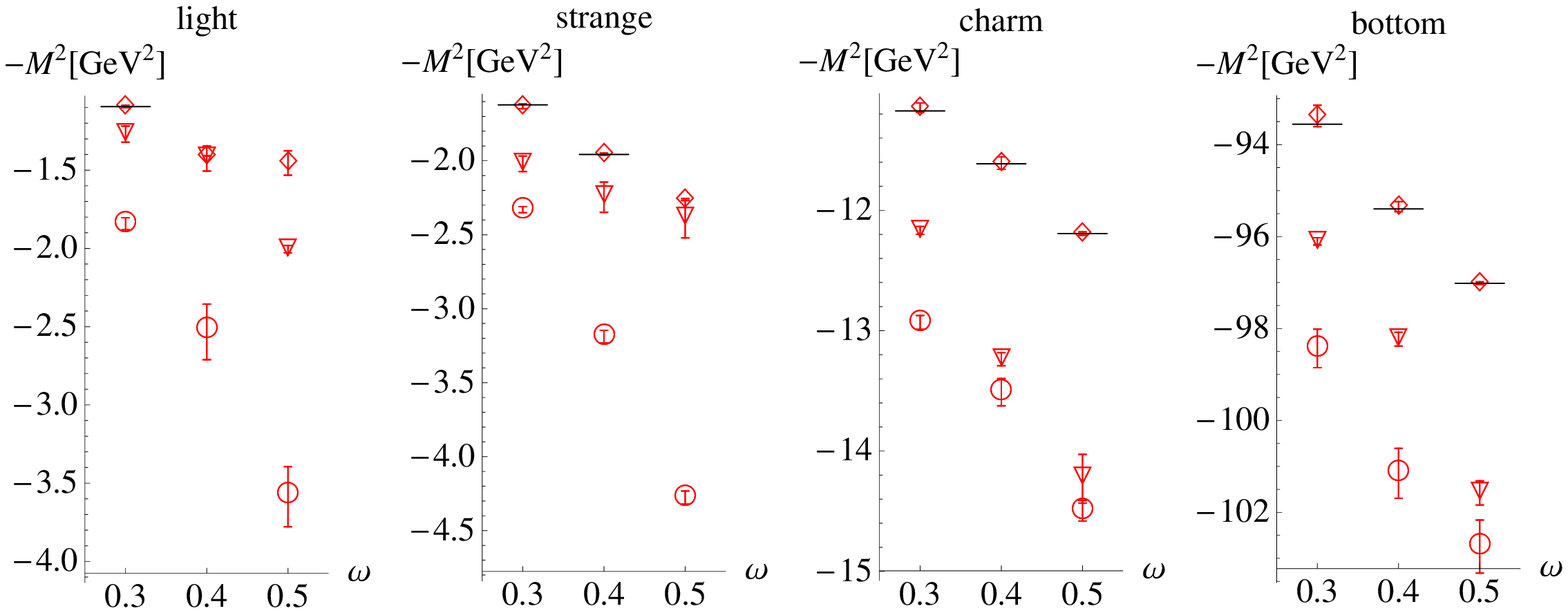}\\ \vspace*{-1cm}}}\vspace*{-0.5cm}
\subfigure[$J^{PC}=1^{++}$]{\parbox{\columnwidth}{\includegraphics[width=\columnwidth]{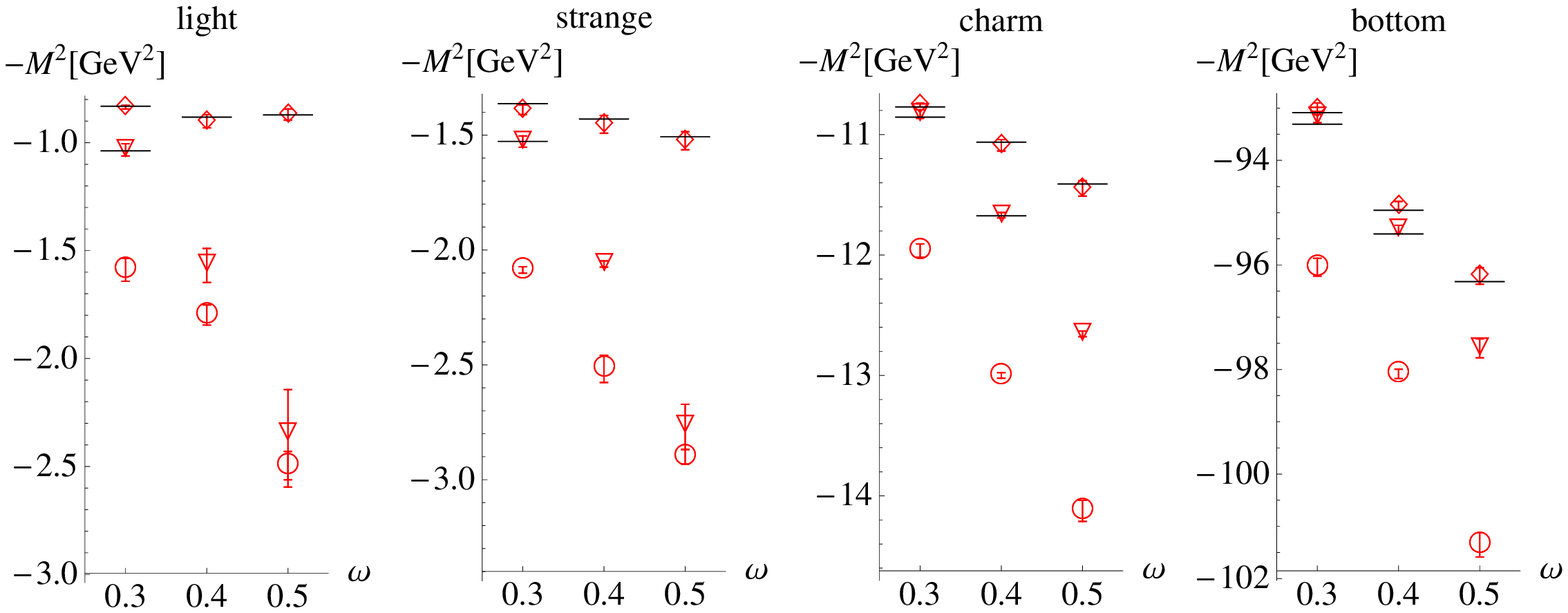}\\ \vspace*{-1cm}}}\vspace*{-0.5cm}
\subfigure[$J^{PC}=1^{+-}$]{\parbox{\columnwidth}{\includegraphics[width=\columnwidth]{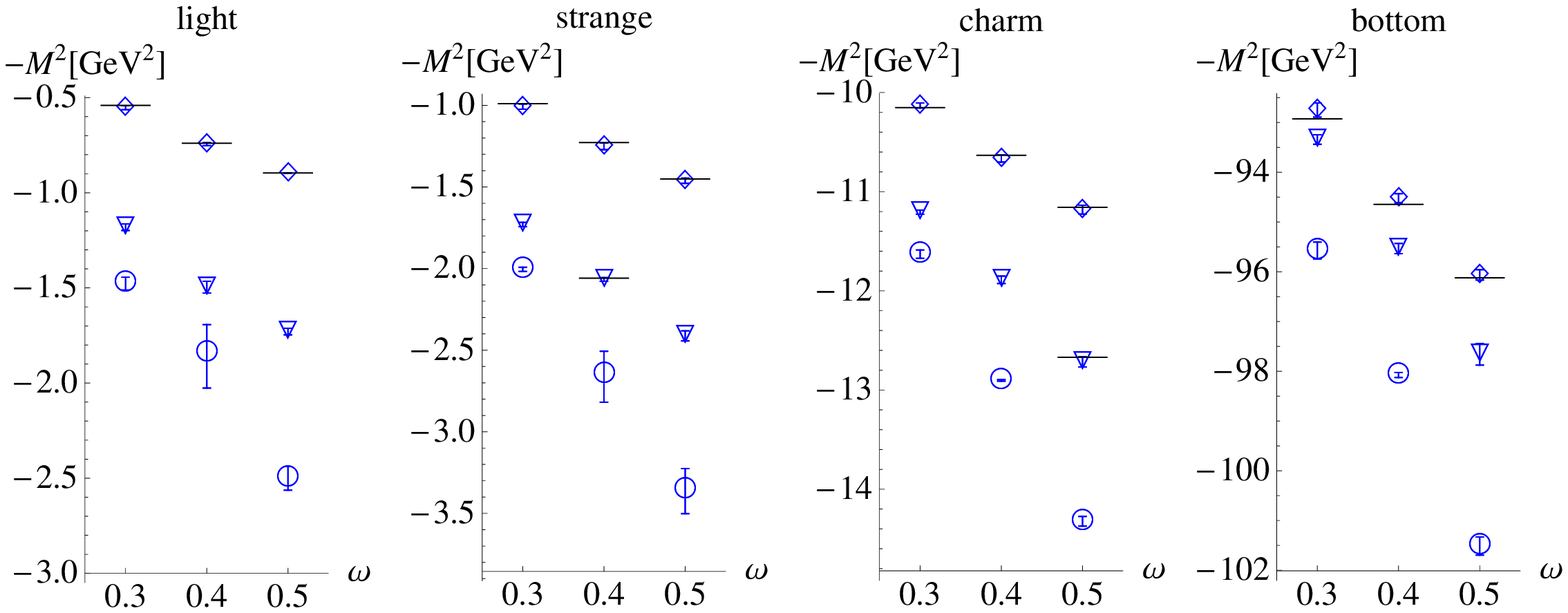}\\ \vspace*{-1cm}}}
\caption[$-M^2$ for the ground state, first and second excitation for $J^{PC}=2^{++}$, $1^{++}$, and $1^{+-}$.]{$-M^2$ for the ground state, first and second excitation for $J^{PC}=2^{++}$ (panel (a)), $1^{++}$ (panel (b)), and $1^{+-}$ (panel (c)), analogous to Fig.~\ref{fig:int1}.\label{fig:int2}}
\end{figure}

In order to test the reliability of this method for extrapolation, we use it to interpolate masses that can be obtained also from a direct calculation, and apply the procedure to the quantum numbers $J^{PC}=0^{-+}$, $0^{++}$, $1^{++}$, $1^{+-}$, $2^{++}$, for quark masses from light to bottom. The result is given in Figs.~\ref{fig:int1} - \ref{fig:int2}, where the symbols denote the fit results, and the lines represent the solutions obtained from a direct calculation. The errorbars of the interpolation are in all cases consistent with the directly obtained values, which are in turn consistent with the results for the ground states given in \cite{Krassnigg:2009zh,Krassnigg:2010mh} and for the first excited states for pseudoscalar and vector quantum numbers in \cite{Krassnigg:2008gd}. This indicates that the extrapolations performed are reliable, and that at least in the cases investigated here the systematics are under control, since deviations from the linear behavior result in larger errorbars of the fit.

\newpage
\section{Ground- and excited-state spectroscopy}\label{sec:spectroscopy}
The results of the previous section allow a stable extrapolation of meson masses beyond the limits given by the pole-pairs in the quark propagator. Thus, we are in the position to study the mass spectrum of quark-antiquark states in the isospin-symmetric limit. We  consider the ground states and first two excitations for the quantum numbers $J^{PC}=0^{-+}$, $0^{++}$, $1^{++}$, $1^{+-}$, $2^{++}$, and the ground state together with three excitation in the $1^{--}$ channel, in the MT model (cf.~App.~\ref{app:models}).

\subsection{Extrapolation in $\omega$}
In order to compare our results to the experimental values, we utilize the dependence of our results on the model parameter $\omega$ while $\omega\,D$ is kept constant. For the ground states in the pseudoscalar, vector and scalar channel this dependence is rather weak \cite{Krassnigg:2009zh}. On the other hand, the masses of the axialvector and tensor ground states change significantly with varying $\omega$ \cite{Krassnigg:2009zh,Krassnigg:2010mh}. This also happens in the case of radial excitations, as can be seen from the plots collected in App.~\ref{app:plots} (cf.~also \cite{Krassnigg:2008gd}). In general, the overall picture indicates better agreement with experimental data for higher values of $\omega$. 

Although the MT model was originally designed to reproduce observables for light pseudoscalar and vector mesons, we investigate the possibility that this model with the accompanying rainbow-ladder truncation is actually well suited to describe heavy mesons (cf.~\cite{Eichmann:2008ae,Eichmann:2008ef}). That picture is supported by recent studies in Coulomb gauge \cite{Popovici:2010mb,Popovici:2010ph}, which show that the RL truncation becomes exact in the limit of infinite quark masses. Thus, we choose the bottomonium system to fix the value for the model parameters. In particular, we focus on the quantum numbers $1^{--}$ where high-precision experimental data are available. Furthermore, as described in \cite{Krassnigg:2009zh}, the ground state of this channel was used to fix the bottom quark mass.

\begin{figure}
 \centering\includegraphics[width=0.38\columnwidth]{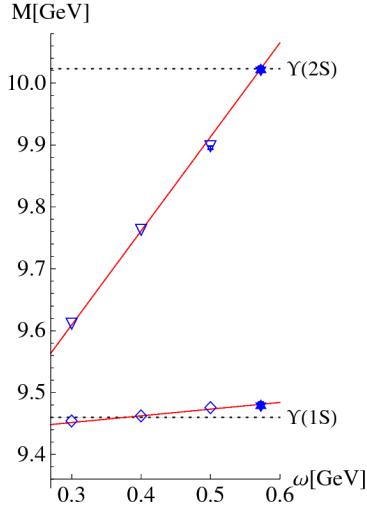}
\caption[$\omega$-dependence of the masses of the ground state and first excitation for the b-\={b}  $J^{PC}=1^{--}$ channel.]{$\omega$-dependence of the masses of the ground state and first excitation for the b-\={b}  $J^{PC}=1^{--}$ channel. The lines represent a linear fit through the calculated values for $\omega=0.3\, ,\;0.4\, ,\;0.5\,$GeV. The first excitation is extrapolated to the experimental value at $\omega=0.572\,$GeV, which is indicated by the star-symbol.\label{fig:omegafit}}
\end{figure}

Fig.~\ref{fig:omegafit} shows the $\omega$ dependence of the masses of the ground state and the first excitation (open symbols). Both show linear behavior, and we use a linear fit to extrapolate the mass of the first excitation to the experimental value. By this procedure, we fix the value of $\omega$ to
\beq
\omega=0.572\,\textrm{GeV}\;,
\eeq
indicated by the star-symbol in the figure. 

It is important to note here that an extrapolation of the masses along the lines of Sec.~\ref{sec:extrapolate} was only necessary for the excited state for $\omega=0.5\,$GeV. A further note concerns the extrapolated value of the $1^{--}$ ground state mass, which does not agree completely with experiment. This small discrepancy is acceptable at this stage, since the aim here is not to fine-tune the bottom quark mass, $D$, and $\omega$ fully but to provide an overall picture with reasonable effort. This is also justified a posteriori by the following discussion.

In the vector channel, we consider a total of four states (ground state and three excitations), in the other channel we calculate three (ground state and two excitations). In order to obtain results at our fixed value of $\omega$, we use the same procedure as described above and extrapolate linearly. The data and the corresponding linear fits are shown in Figs.~\ref{fig:omegaPS} - \ref{fig:omegaTE}, collected in App.~\ref{app:plots}. 

\subsection{Results and discussion}
\begin{figure}
\includegraphics[width=\columnwidth]{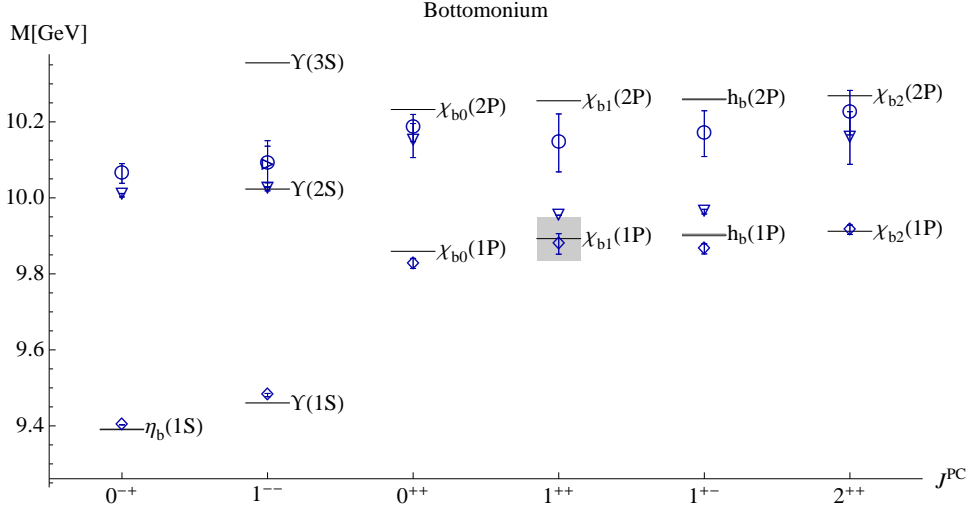}
\caption[Results of the extrapolation for bottomonium compared to the experimental spectra.]{Results of the extrapolation for bottomonium (represented by symbols with errorbars) compared to the experimental spectra \cite{Nakamura:2010zzi} (black lines). The experimental value for the masses of $h_b(1P)$ and $h_b(2P)$ were taken from Refs.~\cite{Lees:2011zp} and \cite{Adachi:2011ji}. The experimental uncertainties are indicated with gray bands (where visible), and the errorbars of our results correspond to the $1\sigma$ confidence interval of the fit in $\omega$, evaluated at $\omega=0.572$ (cf.~Fig.~\ref{fig:omegafit}).\label{fig:bbbar}} 
\end{figure}

\begin{figure}
\includegraphics[width=\columnwidth]{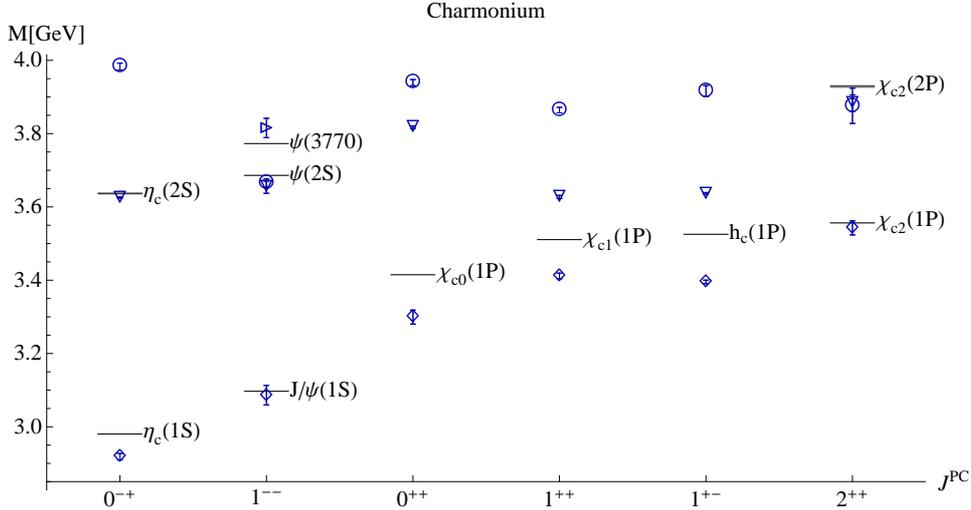}
\caption[Results for charmonium compared to experimental data.]{Our results for charmonium compared to experimental data \cite{Nakamura:2010zzi}, analogous to Fig.~\ref{fig:bbbar}. The experimental value for the mass of $h_c(1P)$ was taken from Ref.~\cite{Ablikim:2010rc}. Note that in the $1^{--}$ channel the first and second excitation are degenerate within errorbars.\label{fig:ccbar}} 
\end{figure}

\begin{figure}
\includegraphics[width=\columnwidth]{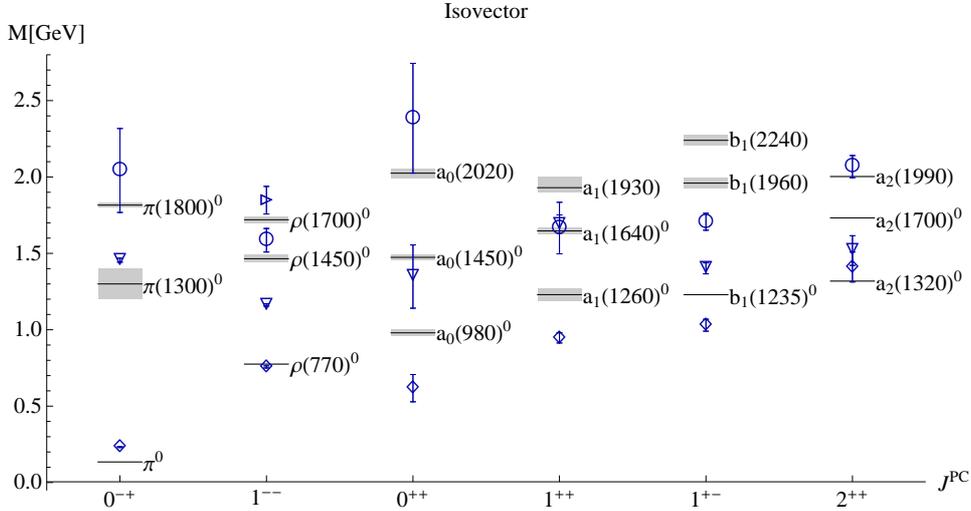}
\caption[Results of the extrapolation in the isovector channel (light quark masses), compared to the experimental spectra.]{Results of the extrapolation in the isovector channel (light quark masses), compared to the experimental spectra \cite{Nakamura:2010zzi}, analogous to Fig.~\ref{fig:bbbar}.\label{fig:isovec}} 
\end{figure}

\begin{figure}
\includegraphics[width=\columnwidth]{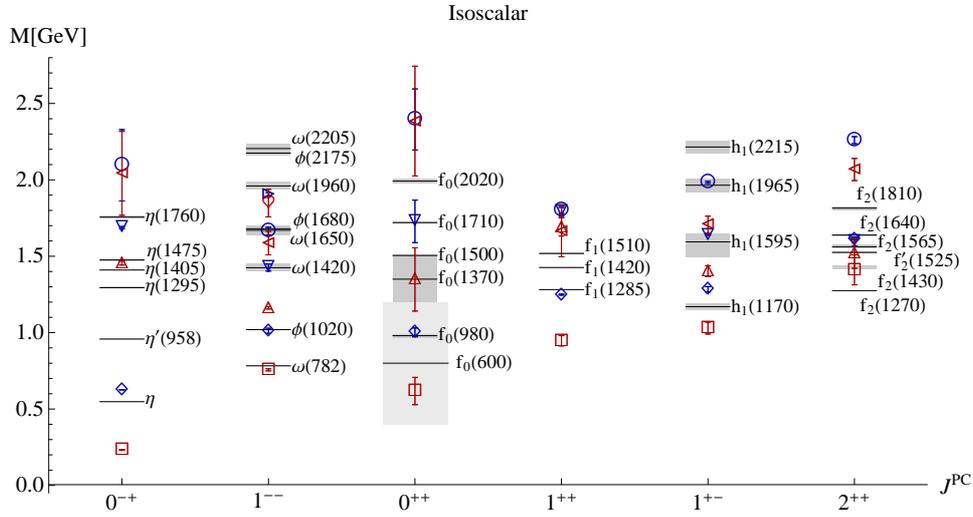}
\caption[Results of the extrapolation in the isoscalar channel, compared to the experimental spectra.]{Results of the extrapolation in the isoscalar channel, compared to the experimental spectra \cite{Nakamura:2010zzi} (black lines), analogous to Fig.~\ref{fig:bbbar}. The blue symbols ($\diamond$, $\triangledown$, $\circ$, $\triangleright$) correspond to s\={s} bound states, the red symbols ($\square$, $\vartriangle$, $\triangleleft$, $\heartsuit$) to n\={n} bound states in our calculation. The experimental error of $f_0(600)$ is shaded in lighter gray, since it overlaps with the error of $f_0(980)$. In the case of the $f_2(1430)$ state, the dotted line is omitted, since in \cite{Nakamura:2010zzi} the mass is given as $M\approx1430\,$MeV, and no experimental error is reported. Note that we do not consider mixing.\label{fig:isosc}} 
\end{figure}

In Figs.~\ref{fig:bbbar} - \ref{fig:isosc} we present the final results of the extrapolations. In all cases, the errorbars correspond to the 1$\sigma$ confidence interval of the linear fit, and reflect the quality of the linear extrapolation combined with the uncertainties in the mass extrapolation. 


The spectrum of b\={b} mesons, given in Fig.~\ref{fig:bbbar}, shows good agreement with experiment for all ground states, which supports the conjecture that the rainbow-ladder truncation is well suited to describe heavy quarks in Landau gauge. In the vector channel the ground state was used to determine the quark mass and the first excitation to fix the value for $\omega$, such that these states are fitted to the experimental values. In the other channels, the mass of the first excited state (where measured) is better described by the second excitation appearing in our calculation. However, it is unclear whether all excitations appearing as the solutions of a BSE correspond to physical excitations of the system under consideration. Hence, we have to allow for the possibility to discard certain so-called `spurious' states in order to achieve agreement with experiment. In the vector channel, this possibly applies to the third and fourth excitation, which are degenerate within errorbars.

In the charmonium system, Fig.~\ref{fig:ccbar}, the pseudoscalar ground state and first excitation are in good agreement with the experimental data. As before, the vector ground state was fitted to the experimental value to fix the quark mass. The first and second excitation in our calculation are degenerate and are both consistent with the $\psi(2S)$ meson, and the $\psi(3S)$ state is described by our third excitation. The situation is similar for $J^{PC}=2^{++}$. In the case of $J^{PC}=0^{++}$, $1^{++}$, and $1^{+-}$, the ground states show deviations from the measured values. This may indicate that due to the lighter quark mass the rainbow-ladder truncation is less adequate to describe these states, and corrections have to be taken into account.

%the mass of the $f_2(1430)$ state is given in \cite{Nakamura:2010zzi} as $M\approx1430$, and no experimental error is reported. This is indicated in the figure by the omission of the dotted line in this case.

This picture also holds for light isovector states, as can be seen from Fig.~\ref{fig:isovec}. Especially in the case of psudoscalar and vector mesons, the ground state and the first two excitations are in good agreement, if the first calculated excitation in the vector channel is discarded. 

Finally, our results for the light isoscalar channel are shown in Fig.~\ref{fig:isosc}. Since we do not consider mixing and work in RL truncation in an isospin-symmetric setup, this channel contains pure n\={n} states (which are degenerate with the corresponding n\={n} isovector states) and the pure s\={s} bound states, where n denotes the light (up/down) quark. In the figure, they are represented by the red symbols ($\square$, $\vartriangle$, $\triangleleft$, $\heartsuit$) and the blue symbols ($\diamond$, $\triangledown$, $\circ$, $\triangleright$), respectively.
For $J^{PC}=1^{--}$ one has ideal $SU(3)$-flavor mixing, and thus our approximation is trustworthy in this case. This also allows a consistent fit of the strange quark mass via the $\phi(1020)$. In order to achieve better agreement with the experimental data in the pseudoscalar channel, one would have to consider mixing effects and explicitly include contributions from anomalous terms. However, such investigations are beyond the scope of this thesis.

\newpage
\section{Applications of the vertex BSE}\label{sec:inhomresults} 
\subsection{On-shell quantities}\label{sec:inhomonshell}
Mesons as bound states are described by pole terms in the quark-antiquark propagator, which allows the construction of the homogeneous BSE to study their on-shell properties, as shown in Sec.~\ref{sec:hom_bse}. However, these pole contributions also translate into the vertex BSE, such that the inhomogeneous BSA, like the quark-antiquark propagator, has poles at the positions of the bound states (cf.~\cite{Blank:2010sn}), as can be seen by, e.g., comparing the spectral representations of the quark-antiquark propagator, Eq.~\eq{spectralG}, and of the inhomogeneous BSA, Eq.~\eq{spectralinhom}. Since the vertex BSE can be solved in a numerically efficient way \cite{Blank:2010bp}, it is interesting to discuss methods to obtain on-shell quantities like masses and decay constants from the inhomogeneous BSA.

\begin{figure}
\centering\includegraphics[width=0.6\columnwidth]{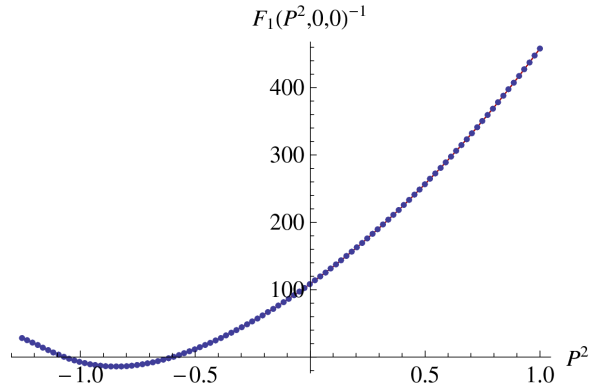}
\caption[The inverse of the first component of the inhomogeneous vector amplitude $1/F_1(P^2,0,0)$, as a function of the square of the total momentum $P^2$.]{\label{fig:massfit}The inverse of the first component of the inhomogeneous vector amplitude $1/F_1(P^2,0,0)$, as a function of the square of the total momentum $P^2$. The zero-crossings give the masses of the ground state and the first excitation.} 
\end{figure}

The most straightforward way to calculate bound state masses in this setup is to locate the positions of the poles in the inhomogeneous BSA with respect to the square of the total momentum, $P^2$. Since the poles appear in all components of the amplitude, this can be done by investigating the zero crossings of any component via an interpolation or a fit \cite{Blank:2010sn,Bhagwat:2007rj}.

As an example, we consider the inhomogeneous BSA for vector quantum numbers and light quarks using the parameter set MT2 (cf.~App.~\ref{app:models}). The correctly renormalized inhomogeneous term $|\Gamma_0\rangle$ in this case is \cite{Maris:1999bh}
\beq
|\Gamma_0\rangle=Z_2 \gamma_\mu\;,
\eeq
with the renormalization constant $Z_2$ from Eq.~\eq{z2z4}. As discussed in App.~\ref{app:cparity}, this choice of $|\Gamma_0\rangle$ has no overlap with states of positive C-parity, such that only non-exotic poles appear.

We calculate (cf.~\cite{Blank:2010sn}) the bound state masses by fitting the inverse of the first component $F_1(P^2,0,0)$ of the inhomogeneous BSA as shown in Fig.~\ref{fig:massfit}, and obtain for the ground and first excited states
\begin{equation}\label{eq:mresults}
P_0^2=-0.5991\;\mathrm{GeV}^2\qquad
P_1^2=-1.0692\;\mathrm{GeV}^2
\end{equation}
which is in agreement with the corresponding results from the homogeneous BSE presented in Sec.~\ref{sec:extrapolate}. 

\begin{figure}
\begin{center}
\includegraphics[width=0.4\columnwidth]{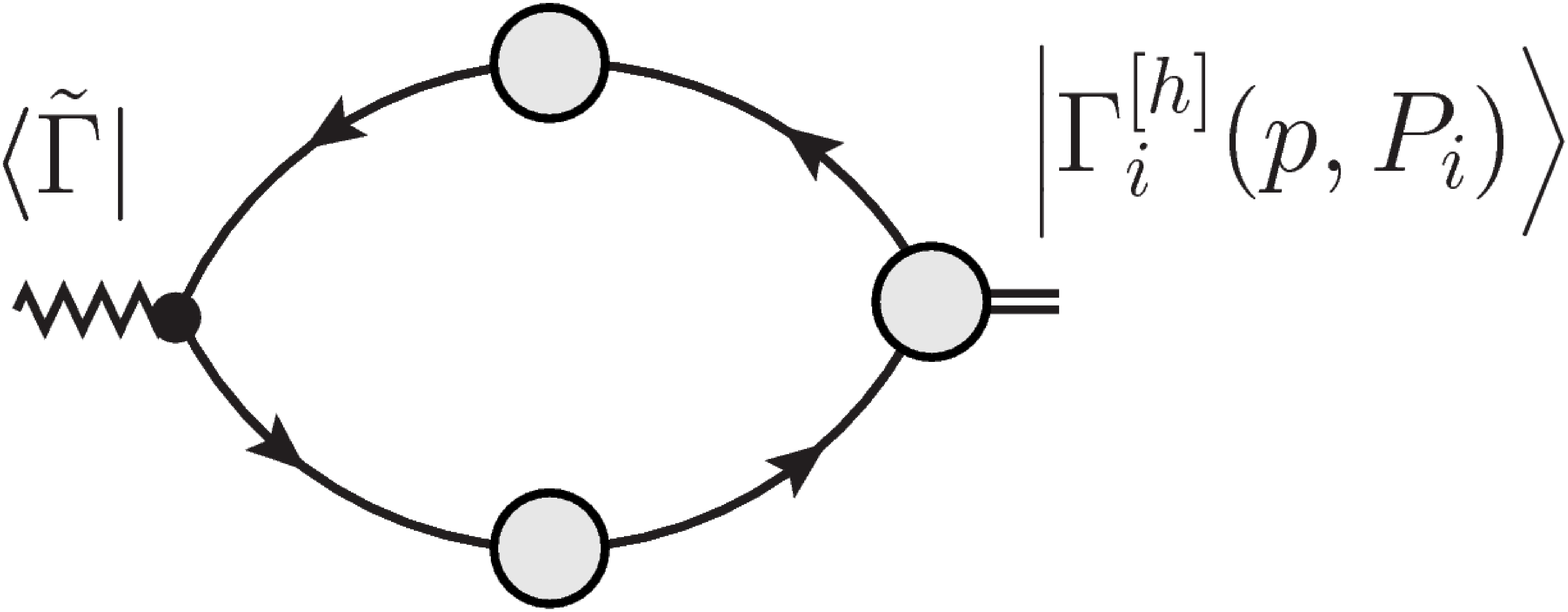} 
\end{center}
\caption[Diagrammatic representation of the equation for a leptonic decay constant.]{\label{fig:decayconst}Diagrammatic representation of Eq.~\eq{decconst}.} 
\end{figure}

In the BSE formalism, mesonic decay constants are related to the coupling $f_{\tilde{\Gamma}}^i$ of a (canonically normalized) on-shell homogeneous BSA to a suitably renormalized current $\langle \tilde{\Gamma}|$ , which in diagrammatic form is given in Fig.~\ref{fig:decayconst}. In the notation employed in this chapter, the equation reads
\beq\label{eq:decconst}
f_{\tilde{\Gamma}}^i=\frac{1}{\sqrt{\mathcal{N}_i}}\langle \tilde{\Gamma}\xr{i}{P_i}{p}\;.
\eeq
For pseudoscalar and vector mesons, $\langle \tilde{\Gamma}|=Z_2\gamma_\mu \gamma_5$ and $\langle \tilde{\Gamma}|=Z_2/3 \gamma_\mu$, respectively \cite{Leutwyler:1973mu,Maris:1997hd,Maris:1999nt}.

As described in \cite{Blank:2010sn}, the projection \eq{decconst} can be extracted also  from the inhomogeneous BSA without resorting to any information from the homogeneous amplitude. Starting from Eq.~\eq{spectralinhom}, we consider the corresponding projection $f^{(ih)}_{\tilde{\Gamma}}(P^2)$ of the inhomogeneous BSA on the current $\langle \tilde{\Gamma}|$,
\beq\label{eq:inhomproj1}
f^{(ih)}_{\tilde{\Gamma}}(P^2)\equiv\langle\tilde{\Gamma}|D(P,p)|\Gamma(P,p)\rangle=\sum_j\frac{1}{1-\lambda_i(P^2)}\langle \tilde{\Gamma}\xr{j}{P_j}{p}\xl{j}{P_j}{q}\Gamma_0\rangle\;.
\eeq
The residue of $f^{(ih)}_{\tilde{\Gamma}}(P^2)$ at the pole $i$ is given by
\beq
\textrm{Res}(f^{(ih)}_{\tilde{\Gamma}}(P^2),P_i)=\frac{1}{\mathcal{N}_i}\langle \tilde{\Gamma}\xr{i}{P_i}{p}\xl{i}{P_i}{q}\Gamma_0\rangle=\frac{1}{\mathcal{N}_i}f_{\tilde{\Gamma}}^i f_{\Gamma_0}^i\;.
\eeq
Thus, in order to calculate $f_{\tilde{\Gamma}}^i$ one in addition has to determine the projection of the homogeneous BSA on the inhomogeneous term, $f_{\Gamma_0}^i$. Using the same arguments as before, this quantity can be obtained via
\beq
\textrm{Res}(f^{(ih)}_{\Gamma_0}(P^2),P_i)=\frac{1}{\mathcal{N}_i}\langle \Gamma_0\xr{i}{P_i}{p}\xl{i}{P_i}{q}\Gamma_0\rangle=\frac{1}{\mathcal{N}_i}\left(f_{\Gamma_0}^i\right)^2\;.
\eeq
In total,
\beq
f_{\tilde{\Gamma}}^i=\textrm{Res}(f^{(ih)}_{\tilde{\Gamma}}(P^2),P_i)/\sqrt{\textrm{Res}(f^{(ih)}_{\Gamma_0}(P^2),P_i)}\;.
\eeq

\begin{figure}
\begin{center}
 \includegraphics[width=0.6\columnwidth]{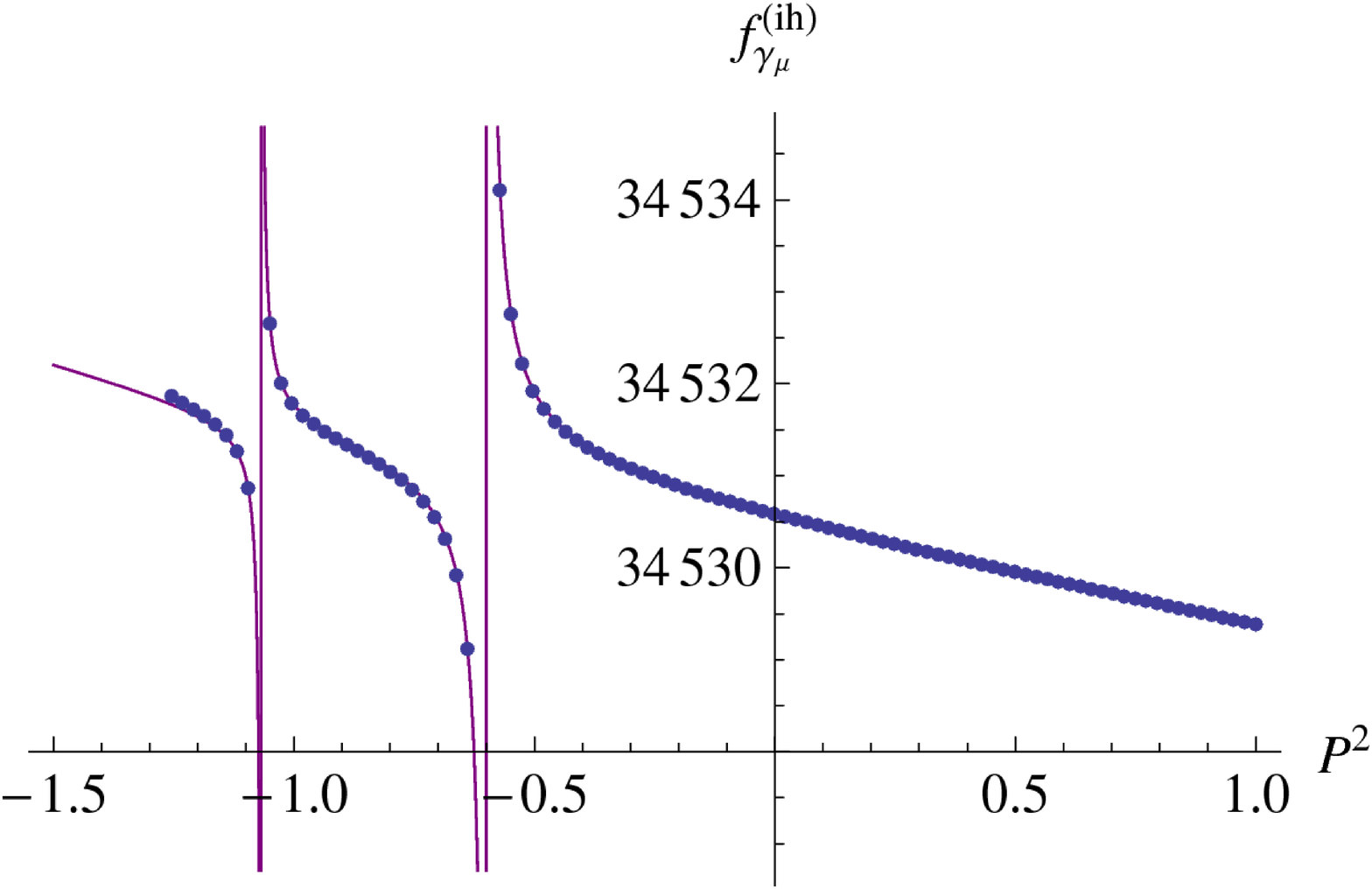}
\end{center}
\caption[The projection $f_{\tilde{\Gamma}}^{(ih)}(P^2)$ for the vector channel for light quarks.]{\label{fig:fvinhom}The projection $f_{\tilde{\Gamma}}^{(ih)}(P^2)$ for the vector channel for light quarks, defined in Eq.~(\ref{eq:inhomproj1}). The line represents a pole fit to the data points, where the pole positions were taken from the fit of Fig.~\ref{fig:massfit}.} 
\end{figure}

In the vector meson channel, the decay constant of the $i$-th state is given by
\beq
f_v=f_{\tilde{\Gamma}}^i/M_i\;,
\eeq
where $M_i=\sqrt{-P_i^2}$ denotes the mass of the respective excitation, and $\tilde{\Gamma}=Z_2/3 \gamma_\mu$, as stated before. Therefore, the inhomogeneous term and the current only differ by a factor of $1/3$, and the decay constant can be calculated using solely the residues of  $f^{(ih)}_{\tilde{\Gamma}}(P^2)$ via
\begin{equation}
 f_v=\frac{1}{M_i}\sqrt{\frac{1}{3} \textrm{Res}(f^{(ih)}_{\tilde{\Gamma}}(P^2),P_i)}\;.
\end{equation}
These residues are extracted from the pole fit shown in Fig.~\ref{fig:fvinhom}, where the masses (\ref{eq:mresults}) are used as input.

For the parameters used here, this gives the decay constants of the $\rho$-meson and its first radial
excitation corresponding to the masses in (\ref{eq:mresults}). The results, as presented in \cite{Blank:2010sn}, are
\begin{equation}\label{eq:fresults}
f_\rho=0.213\;\mathrm{GeV}\qquad
f_{\rho'}=0.079\;\mathrm{GeV}\;,
\end{equation}
which agrees perfectly with the values obtained by using the homogeneous BSE.

\subsection{Projection on bound-state poles}\label{sec:inhomremovepole}
The pole structure of the inhomogeneous BSA is not only determined by the chosen structure of the amplitudes, but also by the choice of the inhomogeneous term $|\Gamma_0\rangle$. Therefore, only poles whose homogeneous BSA has nonvanishing overlap with $|\Gamma_0\rangle$ can appear in the solution of the vertex BSE, which explains why (for momentum partitioning $\eta=1/2$) only poles with the same C-parity as $|\Gamma_0\rangle$ contribute.

This argument, however, can be turned around to construct an inhomogeneous term that removes the contribution of certain poles, provided that the corresponding homogeneous BSAs are known. Consider, for example, the inhomogeneous term
\beq
|\Gamma_0'(p)\rangle=|\Gamma_0\rangle-\gar{i}{P_i}{p}\;,
\eeq
where $\gar{i}{P_i}{p}$ is the homogeneous BSA of state $i$. Inserting it into the spectral representation of the inhomogeneous BSA, Eq.~\eq{spectralinhom}, gives
\beq
|\Gamma'(p,P)\rangle=\sum_j\frac{1}{1-\lambda_j(P^2)}\gar{j}{P}{p}\xl{j}{P}{q}\Gamma_0'(q)\rangle\;.
\eeq
At $P=P_i$, the term corresponding to the $i$th pole cancels, and the inhomogeneous BSA reduces to
\beq
|\Gamma'(p,P_i)\rangle=\sum_{j\neq i}\frac{1}{1-\lambda_j(P_i^2)}\gar{j}{P_i}{p}\xl{j}{P_i}{q}\Gamma_0\rangle\;,
\eeq
which by construction has no contribution from the state represented by $\gar{i}{P_i}{q}$, i.e.~it is regular at $P=P_i$. 

\begin{figure}
\begin{center}
 \includegraphics[width=0.6\columnwidth]{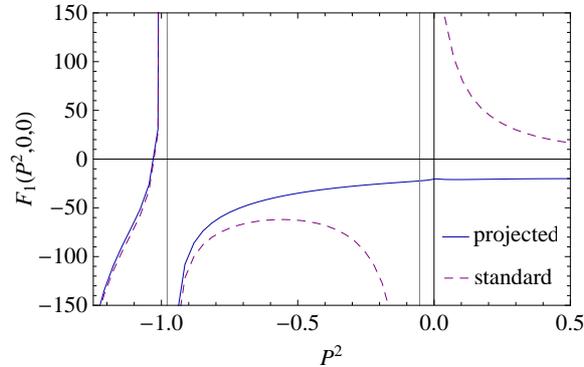}
\end{center}
\caption[Component $F_1(P^2,0,0)$ of the inhomogeneous pseudoscalar amplitude vs. the square of the total momentum $P^2$, with and without projection on the ground state.]{\label{fig:projinhom} Component $F_1(P^2,0,0)$ of the inhomogeneous pseudoscalar amplitude vs. the square of the total momentum $P^2$, with (solid line) and without (dashed line) projection on the ground state. The gray vertical lines mark the pole positions of ground and first excited state. } 
\end{figure}

For a demonstration, we use the parameter set MT2 with light quarks (cf.~App.~\ref{app:models}) in the pseudoscalar channel, where $|\Gamma_0\rangle=Z_4 \gamma_5$ \cite{Maris:1997hd} with $Z_4$ from Eq.~\eq{z2z4}. 
In Fig.~\fig{projinhom}, we show the result of this procedure when applied to the ground state, for the first component $F_1$ corresponding to the pseudoscalar covariant $T_1$ given in Eq.~\eq{tpsN}. As can be seen, the ground state is indeed removed from the inhomogeneous BSA (solid line) as compared to the standard calculation where all poles contribute (dashed line).
%Indeed, as shown in Fig.~\fig{projinhom}, the procedure described above allows to remove the ground state pole from the inhomogeneous BSA (solid line) as compared to the standard calculation (dashed line) where all poles contribute.

If this procedure is applied to all bound states which can be calculated directly, it allows for more stable extrapolations of pole positions based on fits of the inhomogeneous BSA. In general, if more than one pole is considered, these types of fits become unstable and thus complicate the study of excitations. In addition, the method proposed here allows a direct investigation of non-resonant contributions to the inhomogeneous BSA, since only the pole part is removed.

\newpage
\section{Homogeneous BSE and the far infrared}\label{sec:bseir}
The infrared behavior of the gluon and the ghost propagators in Landau gauge Yang-Mills theory is currently debated, and two types of solutions to these propagators which differ only in the far infrared have been proposed.

It is expected that the form of the ghost and gluon propagator in the far infrared has very little impact on phenomenological quantities like meson properties. While this is consistent with the assumption that the different solutions are selected by a gauge choice \cite{Fischer:2008uz}, it is also supported by recent lattice calculations \cite{Glozman:2009cp,Glozman:2010vd}, which find the gauge-invariant contributions to the $\rho$ meson to be only about half a Fermi large.

In the BSE setup, the effect of the far infrared on properties of $\pi$ and $\rho$ mesons has been investigated in \cite{Blank:2010pa}, where the results and analysis described in the following have been presented. 

\subsection{Effective Interaction}
For this study, the rainbow-ladder truncation of the quark DSE and the BSE is used. As discussed in Sec.~\ref{sec:truncation} (cf.~also App.~\ref{app:models}), this setup requires an effective interaction as an input, through which the Yang-Mills sector of the theory enters the equations.

The form of the effective interaction can be based on the QCD running coupling $\alpha(k^2)$ \cite{Maris:1997tm, Maris:1999nt}, where one important ingredient is the perturbative UV behavior. When viewed in this context, the rainbow-ladder truncation amounts to replacing the Lorentz- and Dirac-part of the product of the dressed quark gluon vertex and the dressed gluon propagator by their bare counterparts multiplied by the (modeled) running coupling, such that
\beq\label{eq:model}
\Gamma_\mu(p,q)D_{\mu\nu}(p-q) \rightarrow \gamma_\mu \alpha(k^2) \frac{1}{k^2}\left( \delta_{\mu\nu}-\frac{k_\mu k_\nu}{k^2} \right)\;,
\eeq
where $k=p-q$, and $\Gamma_\mu(p,q)$ and $D_{\mu\nu}(p-q)$ are the full quark-gluon vertex and gluon propagator, respectively. 

In the context of Yang-Mills theory, the coupling can be defined as \cite{Fischer:2008uz,vonSmekal:1997is}
\beq
\alpha(k^2)=\alpha(\mu^2)( G(k^2,\mu^2))^2 Z(k^2,\mu^2)\label{eq:alpha}
\eeq
in the so-called miniMOM (minimal momentum subtraction) scheme \cite{vonSmekal:2009ae}. Herein $G$ and $Z$ are the dimensionless ghost and gluon dressing functions, respectively, which are obtained from the scalar part of the propagators by multiplication with $k^2$. For the scaling-type solution this coupling is infrared finite, while for the decoupling-type solutions it is infrared vanishing like $k^2$ \cite{Fischer:2008uz}. Note that this does not contradict the conjecture that both could be just gauge choices, since the running coupling defined in Eq.~\eq{alpha} is not gauge-invariant.

\begin{figure}
\centering\includegraphics[width=0.6\linewidth]{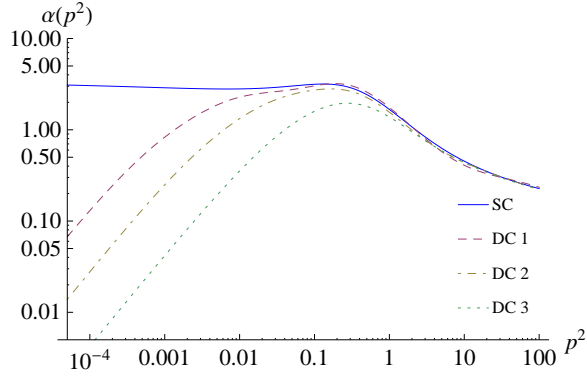}
\caption[The input running coupling corresponding to the scaling and decoupling solutions.]{\label{fig:alpha}The input running coupling \eq{alpha}, where SC denotes the scaling
solution and DC the decoupling solutions. For more details, see \cite{Fischer:2008uz}.}
\end{figure}

The functional form of the input running coupling used here is taken from a fit \cite{maasprivate} to the solutions given in \cite{Fischer:2008uz}. It is shown in Fig.~\ref{fig:alpha}, where it can be seen that the coupling quantitatively depends on the choice of solution below a momentum scale of 1 GeV, and qualitatively below 100 MeV. In contrast to the scaling-type solution (denoted by SC), the decoupling-type solution is possibly not unique \cite{Fischer:2008uz}. For this study, three representatives (denoted by DC 1 - DC 3) are selected, one of them (DC 3) being in rather good quantitative agreement with lattice results \cite{Fischer:2008uz}, in order to study also the dependence on different variants of decoupling-type behavior.

On the other hand, truncation artifacts are introduced by using the bare quark-gluon vertex even if $\alpha(k^2)$ is given via solutions of the Yang-Mills system. In order to correct this mismatch and make contact with phenomenology, we introduce the effective interaction
\beq\label{eq:geff}
\mathcal{G}(k^2)=\alpha(k^2) F(k^2,\omega,D)\;,
\eeq
which replaces $\alpha(k^2)$ in Eq.~\eq{model}. Here,
\beq\label{eq:vertex}
F(k^2,\omega,D)\equiv 1+4 \pi^2 k^2 \frac{D}{\omega^6} \exp^{-\frac{k^2}{\omega^2}}
\eeq
represents a non-trivial momentum dependence of the quark-gluon vertex. In total, this provides a two-parameter model similar to the Maris-Tandy model, Eq.~\eq{mt} \cite{Maris:1999nt}.

The definition \eq{vertex} preserves the IR as well as the UV behavior of the running coupling $\alpha(k^2)$, such that the impact of the IR on meson properties can be studied in a setup where chiral symmetry and its dynamical breaking are implemented correctly (cf.~Sec.~\ref{sec:truncation}). Note that some additional results for the quark-gluon vertex are already available, which focus on its further tensor structures as well as its impact on bound-state calculations (see e.g.~\cite{Alkofer:2008tt,Kizilersu:2006et,Matevosyan:2007cx,Williams:2009ce,Chang:2010xs} and references therein). However, these are either specific to the scaling case \cite{Alkofer:2008tt}, from the lattice for small volumes  \cite{Kizilersu:2006et}, or from more involved calculations using functional methods \cite{Matevosyan:2007cx,Williams:2009ce,Chang:2010xs} and cannot be used in the same way as the gluon input for the comparison aimed at here.

\subsection{Quark propagator}
The quark propagator, consistently given as the solution of the rainbow-truncated gap equation \eq{gapRL}, is a necessary input to the BSE. Its properties are encoded in two scalar dressing functions, which may be written as
\beq
Z_q(p^2)=\frac{1}{A(p^2)}\quad\textrm{and}\quad M(p^2)=\frac{B(p^2)}{A(p^2}
\eeq
in terms of $A(p^2)$ and $B(p^2)$ defined in Eq.~\eq{squark}. $Z_q(p^2)$ and $M(p^2)$ are referred to as wave-function dressing and mass function, respectively. 

In order to calculate these functions, the parameters of the effective interaction as well as the current quark mass $m_q$, which acts as a driving term for the non-perturbative $M$ in the quark DSE, need to be specified. Using the procedure described in Sec.~\ref{sec:renormalization}, we have a value of $m_q=0.0047$ GeV at our renormalization point of $(19\,\textrm{GeV})^2$ and use two degenerate flavors. The parameters $D$ and $\omega$ are fixed by the pion properties as discussed below.

\begin{figure}
\includegraphics[width=0.48\linewidth]{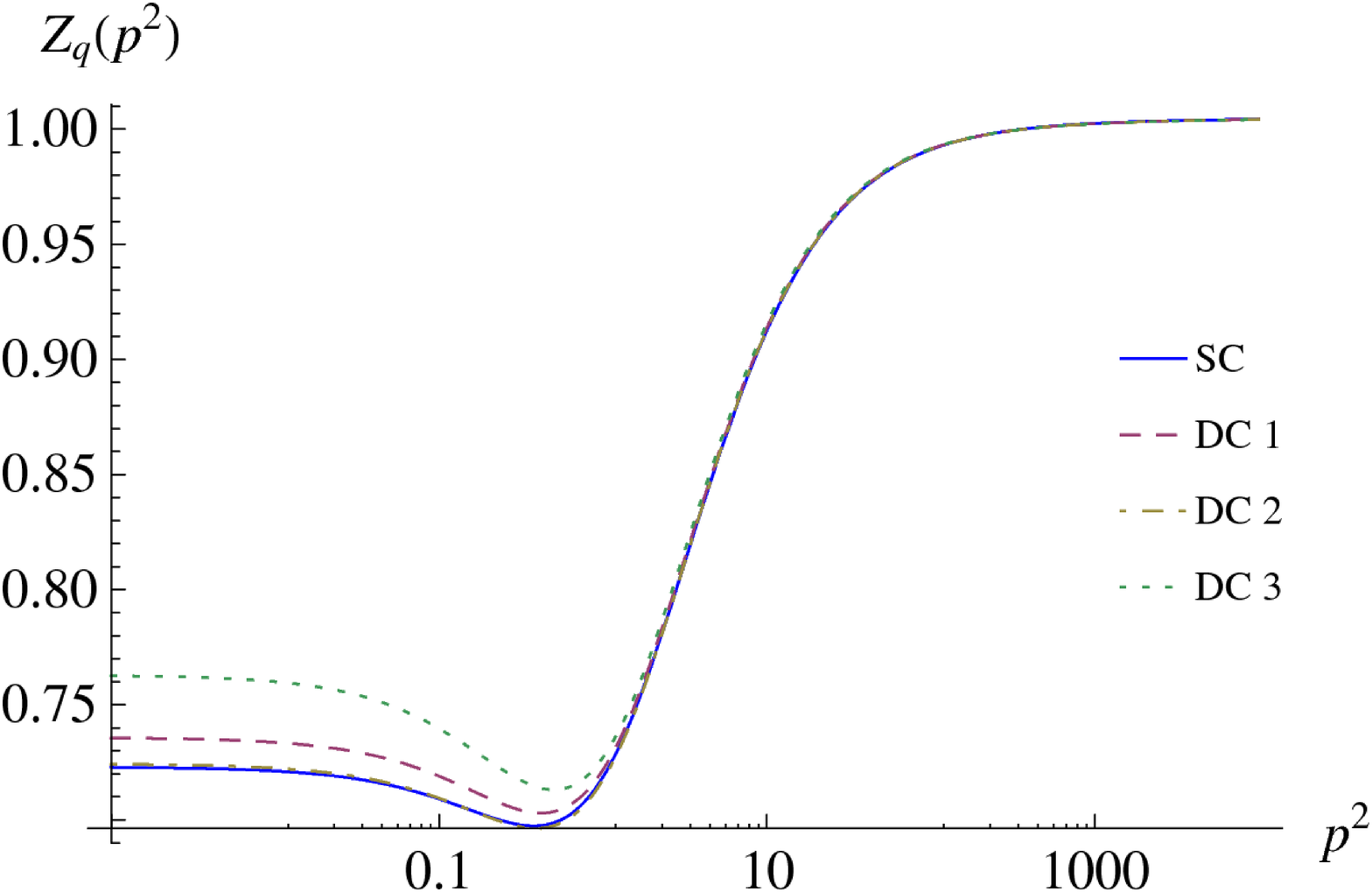}\includegraphics[width=0.48\linewidth]{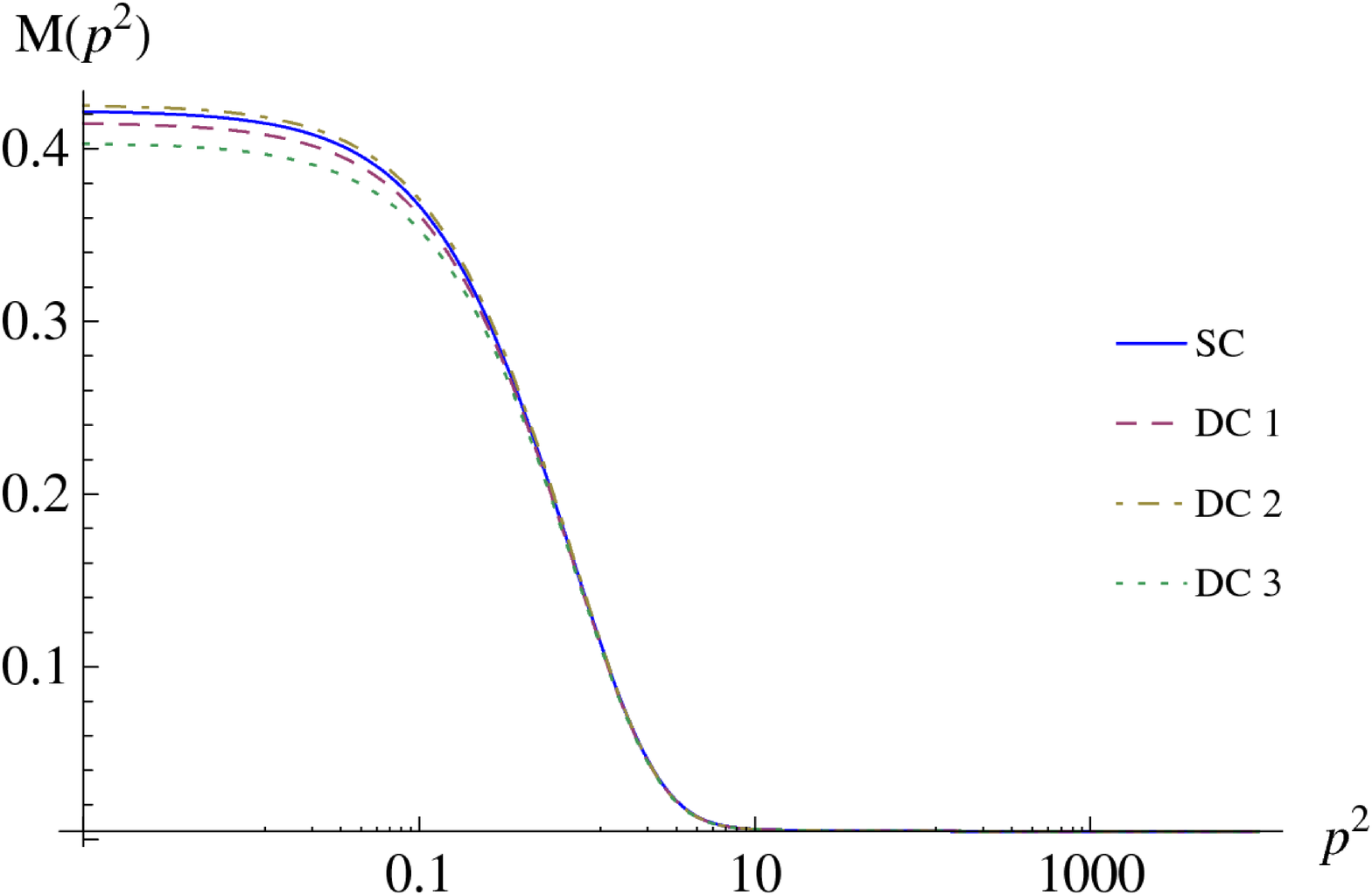}\\
\caption[The quark wave-function dressing and mass function for the four choices of
parameters of the quark-gluon vertex fixed to pion properties.]{\label{fig:quark}The quark wave-function dressing (left) and mass function (right) for the four choices of
parameters of the quark-gluon vertex fixed to pion properties, see Tab.~\ref{tab:fit}.}
\end{figure}

The resulting quark dressing functions are shown in Fig.~\ref{fig:quark}. They no longer show any qualitative difference when comparing results corresponding to the scaling- and decoupling-type inputs; in fact, only a small quantitative difference remains. The qualitative uniformity is a direct consequence of dynamical chiral symmetry breaking, since it makes the quark effectively decouple in the infrared and thus blind to the qualitative differences. The smallness of the quantitative differences among the solutions for $Z_q$ and $M$ turns out to stem from the requirement of accurately reproducing the pion properties.

\subsection{Properties of $\pi$ and $\rho$ mesons}

\begin{figure}
\includegraphics[width=0.5\linewidth]{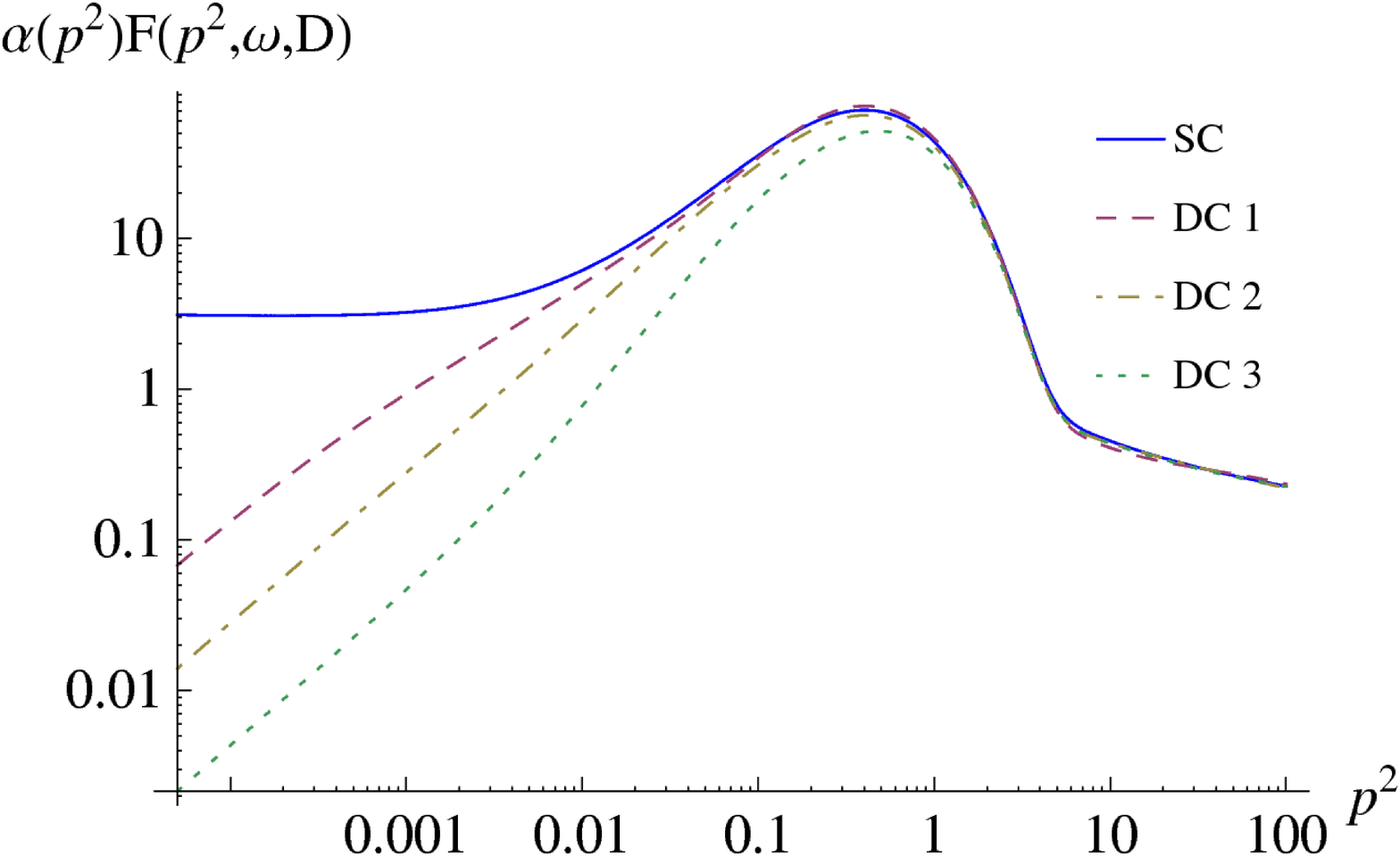}\includegraphics[width=0.5\linewidth]{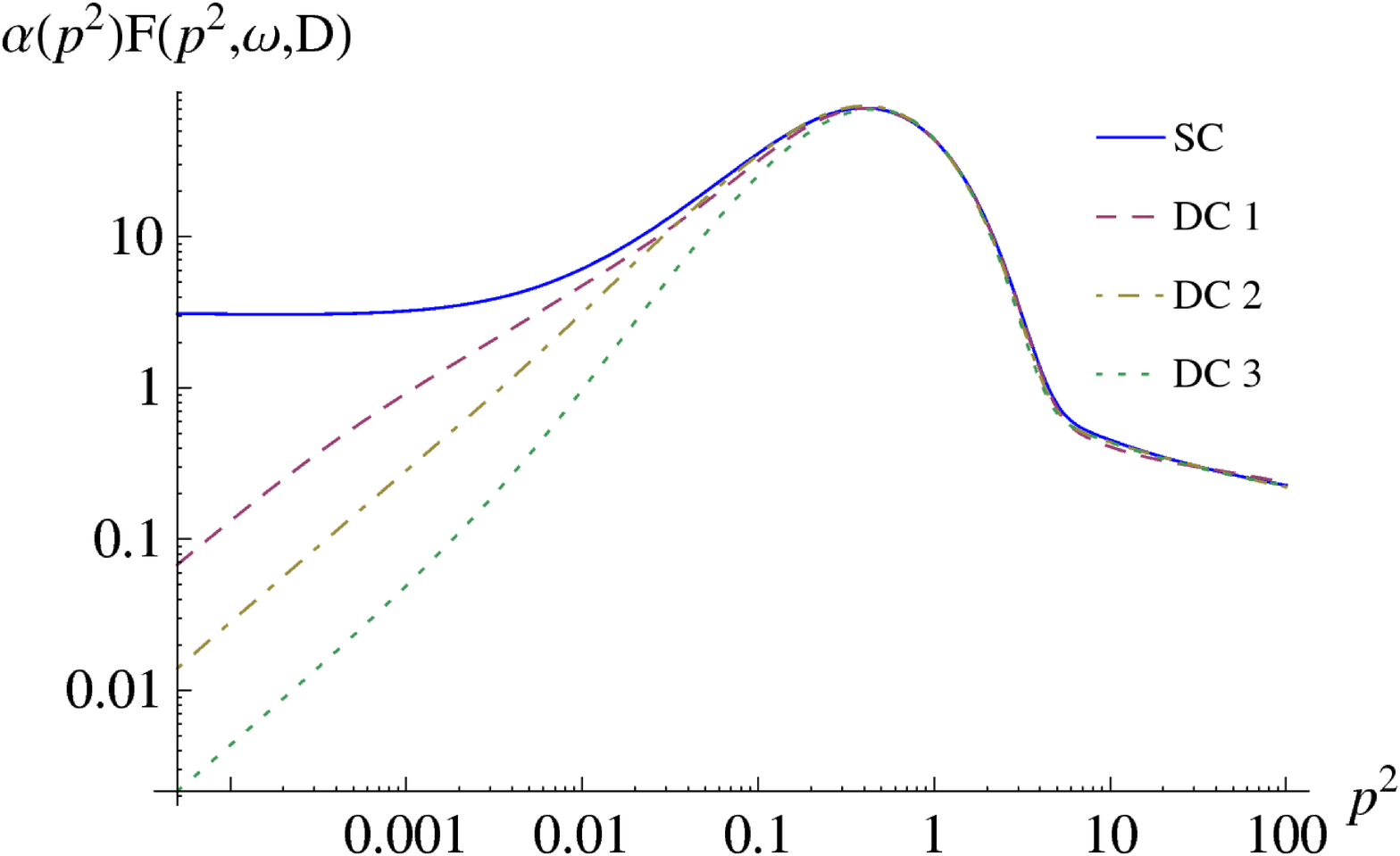}\\
\caption[The momentum dependence of the effective interaction in the scaling and decoupling cases.]{\label{fig:ansatz}The momentum dependence of the effective interaction \eq{geff} for the parameter values $D$ and $\omega$ of Tab. \ref{tab:unfit} (left) and for the optimized values of Tab.~\ref{tab:fit} (right).}
\end{figure}

\begin{table}
\begin{center}
\begin{tabular}{l|c|c|c|c}
  &$m_\pi$& $f_\pi$&$m_\rho$&$f_\rho$\\
\hline
Scaling      & 0.1417&0.0932&0.7443&0.1492\\\hline
Decoupling 1 & 0.1420&0.0956&0.7641&0.1533\\
Decoupling 2 & 0.1401&0.0897&0.7155&0.1439\\
Decoupling 3 & 0.1389&0.0805&0.6321&0.1239\\
\hline
Experiment   & 0.1396&0.0922&0.7755&0.1527
\end{tabular}
\caption{\label{tab:unfit} Results for $\pi$ and $\rho$ mass and decay constant after adjusting the parameters
of the vertex ansatz \eq{vertex} such that the scaling solution best reproduces the pion properties. All quantities
are given in units of GeV; our results are accurate to less than half an MeV.}
\end{center}
\end{table}
We present results for masses and decay constants, where the latter are sensitive to the Bethe-Salpeter amplitude of the meson under consideration. The numbers are compiled in two tables to highlight two different aspects of the investigation. First, we keep the same vertex ansatz for all four different Yang-Mills input sets. We adjust the parameters of the vertex ansatz \eq{vertex} such that the scaling solution reproduces the pion mass and decay constant adequately. The resulting values for $D$ and $\omega$ are 0.8 GeV${}^4$ and 0.8 GeV, respectively. Then we compute the masses $m$ and decay constants $f$, $m_\pi$, $f_\pi$, $m_\rho$, and $f_\rho$ with the same parameters also for the different decoupling-type cases; the corresponding results are listed in Tab.~\ref{tab:unfit} in units of GeV. One can see that the more the input coupling differs from the scaling case, the larger the differences for $f_\pi$, $m_\rho$, and $f_\rho$ become, while the general picture is not spoiled at a qualitative level. The small variation in $m_\pi$ is due to the influence of the axial-vector Ward-Takahashi identity. This increase in discrepancy when only changing the gluonic input has also been observed in the literature \cite{Natale:2009uz,Aguilar:2004td}.

Next, we offer another possibility: irrespective of whether the different inputs correspond to different gauges or are just different truncation assumptions, it is to be expected that the quark-gluon vertex should change accordingly when the running coupling changes. To investigate this scenario, we have fitted the parameters $D$ and $\omega$ for each input individually to reproduce the pion parameters. The corresponding results are shown in Tab.~\ref{tab:fit}. As one can clearly see, already a very moderate change in the parameters without excessive finetuning yields agreement of $m_\pi$, $f_\pi$, $m_\rho$, and $f_\rho$ for all inputs on the percent level. Thus, any consistent change of the quark-gluon vertex and the gluonic sector permits an equally good description of $m_\pi$, $f_\pi$, $m_\rho$, and $f_\rho$. While not guaranteed, it is at least possible that a similar statement holds for arbitrary other observables as well; some confirmation of our observation along these lines has already been found in \cite{Luecker:2009bs,Fischer:2009jm}.

\begin{table}
\begin{center}
\begin{tabular}{l|c|c|c|c|c|c}
  &$D$&$\omega$ &$m_\pi$& $f_\pi$&$m_\rho$&$f_\rho$\\
\hline
Scaling      &0.800&0.800& 0.1417&0.0932&0.7443&0.1492\\ \hline
Decoupling 1 &0.790&0.810& 0.1414&0.0932&0.7443&0.1496\\
Decoupling 2 &0.820&0.785& 0.1411&0.0936&0.7484&0.1502\\
Decoupling 3 &0.870&0.760& 0.1407&0.0935&0.7485&0.1514
\end{tabular}
\caption{\label{tab:fit}As Tab.~\ref{tab:unfit}, but $D$ and $\omega$ have now been adjusted for each input
separately to fit the pion's mass and decay constant. All quantities are given in units of GeV except for $D$
which has units of GeV${}^4$.}
\end{center}
\end{table}

It is interesting to check how the effective interaction \eq{geff} looks for the optimized fits. More precisely, we investigate the combined effect of the coupling and vertex dressing functions on different momentum scales. This is shown in Fig.~\ref{fig:ansatz}. It is nicely visible that the four cases start to differ appreciably only below 1 GeV, but are qualitatively different in the far infrared. This observation supports the argument that the $\rho$ has a gauge-invariant structure which exists at a level of 0.5 fm or below \cite{Glozman:2009cp,Glozman:2010zn}, and is not too sensitive to larger scales.

\chapter{Summary and outlook}\label{chap:summary}
The combined Dyson-Schwinger - Bethe-Salpeter formalism provides a consistent, nonperturbative, continuum approach to QCD phenomenology. This thesis focused on mainly two aspects: the chiral phase transition, which was investigated using the quark gap equation, and the meson spectrum studied by means of the (homogeneous) Bethe-Salpeter equation. 

For the numerical solution of the homogeneous and inhomogeneous vertex BSE, the application of sophisticated matrix algorithms was discussed. For the homogeneous BSE, which is generally solved as an eigenvalue equation, a reliable method to compute several eigenvalues and eigenvectors was investigated, which facilitates studies of excited states in this approach. The inhomogeneous vertex BSE was treated as an inhomogeneous linear system, which could be solved via approximate matrix inversion. This method allowed to resolve not only the ground state pole in the inhomogeneous BSA, but also further poles corresponding to the various excitations. Both of these methods were shown to be numerically advantageous when compared to a more conventional approach.

Inspired by the numerical methods, a spectral representation of the homogeneous BSE, the inhomogeneous vertex BSE, and the BSE for the quark-antiquark propagator was found. Its theoretical consequences allowed to uncover the connections between these types of BSEs, and led to new applications of the inhomogeneous vertex BSE, such as the possibility to calculate on-shell quantities like decay constants without resorting to the corresponding homogeneous BS amplitudes.

In addition to these general considerations, more detailed studies of the properties of quarks and mesons were performed. For these, the rainbow-ladder truncation was applied, which provides a description consistent with the requirements of chiral symmetry. 

In the finite-temperature formalism, the chiral phase transition was studied by means of the quark gap equation. In this context, several different forms of the effective interaction and their impact on the chiral phase transition were investigated. In all cases, the mean-field behavior of the transition in rainbow truncation was confirmed. The transition temperature shows a strong dependence on the particular model used, although no simple overall relationship of its precise value to, e.g., the strength of the interaction in the infrared and mid-momentum regime could be established. However, within one model, a change in the parameters was found to affect the value of the chiral transition temperature in a similar way as hadron properties at vanishing temperature. 

In the investigation of meson properties, the rainbow-ladder truncation allowed to establish a clear connection between the bound-state poles appearing in the quark-antiquark propagator and the behavior of eigenvalues of the homogeneous BSE. From this, a new extrapolation technique for meson masses could be motivated, which considerably enlarges the range of bound state masses accessible by the BSE approach. Combined with the numerical methods proposed, the ground- and excited-state meson spectrum could be calculated for all quark masses from light to bottom, for pseudoscalar, scalar, vector, axialvector and tensor mesons. Especially for the bottomonium system, the results for all ground states showed good agreement with experiment, and also for lighter quark masses the pseudoscalar, vector and tensor ground states were well described in the rainbow-ladder approximation used. A good description of scalar and axialvector states as well as a consistent identification of the physical excitations, however, remains a challenge.

In the rainbow-ladder setup, the effective action can be used to investigate the impact of different treatments of the Yang-Mills system on meson phenomenology. It was shown that the behavior of the effective interaction in the infrared, which originates from different versions of the Landau-gauge ghost and gluon propagators, has little effect on the masses and decay constants of the $\pi$ and $\rho$ mesons.

The results obtained in this thesis open up various possibilities for future research. The extrapolation technique, for example, allows a study of ground state mesons with higher spin, and therefore an investigation of Regge trajectories in the DSE/BSE approach. Another natural extension of this work is a detailed study of excited states, which might lead to a consistent identification of spurious excitations. In addition, the numerical methods discussed here are not limited to either pure quark-antiquark systems or the rainbow-ladder truncation; they allow numerically efficient investigations of baryons, tetraquarks or even more complicated systems, as well as studies in more involved truncations. 
% Extrapolation - higher masses - higher spins - regge trajectories
% numerical methods - pole structure of quark-antiquark propagator - properties of excited states - identification of spurious excitations
% Numerical methods: not limited to mesons and RL - efficient calculation of baryons/tetraquarks/more complicated systems
% together: provide a consistent picture of the applicability of truncations/approximations - hadron phenomenology from QCD

\appendix
\chapter{Euclidean momentum space, kinematics and integration}\label{app:integration}\label{app:bsekinematic}
\section{Euclidean metric}
As originally proposed by Wick \cite{Wick:1954eu}, we consider the Bethe-Salpeter equations and consequently also the quark DSE formulated in Euclidean momentum space. In this setup, the metric is given by $g_{\mu\nu}=\delta_{\mu\nu}$ and the Clifford algebra of the Dirac $\gamma$ matrices has to be modified accordingly. For the in this case hermitian $\gamma$ matrices we use the chiral representation which reads
$$
\gamma_1 =\left(
        \begin{array}{cccc}
0& 0& 0& -\ii\\
0& 0& -\ii& 0\\
0& \ii& 0& 0\\
\ii& 0& 0& 0\\
        \end{array}
\right)\;\;
\gamma_2 =\left(
        \begin{array}{cccc}
0& 0& 0& -1\\
0& 0& 1& 0\\
0& 1& 0& 0\\
-1& 0& 0& 0\\
        \end{array}
\right)
$$
$$
\gamma_3 =\left(
        \begin{array}{cccc}
0& 0& -\ii& 0\\
0& 0& 0& \ii\\
\ii& 0& 0& 0\\
0& -\ii& 0& 0\\
        \end{array}
\right)\;\;
\gamma_4 =\left(
        \begin{array}{cccc}
0& 0& 1& 0\\
0& 0& 0& 1\\
1& 0& 0& 0\\
0& 1& 0& 0\\
        \end{array}
\right)\;.
$$
In this representation, $\gamma_5=\gamma_1\,\gamma_2\,\gamma_3\,\gamma_4$ is diagonal. Using the slash-notation, we denote the scalar product of a four-momentum $q_\mu$ with the vector $\gamma_\mu$ by
\beq
\slashed{q}\equiv\sum_{\mu=1}^4q_\mu \gamma_\mu\;.
\eeq

In this setup, the (timelike) total momentum of a quark-antiquark system $P$ satisfies $P^2<0$, and the corresponding on-shell condition for bound states reads $P^2=-M^2$, where $M\geq 0$ is the mass of the state.

\section{Kinematics and parametrization}\label{sec:parametrization}
Since we work in Euclidean space, it is convenient to use 4-dim. spherical coordinates, such that the momentum integration is written as
\beq
\frac{1}{(2\pi)^4}\int \!\! d^4k\rightarrow \frac{1}{(2\pi)^4}\int_0^\infty \!\! d(k^2) \frac{k^2}{2}\int_{-1}^1 \!\! 
dz \sqrt{1-z^2}\int_{-1}^1\!\! dy \int_0^{2 \pi} \!\! d\phi\;,
\eeq
and a general four-momentum $k_\mu$ is parametrized as
\beq\label{eq:4dspher}
k_\mu = \sqrt{k^2}\left(\sqrt{1-z^2}\sqrt{1-y^2}\sin(\phi),\sqrt{1-z^2}\sqrt{1-y^2}\cos(\phi),y\sqrt{1-z^2} ,z \right)\;.
\eeq

\begin{figure}
\begin{center}
\includegraphics[width=0.6\columnwidth,clip=true]{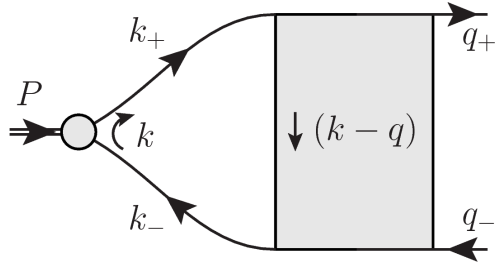}
\caption[Momentum flow of the loop-diagram present in the homogeneous and vertex BSE.]{\label{fig:bse_momentum} Momentum flow of the loop-diagram present in the homogeneous and vertex BSE (cf.~\cite{Blank:2010sn}).}
\end{center}
\end{figure}

The loop term in the homogeneous and vertex BSEs,  Eqs.~(\ref{eq:inhom_bse}) and (\ref{eq:hom_bse}), is a simple triangle diagram, and we use the momentum flow as defined in Fig.~\ref{fig:bse_momentum}. Here, the total momentum $P$ is given by the difference of the (anti-)quark momenta $k_{\pm}= k\pm\eta_{\pm}P$. 
%$\eta_\pm$ represents the momentum partitioning parameters, which satisfy $\eta_++\eta_-=1$. 
The momentum partitioning parameters, denoted by $\eta_\pm$, satisfy $\eta_++\eta_-=1$.
The relative momentum $k$ of the BSA in the loop is therefore given by $k=\eta_- k_++\eta_+ k_-$, and the relative momentum on the left hand side is denoted by $q$ and defined analogously. 

Hence, the BSEs have three relevant momenta: $P$ (total momentum), $q$ (relative momentum), 
and $k$ (loop momentum). Subsequently, using the 4-dim. spherical coordinates introduced in Eq.~\eq{4dspher}, we choose $P$ to be in the rest-frame of the system,
\beq\label{eq:rf1}
P=\left(0,0,0,\sqrt{P^2}\right)\;.
\eeq
The other momenta are chosen accordingly and read
\beq 
q = \sqrt{q^2}\left(0,0,\sqrt{1-z_q^2},z_q \right)
\eeq
and
\beq 
k = \sqrt{k^2}\left(0,\sqrt{1-z^2}\sqrt{1-y^2},y\sqrt{1-z^2} ,z \right)\;.\label{eq:rf2} 
\eeq
In this parametrization, the integration $\int d\phi$ is trivial.

The components of the amplitude on the left hand side of the BSEs Eqs.~(\ref{eq:inhom_bse}) 
and (\ref{eq:hom_bse}) depend on the scalar products $P\dotp q=z_q \sqrt{q^2} \sqrt{P^2}$, 
$P^2$, and $q^2$. Inside the integral, on the right hand side, the components depend on 
$P\dotp k=  z \sqrt{k^2}\sqrt{P^2}$, $P^2$, and $k^2$. Thus, the BSE kernel matrix $\mathcal K$ defined in Sec.~\ref{sec:bsenum}
induces the following mapping on the momentum variables
\beq 
(q^2,z_q) \mapsto (k^2,z)\;, 
\eeq
such that the integration $\int dy$ does not add a dimension to $\mathcal K$, although it is not trivial.

\begin{figure}
\centering\includegraphics[width=0.6\columnwidth]{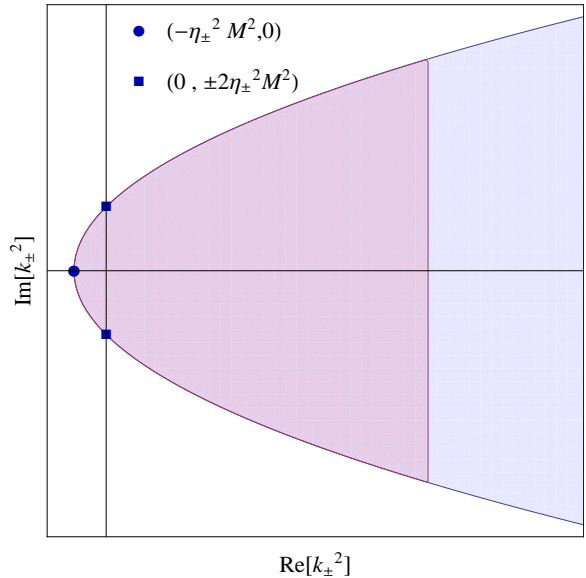} 
\caption[Sketch of the region in the complex $k_\pm^2$-plane spanned by the arguments of the propagator dressing functions.]{\label{fig:parabola} Sketch of the region in the complex $k_\pm^2$-plane spanned by the arguments of the propagator dressing functions. Purple: Bounded region used for solving the gap equation; Blue: region covered by a fit of the solutions on the real axis.}
\end{figure}

The parametrization \eq{rf1}-\eq{rf2}, together with the on-shell condition $P^2=-M^2$ leads to an imaginary total momentum
\beq
P=\left(0,0,0,\ii M\right)\;
\eeq
in the homogeneous BSE and also in the vertex BSE as soon as $P^2<0$. As a consequence the (anti)quark momenta become complex, and the arguments of the dressing functions of the quark propagator $A(k_\pm^2)$ and $B(k_\pm^2)$ read
\beq\label{eq:fullpar}
k_\pm^2=k^2-\eta_\pm^2M^2+2\ii\eta_\pm M z\sqrt{k^2}\;.
\eeq
This defines a parabolic region in the complex $k_\pm^2$-plane on which the dressing functions are needed. The boundary of the region plotted in Fig.~\ref{fig:parabola} is obtained by setting $z=\pm 1$.

As described in Sec.~\ref{sec:gap_complex}, we solve the gap equation for complex momenta by iterating the equation on the boundary of the purple region in Fig.~\ref{fig:parabola}, while the blue region is covered by a fit of the solution on the real axis, which is analytically continued to complex values. The closed contour used in that procedure is given by the union of the curves
\beq
c_a(t)=-\lambda_1+(\lambda_1+\lambda_2)\,t^2-\ii\,t\,2\sqrt{\lambda_1(\lambda_1+\lambda_2)}\quad t\in[-1,1]\;,
\eeq
and
\beq
c_b(t)=\lambda_2+\ii\,t\,2\sqrt{\lambda_1(\lambda_1+\lambda_2)}\quad t\in[-1,1]\;.
\eeq
Here, $\lambda_1=\eta_\pm^2M^2$ gives the vertex of the parabola and $\lambda_2$ defines the point in the UV where the parabola is cut. In our calculations, we use $\lambda_2=10\,\textrm{GeV}^2$.

The maximal size of the parabola (and thus the maximal bound state mass accessible) is in most cases (e.g.~the MT model, cf.~App.~\ref{app:models}) limited by the appearance of complex conjugate poles in the propagator dressing functions. Then, the contour method described in  Sec.~\ref{sec:gap_complex} is no longer applicable since it assumes the dressing functions to be holomorphic, and the integrals in the BSEs are divergent. 

\section{Quadrature rules}
After choosing a parametrization of the momentum integration, the next step is to discretize 
the momentum dependence. In this work, we straightforwardly apply the quadrature method, and replace
\beq 
\int_0^\infty \!\! d(k^2) \frac{k^2}{2}\int_{-1}^1 \!\! dz \sqrt{1-z^2}\int_{-1}^1\!\! dy \rightarrow \sum_{l=1}^{N_k} \sum_{m=1}^{N_z} \sum_{n=1}^{N_y }w[k_l^2] w[z_m] w[y_n]\;, 
\eeq 
where $w[q_l^2]$, $w[z_m]$, $w[y_n]$ denote the quadrature weights and $q_l^2$, $z_m$, $y_n$ the corresponding nodes. 
The factors of $k^2/2$ and $\sqrt{1-z^2}$ are absorbed in the weights.

However, in order to achieve good convergence of the integral $\int_{-1}^1 dz$, one has to choose a quadrature which takes the factor $\sqrt{1-z^2}$ into account explicitly. For solving integrals like
\beq\label{eq:intz}
I=\int d z \sqrt{1-z^2}f(z)\;,
\eeq
we therefore make use of a generalization of the Clenshaw-Curtis quadrature \cite{Clenshaw:1960cc} and expand the integrand $f(z)$ into Chebyshev polynomials of the first kind (denoted by $T_n(z)$), such that
\beq\label{eq:chebyexp}
f(z)=\frac{a_0}{2}+\sum_{n=1}^\infty a_n T_n(z)\;.
\eeq
The $T_n(z)$ are given by
\beq
T_n(z)=\cos(n\arccos(z))\;\;\textrm{if}\;-1\leq z\leq 1
\eeq
for integer $n\geq0$. Inserting into Eq.~\eq{intz} and defining $z=\cos(\theta)$ leads to
\begin{multline}\label{eq:intexpand}
I=\frac{a_0}{2}\int_0^\pi d\theta\,\sin^2(\theta)T_0(\cos(\theta))+\sum_{n=1}^\infty a_n\int_0^\pi d\theta\,\sin^2(\theta)T_n(\cos(\theta))\\
=\frac{a_0}{2}\int_0^\pi d\theta\,\sin^2(\theta)+\sum_{n=1}^\infty a_n\int_0^\pi d\theta\,\sin^2(\theta)\cos(n\theta)\\
=a_0\frac{\pi}{4}-a_2\frac{\pi}{4}\;.
\end{multline}
For a numerical approximation of the integral $I$, the series Eq.~\eq{chebyexp} is truncated at finite order $N$, such that
\beq
f(z)\approx f^{(N)}(z)=\sum_{n=0}^{N}{}'' a_n T_n(z)\;,
\eeq
where $\sum''$ indicates that the first and the last term in the sum have to be halved. In this case, $f^{(N)}(z)$ is a Chebyshev polynomial, and the coefficients $a_n$ are exactly given by \cite{Clenshaw:1960cc}
\beq\label{eq:an}
a_n=\frac{2}{N}\sum_{k=0}^{N}{}''f\left(\cos\left(\frac{k\pi}{N}\right)\right)\cos\left(\frac{n\pi k}{N}\right)\;.
\eeq
Upon inserting Eq.~\eq{an} into Eq.~\eq{intexpand}, we arrive at
\beq
I=\frac{\pi}{2N}\sum_{k=0}^{N}{}'' f\left(\cos\left(\frac{k\pi}{N}\right)\right)\,\left(1-\cos\left(\frac{n\pi k}{N}\right)\right)\;.
\eeq
The above formula gives a quadrature rule $\sum_{k=1}^{N-1}w[z_k]f(z_k)$ with 
\beq
z_k=\cos\left(\frac{k\pi}{N}\right)
\eeq
and
\beq 
w[z_k]=\left(1-\cos\left(\frac{n\pi k}{N}\right)\right)\;.
\eeq
This quadrature has two main advantages: first, its convergence properties are good since it takes the weight function $\sqrt{1-z^2}$ into account explicitly; second, it is ``nested'', which means that it allows, without additional numerical effort, to estimate the errors by comparing the result for a given number of quadrature nodes with the corresponding result for half of the number of quadrature nodes.

\chapter{Models and parameters}\label{app:models}
As mentioned in Sec.~\ref{sec:truncation}, the truncation of the DSE/BSE system that is used for most of our calculations is the rainbow-ladder truncation. The relevant (``interacting'') part of the corresponding effective action is depicted in Fig.~\ref{fig:gamma2}, and leads to the substitution
\beq\label{eq:modelrl}
\Gamma_\mu(p,q)D_{\mu\nu}(p-q) \rightarrow \gamma_\mu  \frac{\mathcal{G}(k^2)}{k^2}\left( \delta_{\mu\nu}-\frac{k_\mu k_\nu}{k^2} \right)\;,
\eeq
where $k=p-q$, and $\Gamma_\mu(p,q)$ and $D_{\mu\nu}(p-q)$ are the full quark-gluon vertex and gluon propagator, respectively. 
Thus, in the quark DSE as well as the BSEs, the product of the full quark-gluon vertex and the gluon propagator is replaced by the corresponding bare quantities, which are then multiplied by the so-called effective interaction $\mathcal G$. 

In the present work we investigate and compare several different forms of this effective interaction. At zero temperature, for the studies of quark-antiquark systems presented in Chap.~\ref{chap:mesons}, we use the interaction defined by Maris and Tandy (referred to as MT, \cite{Maris:1999nt}) in addition to the scaling- and decoupling-type inputs discussed in Sec.~\ref{sec:bseir} (cf.~\cite{Blank:2010pa,Fischer:2008uz}). In the MT-model, the effective interaction is given by
\beq\label{eq:mt}
\frac{\mathcal{G}(k)}{k^2}=D\; \frac{4 \pi^2}{\omega^6} k^2 e^{-k^2/\omega^2} + \mathcal F_{UV}(k^2)\;,
\eeq
where
\beq\label{eq:fuv}
\mathcal F_{UV}(k^2)\equiv\frac{4\pi\;\gamma_m \pi\;\mathcal{F}(k^2) }{1/2 \ln [\tau\!+\!(1\!+\!k^2/\Lambda_\mathrm{QCD}^2)^2]}\;.
\eeq
As given in \cite{Maris:1999nt}, 
$${\cal F}(k^2)= [1 - \exp(-k^2/[4 m_t^2])]/k^2\;.$$
The additional parameters are $m_t=0.5$~GeV, $\tau={\rm e}^2-1$, $N_f=4$, $\Lambda_\mathrm{QCD}^{N_f=4}= 0.234\,{\rm GeV}$, and $\gamma_m=12/(33-2N_f)$.

For our investigation of the chiral transition temperature (which is presented in Ref.~\cite{Blank:2010bz}), we compare the MT interaction to several Ans\"atze, which were proposed by Alkofer, Watson, and Weigel \cite{Alkofer:2002bp}, Maris and Roberts \cite{Maris:1997tm}, and Munczek and Nemirovsky \cite{Munczek:1983dx}, denoted by AWW, MR, and MN, respectively. At finite temperature, they are given by
\begin{eqnarray}\nonumber
\mathcal{G}(s) &=&\\ 
\label{eq:aww}\textrm{AWW}:&=&D\; \frac{4 \pi^2}{\omega^6} s e^{-s/\omega^2}\\ 
\label{eq:mr}\textrm{MR}:&=&D\; \left( \frac{4 \pi^3}{T} \delta^3(\vec{k}) \delta_{k-l,0} +\frac{4 \pi^2}{\omega^6} s e^{-s/\omega^2}\right) + \mathcal F_{UV}(s),\\ 
\label{eq:mn}\textrm{MN}:&=&D\;\frac{4 \pi^3}{T} \delta^3(\vec{k}) \delta_{k-l,0}\;,
\end{eqnarray}
where $s:= \vec{k}\,^2 + \Omega^2 + m_g^2$, and the bosonic Matsubara frequency $\Omega$ of the gluons is given by the difference of the fermionic ones of the quarks, $\Omega=\omega_k-\omega_l$, as mentioned in Sec.~\ref{sec:fintmodel}. The continuation of the MT model to finite temperature is straightforwardly implemented by replacing $k^2\rightarrow s$. For the MR model, the function $\mathcal F_{UV}(k^2)$, given in Eq.~\eq{fuv}, is the same as in the case of MT.

\begin{table}
\caption{\label{tab:models}Parameter sets $D$ and $\omega$ used for the different interactions
MT \cite{Maris:1999nt}, Eq.~(\ref{eq:mt});
AWW \cite{Alkofer:2002bp}, Eq.~(\ref{eq:aww});
MR \cite{Maris:1997tm}, Eq.~(\ref{eq:mr});
MN \cite{Munczek:1983dx}, Eq.~(\ref{eq:mn});
together with corresponding observables in $T=0$ meson studies, where available.
Numbers are in GeV except for $D$ (GeV${}^2$) and the chiral condensate,
whose dimension is given explicitly. For the MT model we use the quark masses of Ref.~\cite{Krassnigg:2009zh}.}
\centering\begin{tabular}{l|c c c c c}
Model&$\omega$&$D$&$-\langle \bar{q}q\rangle_0$&$m_\pi$&$m_\varrho$\\
\hline
MT1&0.3&1.24&  $(.243\, \mbox{GeV})^3$&0.230&0.780\\
MT2&0.4&0.93&  $(.242\, \mbox{GeV})^3$&0.230&0.774\\
MT3&0.5&0.744& $(.239\, \mbox{GeV})^3$&0.232&0.762\\ 
AWW1&0.3&1.47 &$(.245\, \mbox{GeV})^3$&0.135&0.745\\
AWW2&0.4&1.152&$(.246\, \mbox{GeV})^3$&0.135&0.748\\
AWW3&0.5&1.0&  $(.251\, \mbox{GeV})^3$&0.137&0.758\\
MR1&0.3&0.78&  $(.241\, \mbox{GeV})^3$&0.139&\\
MR2&0.4&0.78&  $(.250\, \mbox{GeV})^3$&0.139&\\
MR3&0.5&0.78&  $(.255\, \mbox{GeV})^3$&0.139&\\
MN&&0.5618&    $(.115\, \mbox{GeV})^3$&0.14 &0.77
\end{tabular}
\end{table}

At vanishing temperature, all of these Ans\"atze provide the correct amount of dynamical chiral symmetry breaking as well as quark confinement via the absence of a Lehmann representation for the dressed quark propagator. In addition, MT and MR produce the correct perturbative limit of the QCD running coupling, i.e.~they preserve the one-loop renormalization-group behavior of QCD for solutions of the quark DSE. Note that only for MT and MR the integrals occurring in the gap equation need to be renormalized (cf.~Sec.~\ref{sec:renormalization}). 

All of the effective interactions introduced above contain the parameter $D$, which describes the strength of the interaction. The other parameter $\omega$ represents an effective inverse range of the interaction, which is not present in the MN model. For each model, the parameter sets used are given in Tab.~\ref{tab:models}.

In order to study mesons, the value of the current quark mass has to be fixed in addition to the parameters $D$ and $\omega$. For the calculations presented in Chap.~\ref{chap:mesons} using the MT model, we use the values of Ref.~\cite{Krassnigg:2009zh} which translate, at our renormalization scale of $\mu^2=19\,\textrm{GeV}^2$, to $m_{u/d}=0.01\textrm{GeV}$, $m_s=0.0685\textrm{GeV}$, $m_c=0.828\textrm{GeV}$, and $m_b=3.71\textrm{GeV}$ for light (up/down), strange, charm, and bottom quarks, respectively.

\chapter{Dirac structure of BS amplitudes}\label{app:basis}
The basic ingredients for the construction of the covariant structure of the Bethe-Salpeter amplitudes are the four-vectors occurring in the equation, namely $\gamma_\mu$, $q_\mu$, $P_\mu$. Out of those vectors, the four covariants that constitute the most general scalar amplitude read
\beq\label{eq:tsc}
\mathbf{1}\;,\;\;\slashed{P}\;,\;\;\slashed{q}\;,\;\;\ii/2[\slashed{q},\slashed{P}]\:.
\eeq
For computational convenience, we would like to construct covariants which satisfy the orthogonality relation
\beq\label{eq:orthonormality}
\Tr[T_i \cdot T_j]=0\;\;\textrm{for}\;i\neq j
\eeq
where the product implies a summation over all Lorentz indices. To achieve this, we apply the Gram-Schmidt procedure to the covariants (\ref{eq:tsc}), and obtain
\beq\label{eq:tscN}
T_1=\mathbf{1}\;,\;\;T_2=\slashed{P}\;,\;\;T_3=\slashed{q}-\slashed{P}\frac{P\cdot q}{P^2}\;,\;\;T_4=\ii/2[\slashed{q},\slashed{P}]\:.
\eeq
Note that these covariants describe a scalar quark-antiquark system, i.e., a state of positive parity. In order to obtain a basis for a pseudoscalar state, the above covariants have to be multiplied by $\gamma_5$, resulting in
\beq\label{eq:tpsN}
T_1=\frac{\gamma_5}{2}\;,\;\;T_2=\frac{\gamma_5 (\slashed{P})}{2 \sqrt{-P^2}}\;,\;\;T_3=\frac{\gamma_5 ( \slashed{q}- \frac{\slashed{P} (P\cdot q)}{P^2})}{ 2 \sqrt{\frac{(P\cdot q)^2}{P^2}-q^2}}\;,\;\;T_4=\frac{\frac{1}{2} \ii \gamma_5(\slashed{q} \slashed{P}  - \slashed{P} \slashed{q})}{2 \sqrt{P^2 q^2-(P\cdot q)^2}}\:,
\eeq
where all $T_i$ have been normalized to satisfy Eq.~\eq{orthonormality}. The structure of the covariants, however, does not restrict the C-parity of the resulting state, as discussed in detail in App.~\ref{app:cparity}.

To consider mesons of higher spin, one has to construct Lorentz-tensors of the appropriate rank which are symmetric, transverse in all open indices and Lorentz-traceless \cite{Corson:1982wi,Dai:1993qr}. To add open indices in a way which automatically satisfies Eq.~(\ref{eq:orthonormality}), we define
\begin{eqnarray}
q_\mu^T&:=&q_\mu-P_\mu \frac{P\cdot q}{P^2}\;,\\ 
\gamma_\mu^{TT}&:=&\gamma_\mu-P_\mu \frac{\slashed{P}}{P^2} - q_\mu^T \frac{\slashed{q^T}}{(q^T)^2}\;,
\end{eqnarray}
such that the vectors $\{P_\mu\;,\;\;q_\mu^T\;,\;\;\gamma_\mu^{TT}\}$ are orthogonal to each other. Constructing orthogonal (in the sense of Eq.~(\ref{eq:orthonormality})) covariants for a vector meson is now straightforward: The scalar covariants are multiplied by $q_\mu^T$ to give the first four, and by $\gamma_\mu^{TT}$ to give the second four covariants. By construction, they also satisfy the transversality condition, since $q_\mu^T$ as well as $\gamma_\mu^{TT}$ are orthogonal to $P_\mu$. Note again that a basis of the axialvector states is given via a multiplication by $\gamma_5$.

To proceed to the tensor meson (quantum numbers $J^P=2^+$), we first need all symmetric combinations of $q_\mu^T$ and $\gamma_\mu^{TT}$ of rank two, namely \cite{Krassnigg:2010mh}
\begin{eqnarray}
M_{\mu\nu}&=&\gamma_\mu^{TT}q_\nu^T + q_\mu^T\gamma_\nu^{TT}\;\;\textrm{and}\\
N_{\mu\nu}&=&q_\mu^T q_\nu^T\;.
\end{eqnarray}
These quantities are by construction transverse in all indices as well as orthogonal. To implement the traceless-condition, they have to be orthogonalized with respect to the transversely projected metric tensor, $g^T_{\mu\nu}=\delta_{\mu\nu}-\frac{P_\mu P_\nu}{P^2}$. Finally, we arrive at
\begin{eqnarray}
M_{\mu\nu}-g^T_{\mu\nu} \frac{M_{\rho\sigma}g^T_{\rho\sigma}}{(g^T)^2}\\
N_{\mu\nu}-g^T_{\mu\nu} \frac{N_{\rho\sigma}g^T_{\rho\sigma}}{(g^T)^2}\;,
\end{eqnarray}
which - by multiplication with the four scalar covariants - give a full orthogonal set of covariants for a meson of spin 2  and positive parity.

Note that in the numerical implementation all covariants are normalized according to $\hat{T}_i=T_i/\sqrt{\Tr[T_i \cdot T_i]}$.

\chapter{C-parity and symmetry}\label{app:cparity}
If the quark and antiquark that constitute a meson have the same flavor, then the state has the quantum number of charge-conjugation parity (C-parity) in addition to spin and parity. In mathematical terms, the BSAs $\Gamma(p,P_i)$ are eigenstates of the operation of charge-conjugation, defined via
\beq\label{eq:cconjg}
\bar \Gamma(q,P_i)\equiv[C\: \Gamma(-q,P_i)\: C^{-1}]^t=\eta_c \Gamma(q,P_i)\;,
\eeq
where $C=\gamma_2\gamma_4$ is the charge-conjugation matrix and the superscript $t$ denotes the matrix transpose. The eigenvalue $\eta_c=\pm 1$ gives the C-parity of the state. 

As mentioned in Sec.~\ref{sec:bsenum}, the BSA can be decomposed into covariants and components (Eq.~\eq{decomp}),
$$\Gamma(q,P_i) = \sum_{n=1}^N T_n(q,P_i) \;F^n(P_i^2,q^2,q\dotp P_i)\;.$$
If the covariants are constructed such that they have a definite C-parity $\zeta_n$, 
\beq\label{eq:ticpar}
\equiv[C\: T_n(-q,P_i)\: C^{-1}]^t=\zeta_n T_n(q,P_i)\;,
\eeq
the C-parity is manifest in the symmetry of the components  with respect to $q\dotp P_i$. To see this, we first insert the decomposition \eq{decomp} and Eq.~\eq{ticpar} into Eq.~\eq{cconjg},
\begin{multline}
\bar \Gamma(q,P_i)=\sum_{n=1}^N F^n(P_i^2,q^2,-q\dotp P_i)[C\: T_n(-q,P_i)\: C^{-1}]^t\\
=\sum_{n=1}^N F^n(P_i^2,q^2,-q\dotp P_i) \zeta_n T_n(q,P_i)\;.
\end{multline}
For the (anti)symmetry of the components with respect to $q\dotp P_i$ we introduce the notation 
\beq
F^n(P_i^2,q^2,-q\dotp P_i)=\tilde{\zeta}_nF^n(P_i^2,q^2,q\dotp P_i)\;,
\eeq
with $\tilde{\zeta}_n=\pm 1$, and arrive at
\beq
\eta_c \sum_{n=1}^N T_n(q,P_i) \;F^n(P_i^2,q^2,q\dotp P_i)=\sum_{n=1}^N (\zeta_n \tilde{\zeta}_n) T_n(q,P_i) \;F^n(P_i^2,q^2,q\dotp P_i)\;.
\eeq
Thus, the C-parity $\eta_c$ is given by $\zeta_n \tilde{\zeta}_n$, which is the same for all $n$ if the state under consideration is indeed an eigenstate of charge conjugation. Therefore, in order to produce a state with $\eta_c=+1$, the product $\zeta_n \tilde{\zeta}_n$ has to be always positive, and if one covariant has negative C-parity, $\zeta_n=-1$, the corresponding component must be antisymmetric with respect to $q\dotp P_i$.

If the system is parametrized in its rest frame according to Eqs.~\eq{rf1}-\eq{rf2}, the scalar product $q\dotp P_i$ is given by
\beq
q\dotp P_i=q z_q \sqrt{P_i^2}=\ii\:q z_q M_i\;,
\eeq
and the (anti)symmetry of the components in $q\dotp P_i$ is mapped onto an (anti)symmetry with respect to the angular variable $z_q$. By additionally choosing the momentum partitioning $\eta_\pm=0.5$ (cf.~Sec.~\ref{sec:parametrization}), the components become either real or purely imaginary (see, e.g.~\cite{Smith:1969az}).

These properties allow to restrict the homogeneous BSE in such a way that only states of one C-parity appear as solutions. In most cases, this is done by expanding the components into Chebyshev polynomials of even or odd order depending on the desired symmetry properties (see, e.g., \cite{Maris:1997tm,Maris:1999nt,Krassnigg:2009zh}).

In the case of the vertex BSE, the inhomogeneous BSA has the same symmetry properties as the homogeneous, with the addition that only states which have overlap with the inhomogeneous term $\Gamma_0$ can contribute, as exemplified in Sec.~\ref{sec:inhomremovepole}. Thus, if the momentum partitioning $\eta_\pm=0.5$, only poles corresponding to homogeneous amplitudes which have the same $z$-symmetry as $\Gamma_0$ appear in the inhomogeneous BSA.

\chapter{Dependence of meson masses on $\omega$}\label{app:plots}
\begin{figure}
\vspace{-1.5cm}
\includegraphics[width=0.9\columnwidth]{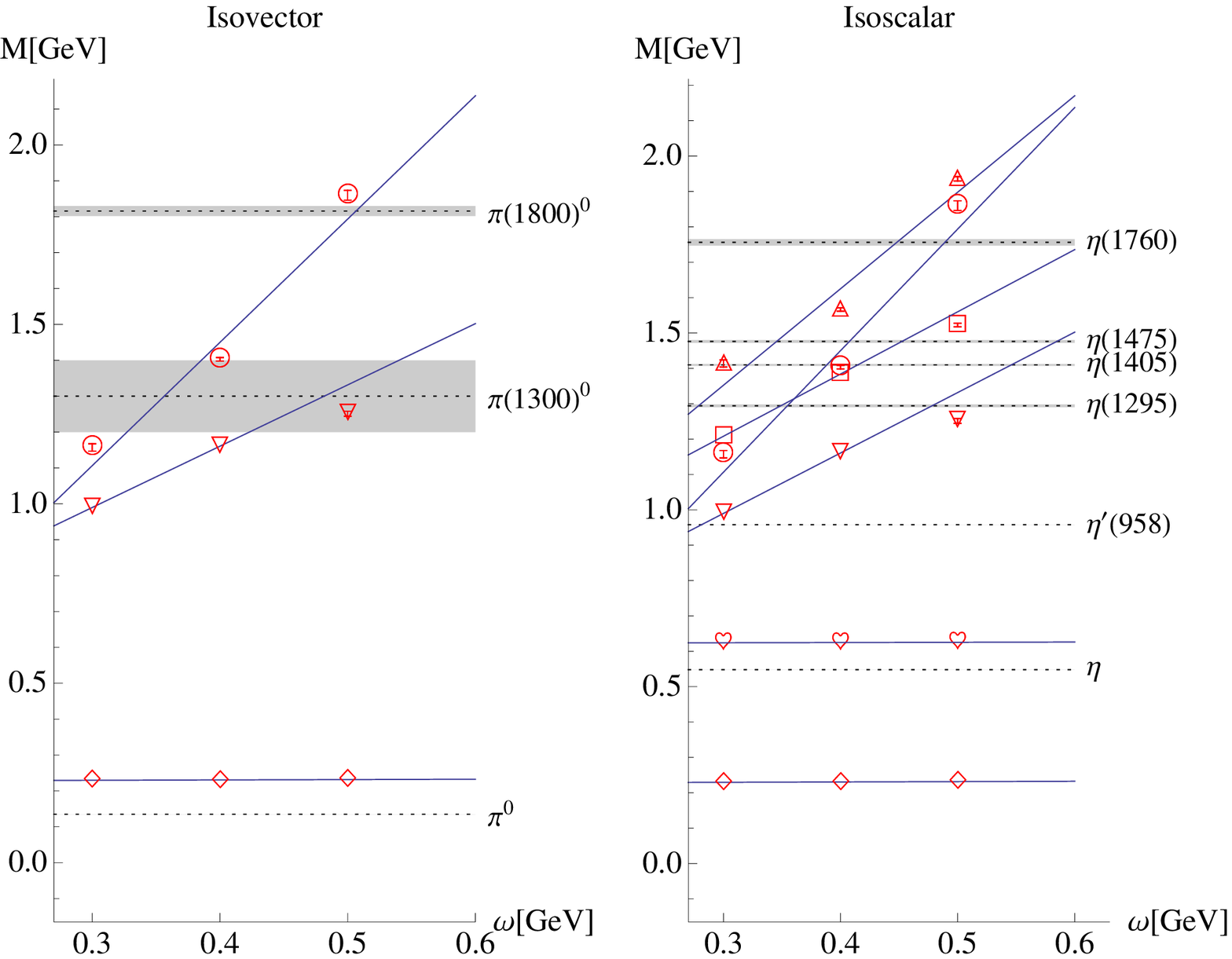}\\[-2cm] 
\includegraphics[width=0.9\columnwidth]{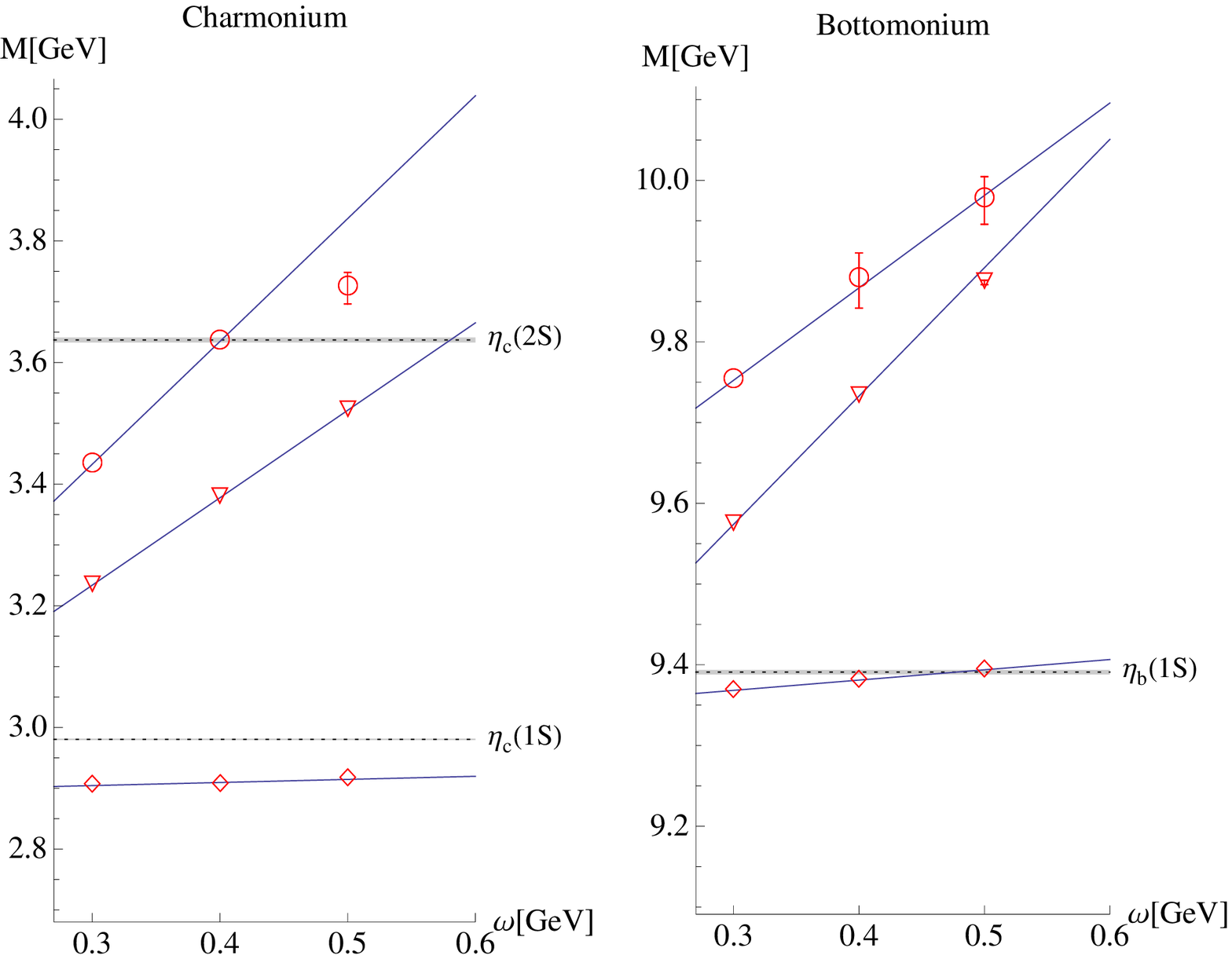}\vspace{-0.5cm}
\caption[$\omega$ dependence of the ground- and excited-state masses for $J^{PC}=0^{-+}$.]{$J^{PC}=0^{-+}$, see text.\label{fig:omegaPS}}
\end{figure}

\begin{figure}
\vspace{-1.5cm}
\includegraphics[width=0.9\columnwidth]{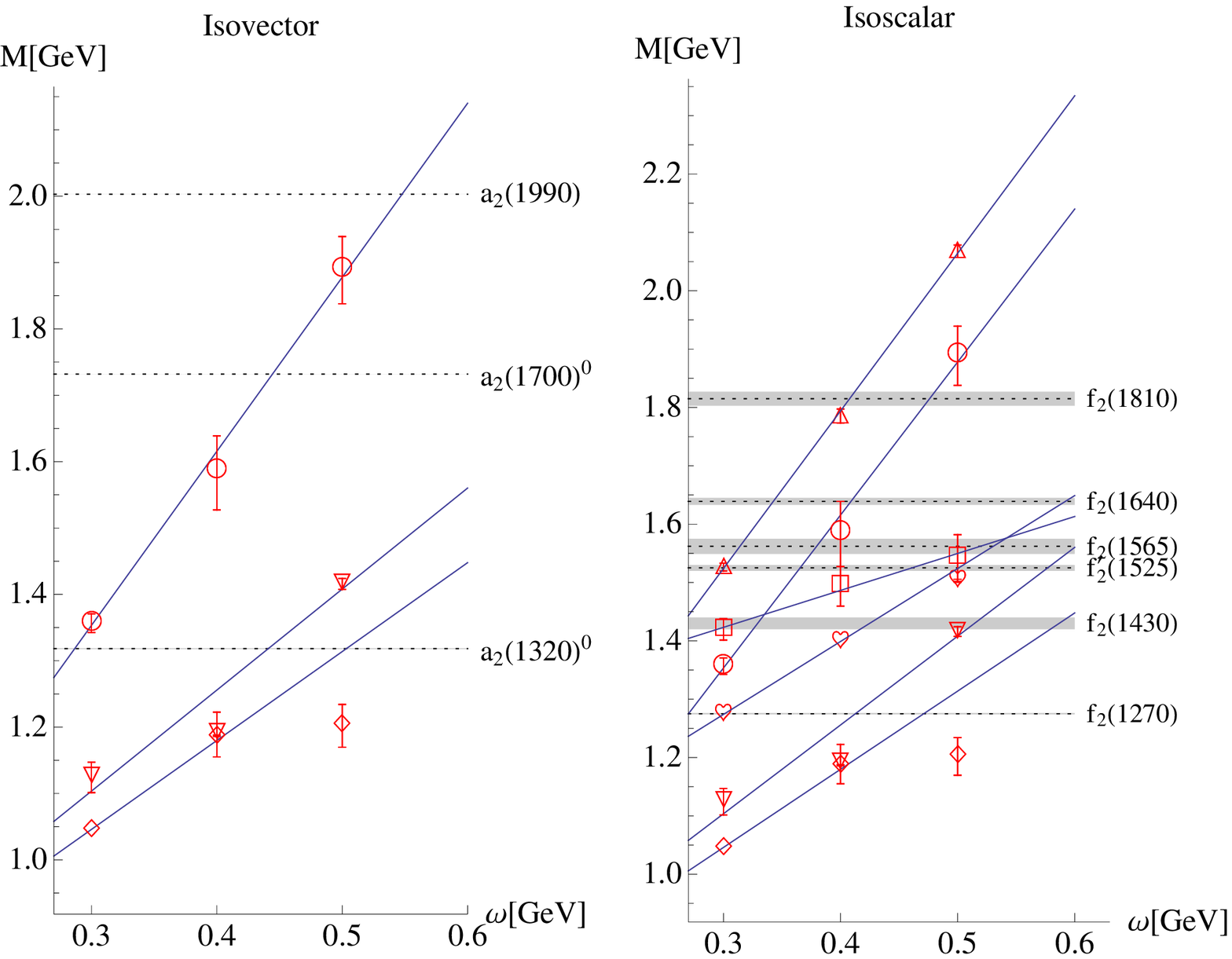} \\[-2cm] 
\includegraphics[width=0.9\columnwidth]{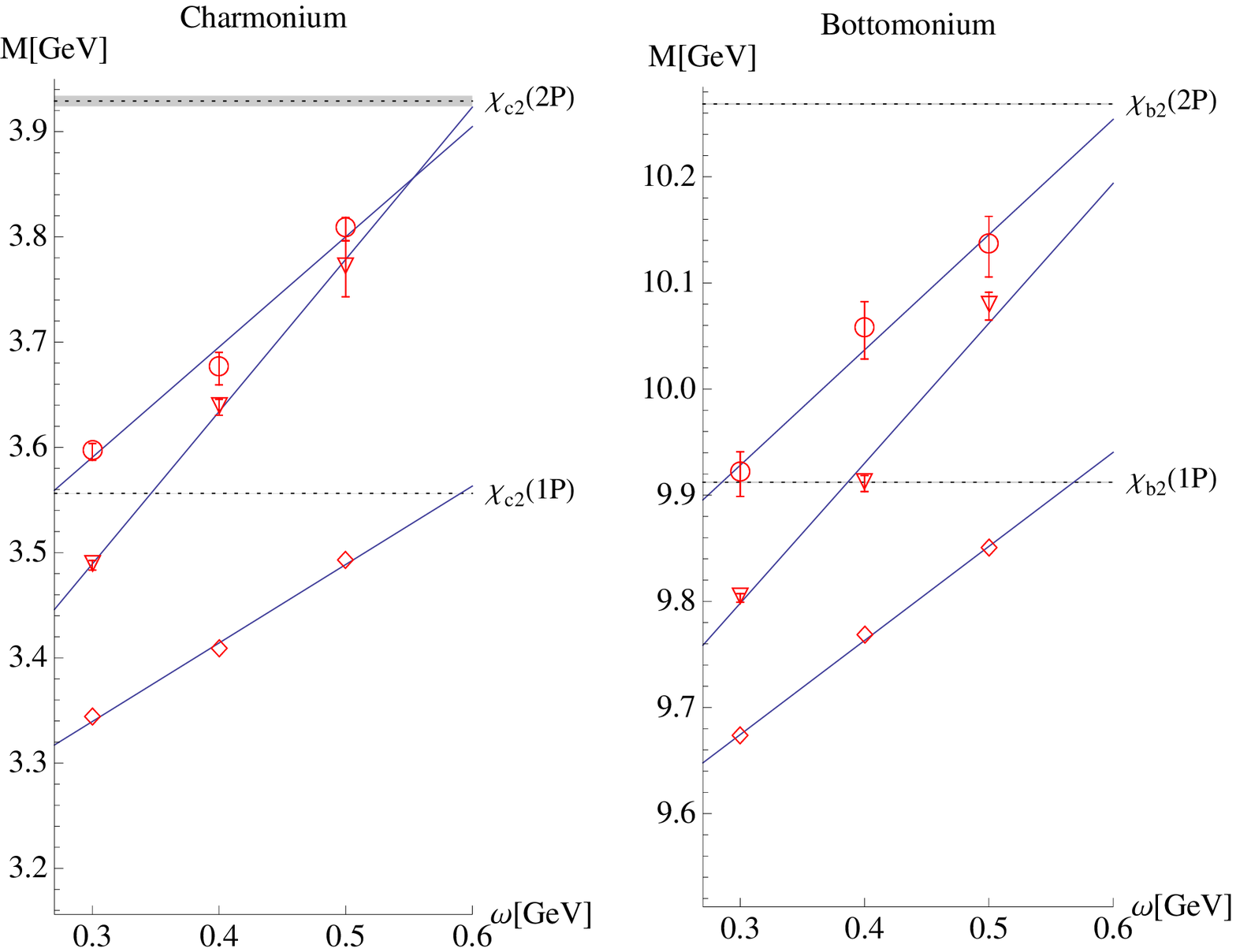}\vspace{-1cm}
\caption[$\omega$ dependence of the ground- and excited-state masses for $J^{PC}=2^{++}$.]{$J^{PC}=2^{++}$, see text.The mass of $f_2(1430)$ (isoscalar) is given in \cite{Nakamura:2010zzi} as $M\approx1430\,$MeV, and no experimental error is reported. This is indicated by omitting the dotted line.\label{fig:omegaTE}}
\end{figure}

In Figs.~\ref{fig:omegaPS} - \ref{fig:omegaTE}, we show the dependence of all states investigated in Sec.~\ref{sec:spectroscopy} on the MT parameter $\omega$, defined via Eq.~\eq{mt}, in comparison to the experimental values of the masses. The experimental data is represented by black dotted lines, the corresponding errors are indicated by the grey bands. The symbols without errorbars are obtained by direct calculations, i.e.~the masses are within the numerically accessible region, while the symbols with errorbars are the results of extrapolations of the eigenvalue curves, as described in Sec.~\ref{sec:extrapolate}. 

The solid lines are linear fits in $\omega$ to the data points, weighted by the appropriate error. In the case of the direct caclulation, the error is determined by the numerical precision of the eigenvalue, and estimated by $10^{-7}$. We consider a total of three states (ground state and two excitations) for each $J^{PC}$ for all quark masses listed in App.~\ref{app:models}, except for $J^{PC}=1^{--}$ where three excitations are calculated.

In the isoscalar case, six states are shown (eight for $J^{PC}=1^{--}$), which correspond to the pure s\={s} states (symbols $\heartsuit$, $\square$, $\vartriangle$, $\triangleleft$) and the pure n\={n} states (symbols $\diamond$, $\triangledown$, $\circ$, $\triangleright$), since we do not consider mixing.

\begin{figure}
\vspace{-1.5cm}
\includegraphics[width=0.9\columnwidth]{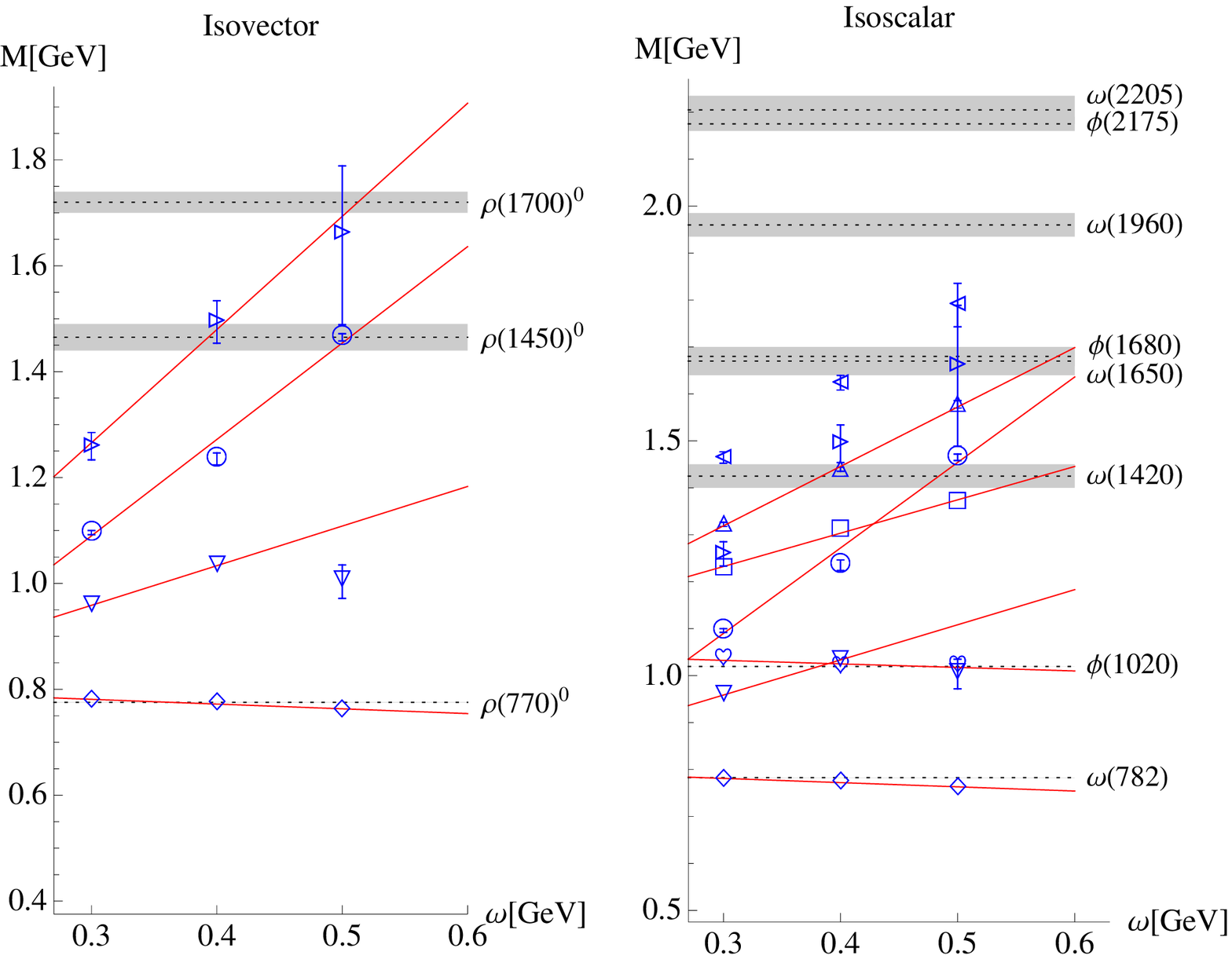} \\[-2cm] 
\includegraphics[width=0.9\columnwidth]{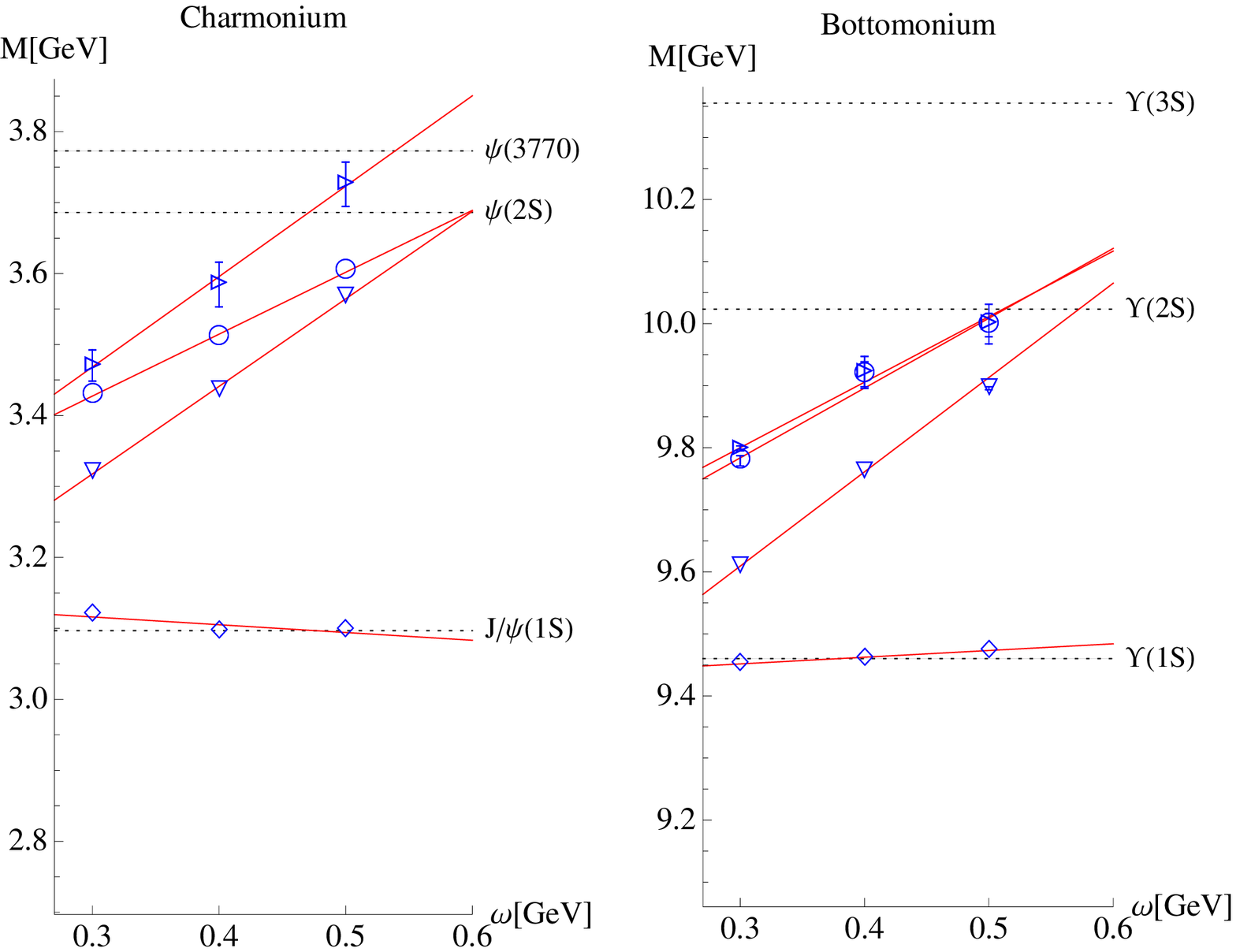}\vspace{-0.5cm}
\caption[$\omega$ dependence of the ground- and excited-state masses for $J^{PC}=1^{--}$.]{$J^{PC}=1^{--}$, see text.}
\end{figure}
\begin{figure}
\vspace{-1.5cm}
\includegraphics[width=0.9\columnwidth]{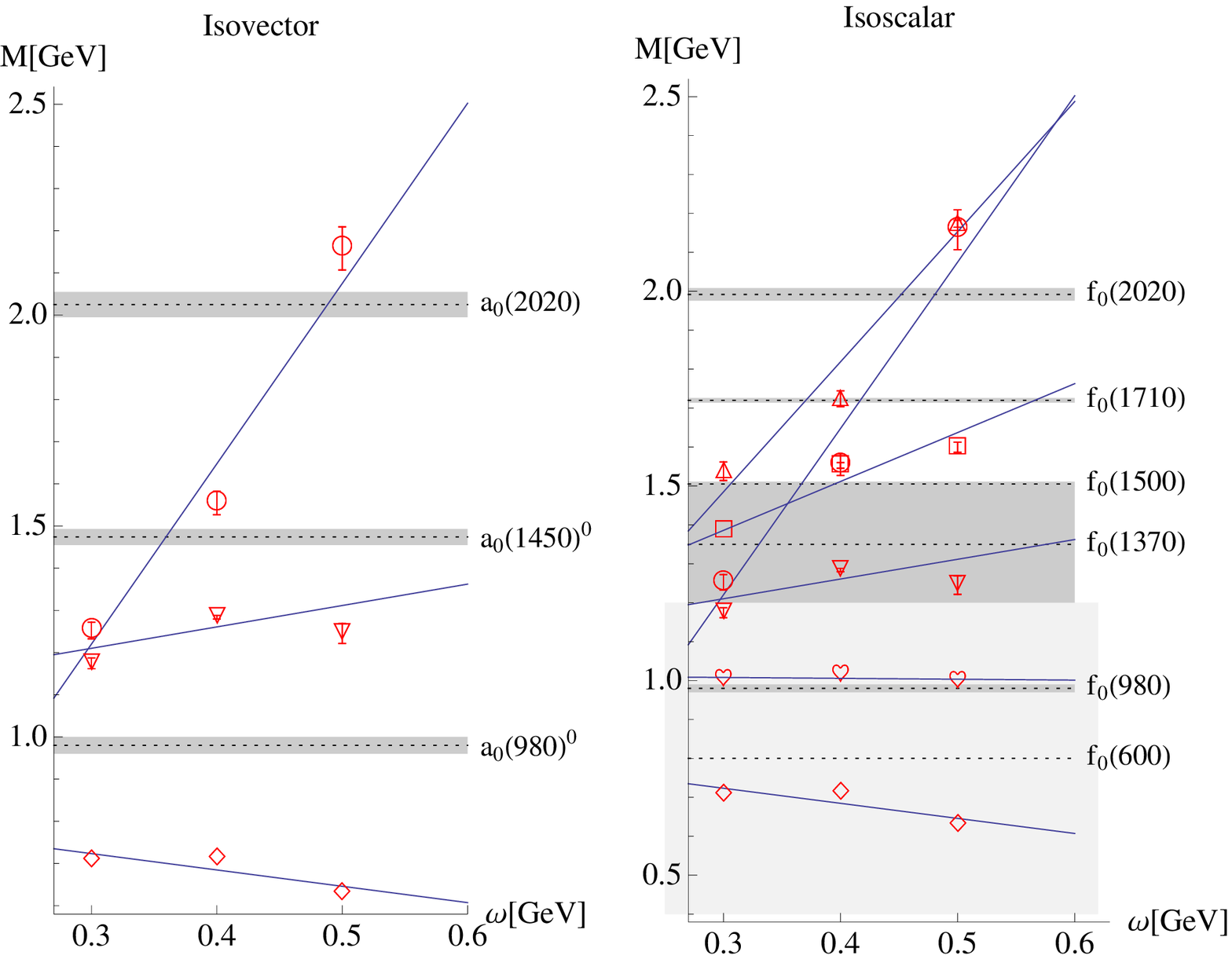} \\[-2cm] 
\includegraphics[width=0.9\columnwidth]{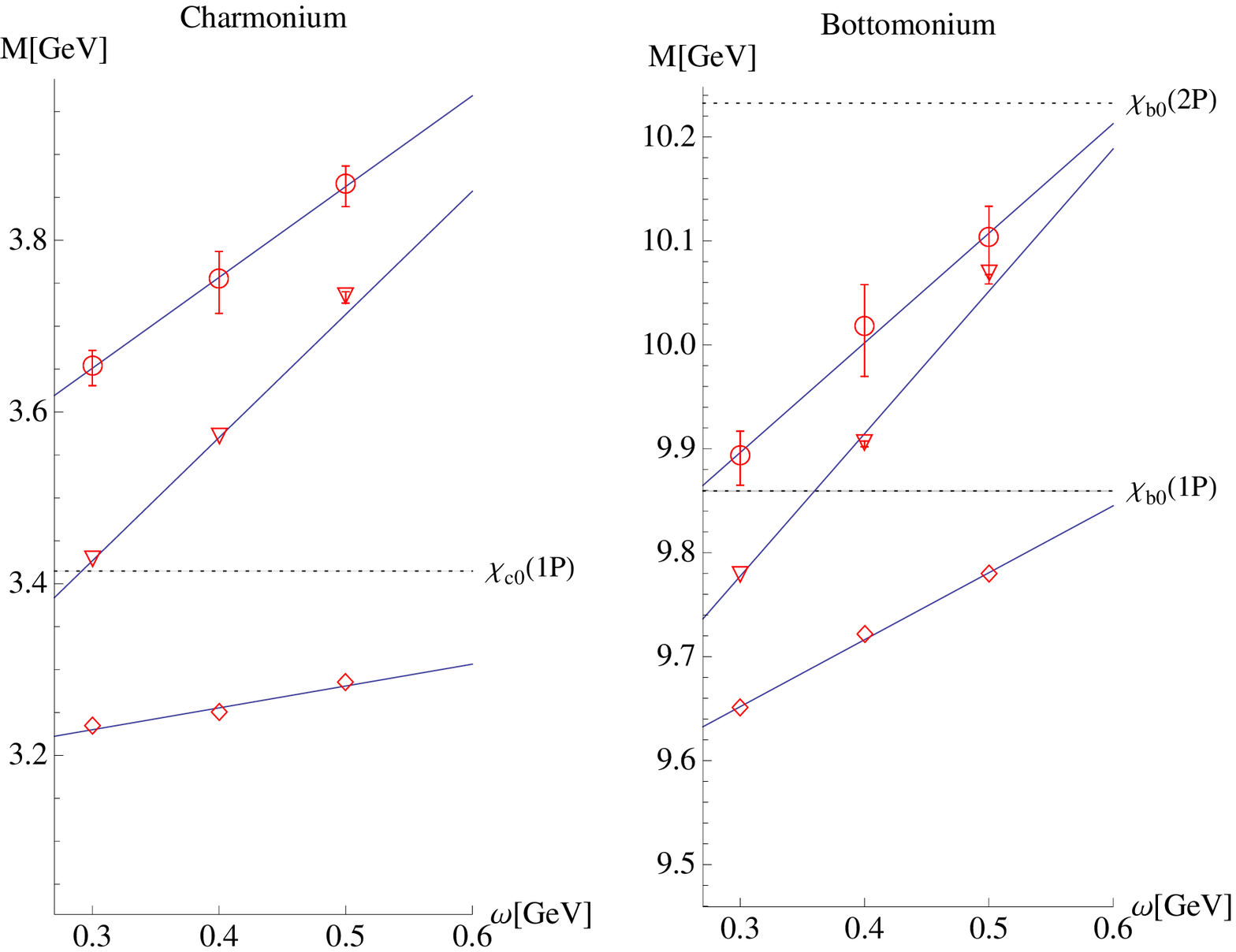}\vspace{-1cm}
\caption[$\omega$ dependence of the ground- and excited-state masses for $J^{PC}=0^{++}$.]{$J^{PC}=0^{++}$, see text. In the isoscalar channel, the experimental error of $f_0(600)$ is shaded in lighter gray, since it overlaps with the error of $f_0(980)$.}
\end{figure}
\begin{figure}
\vspace{-1.5cm}
\includegraphics[width=0.9\columnwidth]{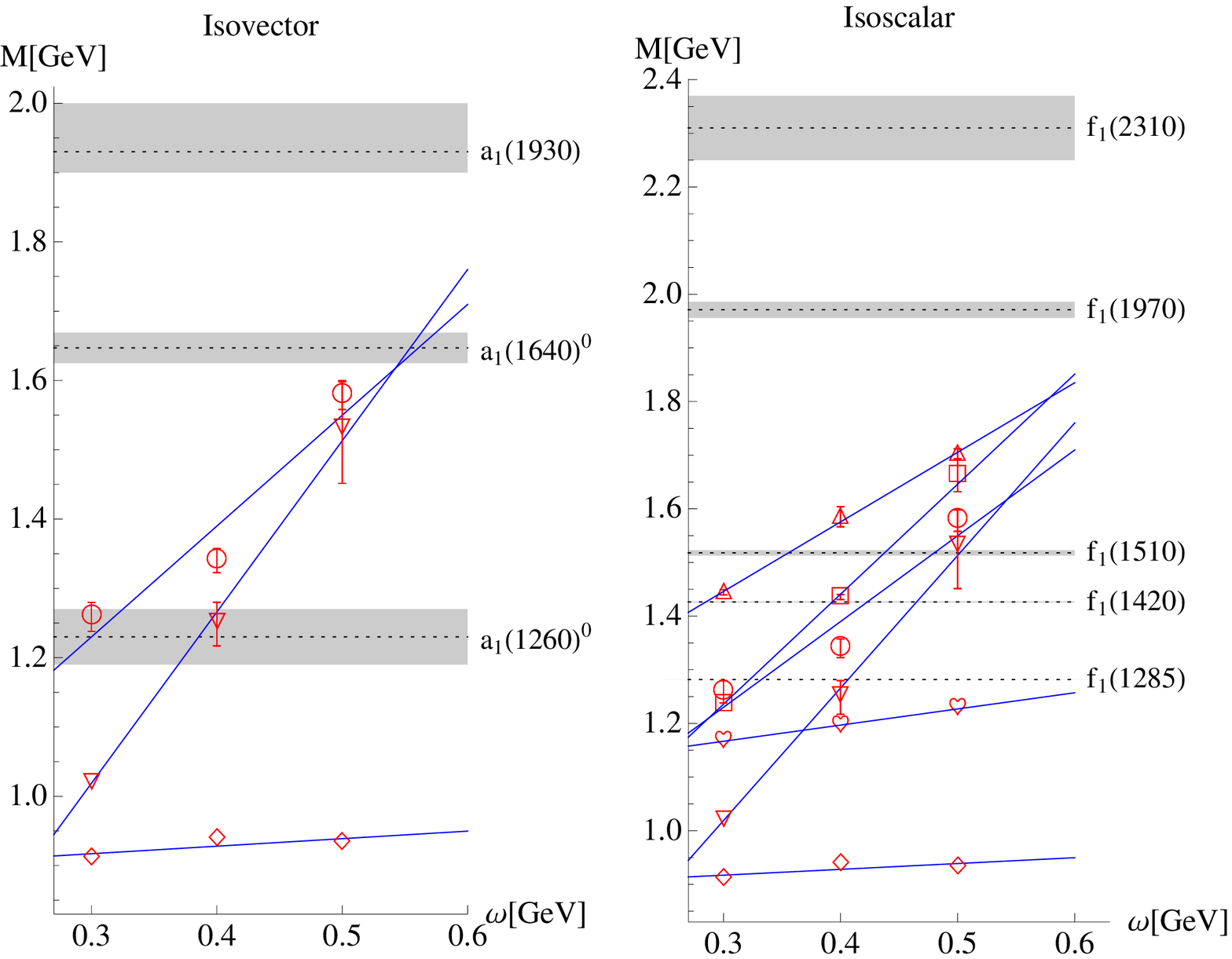} \\[-2cm] 
\includegraphics[width=0.9\columnwidth]{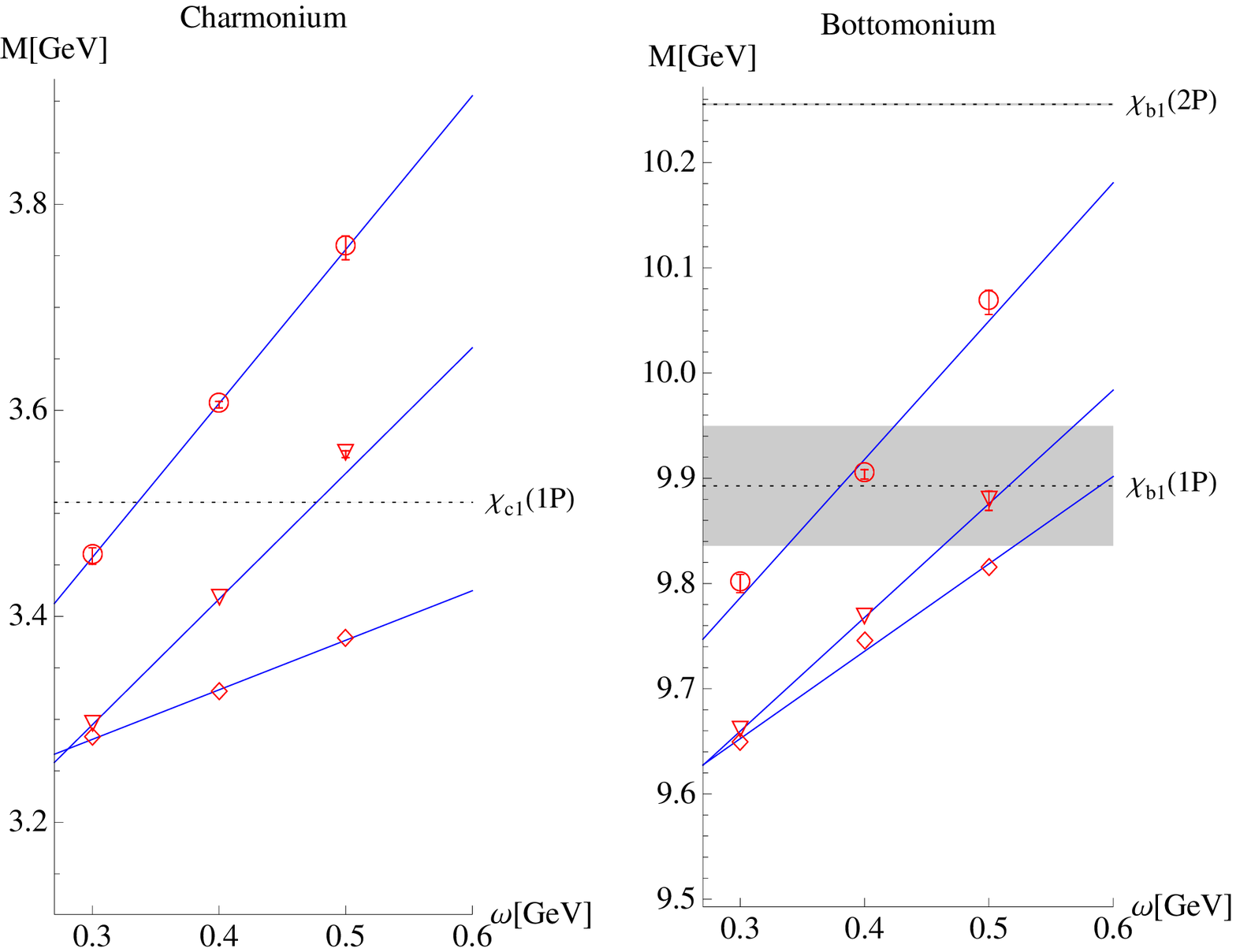}\vspace{-0.5cm}
\caption[$\omega$ dependence of the ground- and excited-state masses for $J^{PC}=1^{++}$.]{$J^{PC}=1^{++}$, see text.}
\end{figure}
\begin{figure}
\vspace{-1.5cm}
\includegraphics[width=0.9\columnwidth]{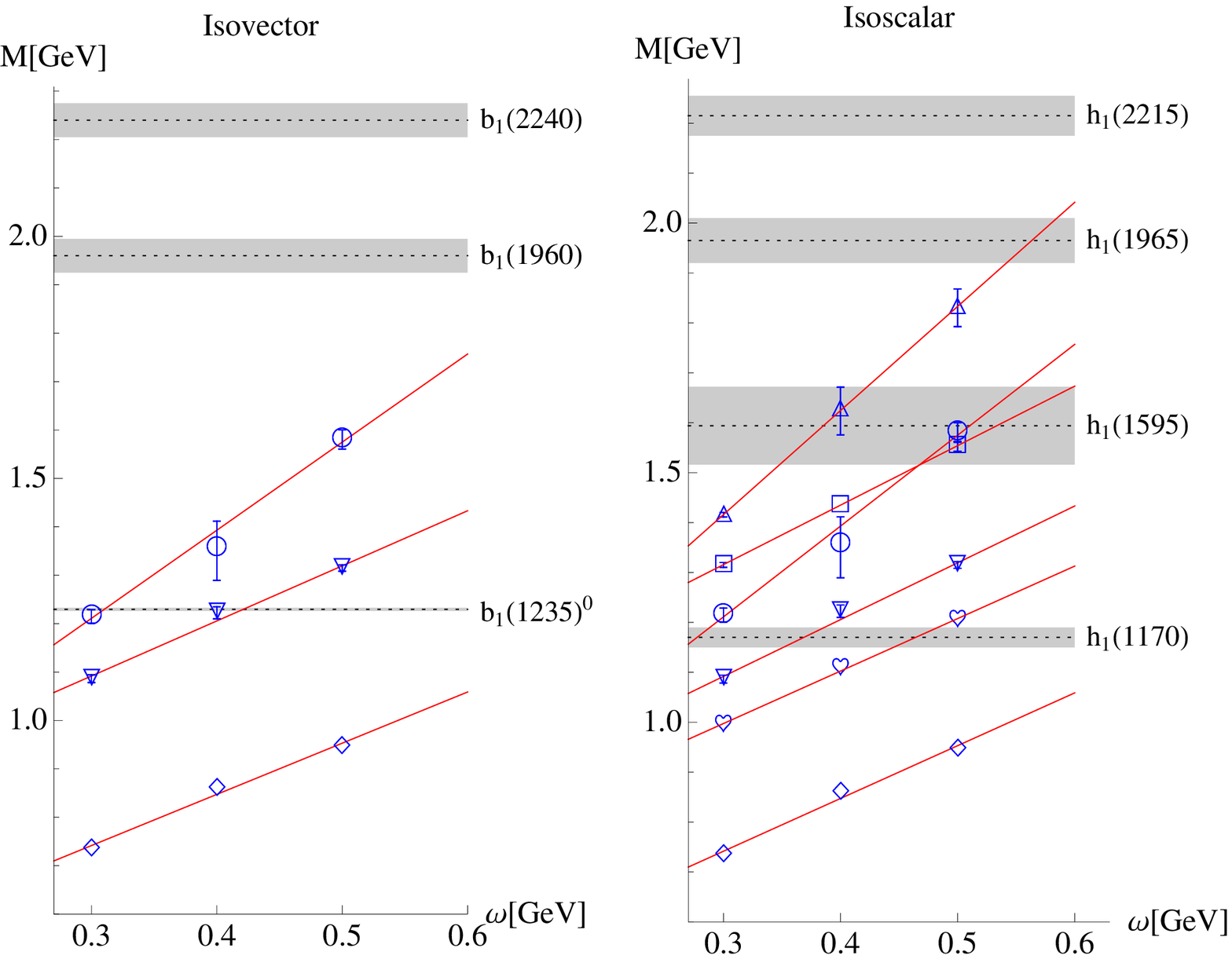} \\[-2cm] 
\includegraphics[width=0.9\columnwidth]{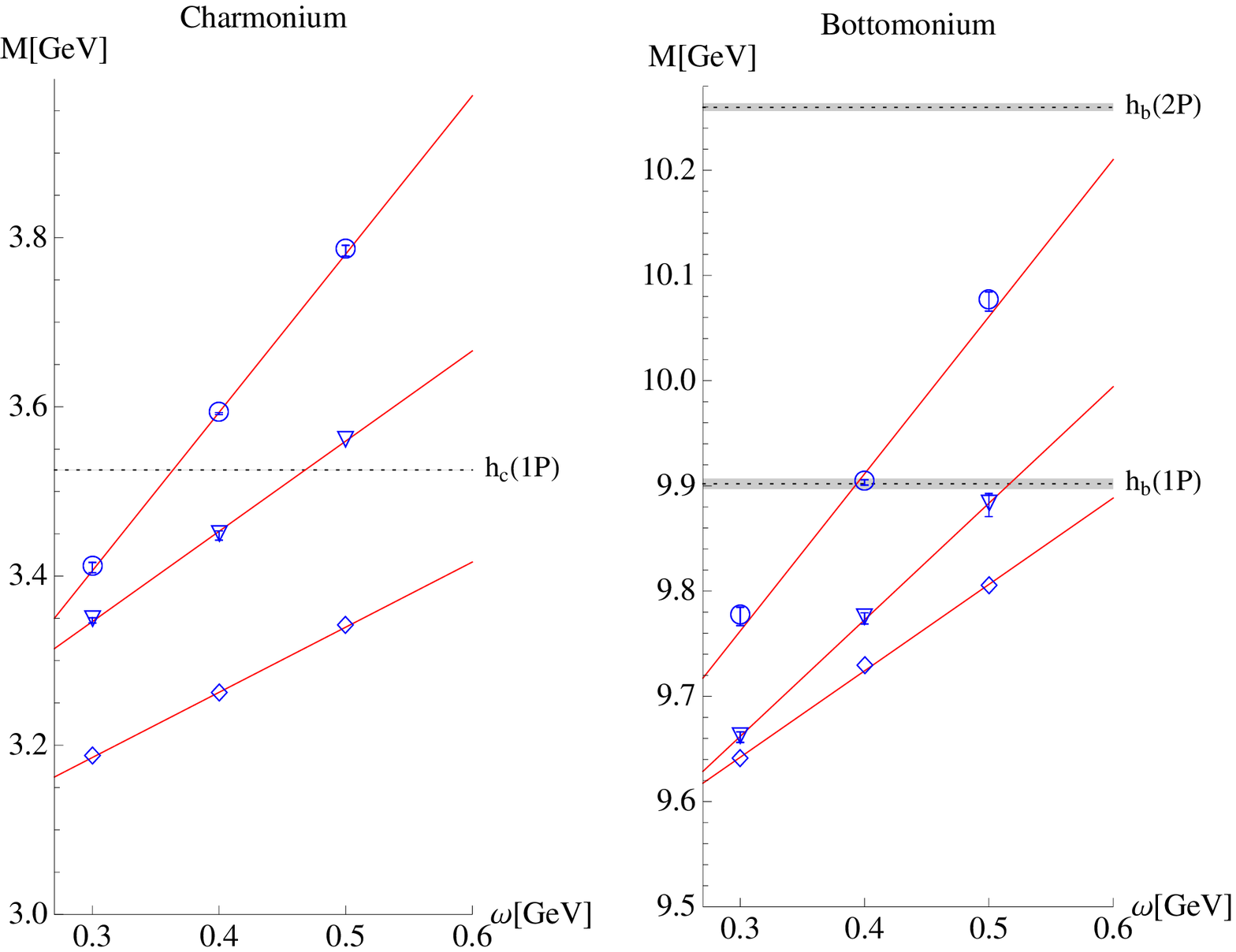}\vspace{-1cm}
\caption[$\omega$ dependence of the ground- and excited-state masses for $J^{PC}=1^{+-}$.]{$J^{PC}=1^{+-}$, see text. The experimental value for the mass of $h_c(1P)$ was taken from Ref.~\cite{Ablikim:2010rc}, and those of $h_b(1P)$ and $h_b(2P)$ from Refs.~\cite{Lees:2011zp} and \cite{Adachi:2011ji}. }
\end{figure}

\clearpage

\backmatter
\chapter{Acknowledgements}
I am greatly indebted to my advisor Andreas Krassnigg, for giving me the opportunity to work on this project, sharing his ideas, and supporting me in any possible way. It was a great pleasure for me to work in such an ideal and stimulating environment, where any question was taken seriously and discussed with great care.

I also want to thank Reinhard Alkofer, for supporting this work and sharing his broad knowledge as well as valuable advice. In addition, via the SIC!QFT research group, he provided the framework that allowed me to develop my own ideas in the context of the newest scientific developments.

In the course of this work, I could collaborate with Sanjin Beni\'c, Davor Horvati\'c, Dubravko Klabu\v{c}ar, Axel Maas, and Valentin Mader, whom I would like to thank for many fruitful discussions and ideas that greatly added to this thesis.

I am grateful to Dubravko Klabu\v{c}ar and Davor Horvati\'c for their support and hospitality during my stay in Zagreb, and to Wolfgang Lucha for his invitation to a short-term visit to the Istitute of High Energy Physics in Vienna, and the accompanying discussions. 

My thanks also go to K.~S.~Choi, J.~Danzer, Y.~Delgado-Mercado, G.~Eichmann, G.~Engel, T.~Herbst, M.~Q.~Huber, C.\ B.~Lang, G.~Lassnigg, K.~Lichtenegger, M.~Limmer, M.~Mitter, D.~Mohler, D.~Nicmorus, C.~Popovici, H.~Sanchis-Alepuz, B.-J.~Schaefer, M.~Schwinzerl, P.~Watson, R.~Williams, and all members of the SIC!QFT research group; they substantially contributed to this work by sharing their knowledge and viewpoints in many discussions. 

During my studies, I could greatly profit from the excellent lectures at the University of Graz, as well as the seminars and workshops provided via the FWF PhD program `Hadrons in vacuum, nuclei, and stars', which allowed me to get in touch with many different aspects of physics beyond my area of research.

In addition, I want to thank Carina Popovici for proof-reading the thesis, and the referee of Ref.~\cite{Blank:2010bp} for greatly improving the quality of the material contained in Secs.~\ref{sec:bsenum} - \ref{sec:hominhomcompare}.

My special thanks go to Klaus Lichtenegger, for long hours of scientific and non-scientific discussions, encouragement, and all the coffee and ideas that we could share.

Last but not least I want to thank my family, for their continuous support during all the stages of this work. In particular, I want to thank my husband Werner for his love and understanding.

This work was financially supported by the Austrian Science Fund \emph{FWF} under project no.\ P20496-N16,  and
was performed in association with and supported in part by the \emph{FWF} doctoral program no.\ W1203-N08, `Hadrons in vacuum, nuclei, and stars'.

\thispagestyle{plain}
\newpage

\phantomsection
\bibliographystyle{phd-doi-new}
\addcontentsline{toc}{chapter}{Bibliography}
\bibliography{had_nucl_graz}
\clearpage
\phantomsection
\addcontentsline{toc}{chapter}{List of figures}
\listoffigures

\end{document}